\documentclass[journal,10pt]{IEEEtran}
\IEEEoverridecommandlockouts
\usepackage[T1]{fontenc}
\usepackage{algorithm}
\usepackage{algpseudocode}
\usepackage{amsmath}
\usepackage{amssymb}
\usepackage{amsfonts}
\usepackage{graphicx}
\usepackage{epsfig}
\usepackage{psfrag}
\usepackage{cite}
\usepackage{latexsym}
\usepackage{lettrine}
\usepackage{url}
\usepackage{color}
\usepackage{caption}
\usepackage{multirow}
\usepackage{subfigure}
\usepackage{yhmath}
\usepackage{booktabs}
\usepackage{threeparttable}
\usepackage{makecell}
\usepackage{bbding}
\usepackage{bm}
\usepackage{indentfirst}

\usepackage{verbatim}
\usepackage{array}
\usepackage{hyperref}					
\hypersetup{colorlinks,
	linkcolor=blue,%
	anchorcolor=blue,
	citecolor=blue}

\makeatletter
\renewcommand{\maketag@@@}[1]{\hbox{\m@th\normalsize\normalfont#1}}%
\makeatother
\PassOptionsToPackage{bookmarks={false}}{hyperref}

\begin{document}
	
	\captionsetup[figure]{name={Fig.},labelsep=period,singlelinecheck=off} 
	
	\title{Integrated Sensing and Communications: \\ Recent Advances and Ten Open Challenges}
	
	\author{Shihang Lu, Fan Liu,~\IEEEmembership{Senior Member,~IEEE}, Yunxin Li, Kecheng Zhang, Hongjia Huang, Jiaqi Zou, Xinyu Li, Yuxiang Dong, Fuwang Dong, Jia Zhu, Yifeng Xiong,~\IEEEmembership{Member,~IEEE}, Weijie Yuan,~\IEEEmembership{Member,~IEEE}, Yuanhao Cui,~\IEEEmembership{Member,~IEEE} and Lajos Hanzo, ~\IEEEmembership{Life Fellow,~IEEE}
		\thanks{(\textit{Corresponding authors: Fan Liu and Yuanhao Cui.})}
		\thanks{S. Lu, F. Liu, Y. Li, K. Zhang, Y. Dong, F. Dong, W. Yuan and Y. Cui are with the School of System Design and Intelligent Manufacturing (SDIM), Southern University of Science and Technology, Shenzhen 518055, China (email: \{lush2021@mail, liuf6, liyx2022@mail, zhangkc2022@mail, dongyx2021@mail\}.sustech.edu.cn; \{dongfw, yuanwj, cuiyh\}@sustech.edu.cn).}
		\thanks{H. Huang is with the Department of Electronic and Information Engineering, The Hong Kong Polytechnic University, Hong Kong 999077, China (e-mail: hongj.huang@connect.polyu.hk).}
		\thanks{X. Li is with the School of Information Science and Engineering, Southeast University, Nanjing 210096, China (e-mail: xinyuli@seu.edu.cn).}
		\thanks{J. Zou, J. Zhu, and Y. Xiong are with the School of Information and Communication Engineering, Beijing University of Posts and Telecommunications, Beijing 100876, China (e-mail: \{jqzou, zhujia, yifengxiong\}@bupt.edu.cn).}
		\thanks{L. Hanzo is with the School of Electronics and Computer Science, University of Southampton, Southampton SO17 1BJ, U.K. (e-mail:lh@ecs.soton.ac.uk).}
	}
	
	\maketitle
	\begin{abstract}
		It is anticipated that integrated sensing and communications (ISAC) would be one of the key enablers of next-generation wireless networks (such as beyond 5G (B5G) and 6G) for supporting a variety of emerging applications. In this paper, we provide a comprehensive review of the recent advances in ISAC systems, with a particular focus on their foundations, physical-layer system design, networking aspects and ISAC applications. Furthermore, we discuss the corresponding open questions of the above that emerged in each issue. Hence, we commence with the information theory of sensing and communications (S$\&$C), followed by the information-theoretic limits of ISAC systems by shedding light on the fundamental performance metrics. Next, we discuss their clock synchronization and phase offset problems, the associated Pareto-optimal signaling strategies, as well as the associated super-resolution physical-layer ISAC system design. Moreover, we envision that ISAC ushers in a paradigm shift for the future cellular networks relying on network sensing, transforming the classic cellular architecture, cross-layer resource management methods, and transmission protocols. In ISAC applications, we further highlight the security and privacy issues of wireless sensing. Finally, we close by studying the recent advances in a representative ISAC use case, namely the multi-object multi-task (MOMT) recognition problem using wireless signals.      
	\end{abstract}
	

	\newcommand{\mv}[1]{\mbox{\boldmath{$ #1 $}}}
	\setlength\abovedisplayskip{6pt}
	\setlength\belowdisplayskip{6pt}
	
	\section{Introduction}
	\subsection{Potential Drive of Integrated Sensing and Communications}
	\IEEEPARstart{G}{iven} the rapid roll-out of the 5G network, numerous beyond-5G (B5G) and 6G concepts have emerged \cite{liufTCOM2020joint,IoTJwang2022integrated,nguyen20216g,leyva2021cooperative,dangshuping20206Gshould,zhuguangxu2022ISCC}. There is a flood of emerging applications as well, such as Internet of Things (IoT) networks \cite{IoTJstankovic2014research,ma2022wideband,liu2022performance,ren2023towards,ma2021covert}, vehicle-to-everything (V2X) communications \cite{Overview_bilik2019rise,overview_OJCOMS_osorio2022towards,liu2020radar,wang2023integrated}, connected autonomous systems \cite{xuiotJ2022wireless}, autonomous driving \cite{overview_zheng2015reliable}, human activity sensing \cite{lixinyu2022humanactivity} as well as smart home and unmanned aerial vehicle (UAV) networks \cite{Overview_wuqingqing2021comprehensive,jing2022efficient} that rely on sensing functionalities, such as wireless localization in support of compelling services \cite{liuliang2022networkedsesning, Overview_fengzhiyong2020joint,sturm2011waveform,hassanien2016signaling,tan2018exploiting,zhengle2019RCC,cui2021integrating}. Consequently, having a sensing functionality is envisioned to become one of the basic services in the B5G/6G networks as the next evolutionary stage beyond the existing communication-only scenarios \cite{Andewzhang2021overview,liu_an2022survey,JSAC_liu2022integrated,zhou2022integrated,mishra2019toward,Overview_akan2020internet,liurang2022integratedris,wang2020thirty}.
	
\begin{figure}[t!]
		\centering
		\epsfxsize=1\linewidth
		\includegraphics[width=\columnwidth]{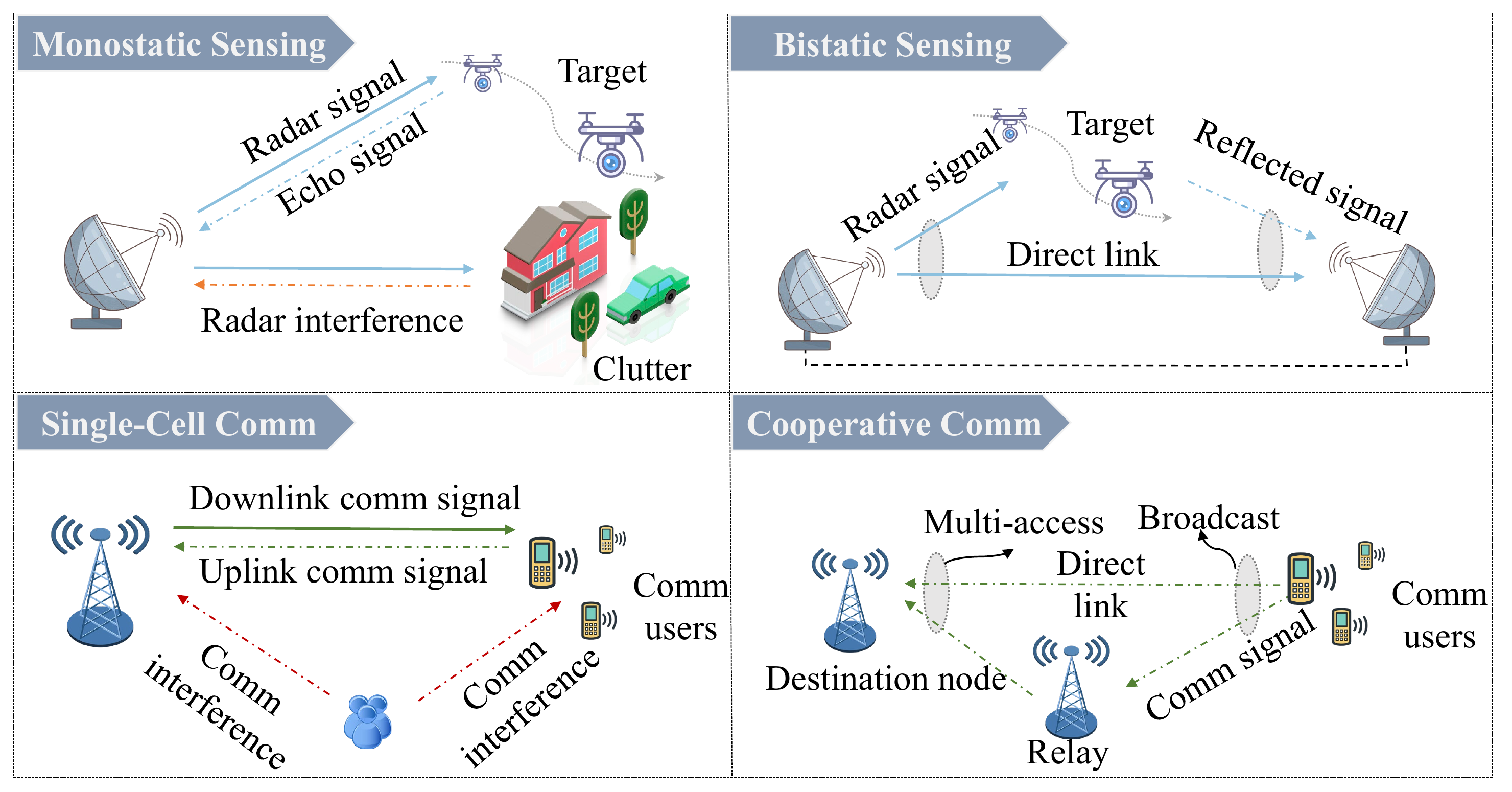}
		\caption{ {The topology of monostatic/bistatic deployments and single-cell/cooperative scenarios.}}
		\label{mono_bi_static}
	\end{figure}

	In the meantime, given the smooth evolution of wireless communication systems, the increasingly congested spectral resources tend to limit the throughput of future wireless systems \cite{saad2019vision,chen2023simultaneous}. As a remedy, the radar bands set aside for sensing can be harnessed as one of the potential alternative future bands. This exciting prospect has promoted the initial integration of spectral resources of radar and communication systems, leading to the concept of radar-communication coexistence (RCC) \cite{zhengle2019RCC}. On the other hand, the millimeter wave (mmWave) and terahertz (THz) bands envisaged for next generation networks could also be exploited for sensing in future cellular networks \cite{du2022integrated,chen2022enhancing,lijie2022indoor,huang2021mimo,sarieddeen2020next,chen2021terahertz}. To this context, as illustrated in Fig. \ref{mono_bi_static}, the generic radar sensing topology can be categorized into {\it monostatic and bistatic} scenarios \cite{yuxianxiang2020mimo,cuiguolong2015signal,zhangtong2022accelerating}, which are similar to the single-cell scenario \cite{zhougui2021rate,maoyijie2022rate} and cooperative communication scenarios of wireless systems \cite{chenxiaoming2014cooperative,liu2022performance}, respectively. 
	
	\begin{figure*}[ht]
		\centering
		\epsfxsize=1\linewidth
		\includegraphics[width=\textwidth]{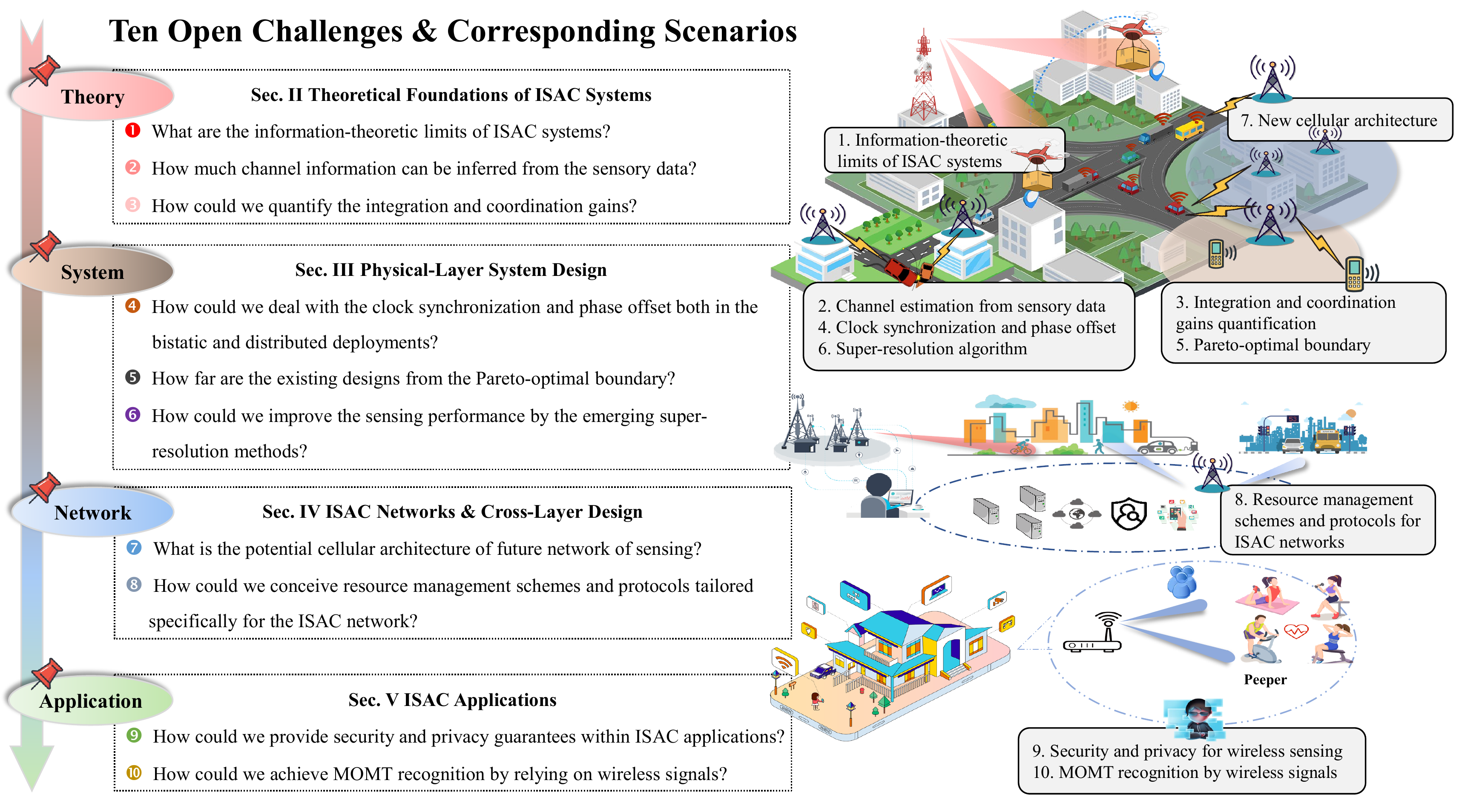}
		\caption{ {Structure of this paper and the relevant ISAC scenarios.}}
		\label{PaperFramework}
	\end{figure*}
	
	As a benefit of intense research, sensing and communications (S\&C) tend to become more integrated both in terms of their hardware architecture and signal processing algorithms, making the co-design of these systems more realistic, even though in the past they have been considered as two isolated fields \cite{liu2022seventy}. Given the shared use of both the hardware and wireless resources, radar and communication systems evolve towards integrated sensing and communications (ISAC) systems \cite{cui2021integrating,liu_an2022survey,JSAC_liu2022integrated}. In other words, the joint exploitation of the limited hardware, spectral as well as energy resources, results in a beneficial {\it integration gain} \cite{cui2021integrating}. Moreover, sharing resources between S$\&$C leads to the compelling concept of sensing-assisted communications \cite{liu2020radar}, thus offering an attractive {\it coordination gain}. Inspired by the aforementioned benefits of integration, sensing-assisted communication and communication-assisted sensing have attracted tremendous research attention in recent years \cite{liurang2021dual, 8386661,2021chenli_Pareto,liuxiang2020joint,Overview_huang2020majorcom,yuxianxiangTSP2022integrated,xiaozhiqiangSI2022waveform,xu2021rate,qian2023enhancement}.
	
	 {As illustrated in Fig. \ref{PaperFramework}, recent advances of ISAC facilitate the support of emerging IoT scenarios, such as smart cities and smart homes \cite{JSAC_liu2022integrated,IoTJwang2022integrated,lixiaoyang2022over}. Moreover, embracing ISAC leads to a transformative shift in IoT architectures, elucidated in the following two aspects..}

	 {
		\textbf{1) Leveraging ISAC equips IoT devices with the ability to ``perceive and connect'' with the world.} This transformation enables IoT devices to not only sense their environment but also to establish connections with other devices and networks, facilitating seamless data sharing and decision-making \cite{nguyen20216g}. As IoT ecosystems continue to evolve, the ability to ``perceive and connect'' becomes a critical enabler for the successful implementation of IoT applications \cite{JSAC_liu2022integrated}. To be more specific, the resilient sensing capacity of the distributed IoT nodes plays a vital role in improving the performance of networked sensing. Subsequently, networked collaborative IoT nodes may be flexibly deployed to enhance environmental sensing capabilities through decentralized transmission and processing \cite{cui2021integrating}. The multiple-source data fusion significantly reduces sensing inaccuracies, thereby elevating the efficiency of distributed sensing. To this end, sensing assignments can be effectively distributed across multiple nodes, and the exchange of ``sensing-with-communication'' findings promotes efforts to reach a consensus regarding the surrounding environment \cite{weizhiqing_2023}. Finally, IoT devices are equipped with the ability to ``perceive and connect'' with the physical world within ISAC modes \cite{cui2021integrating}.}
	
	 {
		\textbf{2) Utilizing ISAC meets the diverse demands of IoT scenarios to support future networks.} In future networks supporting vertical industries, a multitude of IoT devices like sensors and robots will be prevalent \cite{IoTJstankovic2014research}. These devices rely on precise sensing and positioning information to make decisions and enable intelligent operations and management \cite{liu2022performance}. In a large-scale connected sensing architecture, it is essential not only to sense and detect the physical environment information of target nodes but also to integrate this sensory data with digital information \cite{zhuguangxu2022ISCC}. Traditional wireless sensor networks may struggle to meet this demand. Fortunately, ISAC allows the convergence of sensing and communication functions by the unified signals, addressing the limitations of conventional wireless sensor networks and achieving a tight coupling between the physical and digital realms of network nodes \cite{liu_an2022survey,Andewzhang2021overview,JSAC_liu2022integrated,cui2021integrating}.
	}
	
	 { Note that the traditional IoT data processing flow falls in the ``communication-after-sensing'' approach, while ISAC-enabled IoT devices can sense environmental data and reduce data-exchange time delays as well as the processing cost, resulting in ``sensing-with-communication'' \cite{cui2021integrating}. In large-scale connected IoT scenarios, ISAC-enabled IoT devices surpass the limitations of traditional wireless sensor networks, tightly coupling the physical and digital aspects of IoT nodes \cite{weizhiqing_2023,xuiotJ2022wireless,jing2022efficient,ma2022wideband}. However, numerous ISAC challenges systems still exist, as detailed below.}

	\subsection{Ten Open Challenges in ISAC Systems}
	 {Recently, ISAC has garnered recognition from the International Telecommunication Union (ITU) as a pivotal element within the 6G landscape, serving as one of its six key usage scenarios \cite{ITU2023}. This acknowledgment underscores the growing significance of ISAC for the evolution of next-generation wireless systems.}
	Although substantial advances have been made, ranging from determining the degrees of freedom tradeoffs, the fundamental limits, coding design, resource allocation as well as signaling strategies of ISAC systems \cite{lu2022degrees,liu2017adaptiveOFDM,pezeshki2008doppler,xiong2022flowing,c1,shi2019joint,wangxinyi2022partially,liuyongjun2017adaptive,mengwcl2022uav,mengTWC2022multi,mengkaitao2022throughput}, there still remains a host of open questions in the fields of the theoretical foundation of ISAC, physical-layer system design, ISAC networking and ISAC applications, which motivates the conception of this paper. 
	
	\begin{table*}[ht]
		\caption{Recent advances and ten open questions in ISAC systems.} 
		\label{Framework}
		\resizebox{\textwidth}{!}{
			\begin{threeparttable}
				\begin{tabular}{|l|l|l|c|}
					\specialrule{0em}{2pt}{2pt} 
					\hline \hline
					\textbf{\qquad \qquad \qquad \qquad Open Challenges}                                                                 & \multicolumn{1}{c|}{ \textbf{Recent Advances and  Efforts}}                                                                  & \multicolumn{1}{c|}{ \textbf{Future Directions \& Potential Solutions}} 
					& \multicolumn{1}{c|}{\textbf{ Scenarios}}                                                                                                                                         \\ \hline
					\begin{tabular}[l]{@{}c@{}} What are the information-theoretic limits of ISAC systems? \end{tabular}               
					& \begin{tabular}[c]{@{}l@{}}
						$\bullet$ MMSE and MI for radar sensing \cite{MMSE_MIyang2007mimo,MI_Sensing_tang2010mimo}\\
						$\bullet$ Wireless localization \cite{YuanShen2010}\\ 
						$\bullet$ I-MMSE equation \cite{IMMSE_guo2005mutual} \\ 
						$\bullet$ Distortion-capacity \cite{kobayashi2018joint,liuyao2022generalized}\\ 
						$\bullet$ CRB-rate region \cite{xiong2022flowing,hua2022mimo,CRBrate_ren2022fundamental} \\
						$\bullet$ MI-based ISAC framework \cite{ouyang2022performance,ouyang2022integrated,MI_Sensing_yuan2020spatio,lijinMI2022framework}\\
						$\bullet$ Joint communication and binary state\\~~ detection \cite{KLD_joudeh2022joint}
					\end{tabular}                                                       & \begin{tabular}[c]{@{}l@{}}
						$\bullet$ The first principle in ISAC information \\~~~theory\\ 
						$\bullet$ Proper strategy and coding approach \\ 
						$\bullet$ KLD measure vs. mutual information\\~~tradeoff\\ $\bullet$ Ergodic CRB and PCRB
					\end{tabular} 
					& \begin{tabular}{c|c}
						\\ \\  \\    {$\boldsymbol{\odot}$} 
						&  {$\bigstar$}  \\  
						 {$\boldsymbol{\clubsuit}$}
						\\ \\ \\ \\
					\end{tabular} 
					\\ \hline
					\begin{tabular}[l]{@{}l@{}}How much channel information can be inferred from the sensory data? \end{tabular}             & \begin{tabular}[c]{@{}l@{}}
						$\bullet$ OTFS vs. OFDM in ISAC systems \cite{yuan2021integrated,wukai2022integrating}\\ 
						$\bullet$ EKF and deep learning methods \cite{liu2020radar,huang2021mimo} \\ 
						$\bullet$ Two-stage channel estimation protocol \cite{huang2022joint} \\
						$\bullet$ Real-world S$\&$C dataset \cite{demirhan2022radar,OJCOMS_wu2022blockage,alkhateeb2021deepsense}
					\end{tabular}                        
					& \begin{tabular}[c]{@{}l@{}}
						$\bullet$ The relationship between S$\&$C\\~~channels\\ 
						$\bullet$ CSI inferred from the sensory data  \\ 
						$\bullet$ Mutual information method and factor\\~~~graph method\end{tabular}
					& \begin{tabular}{c|c}
						\\  {$\boldsymbol{\odot}$} &  {$\bigstar$} \\ 
						 {$\boldsymbol{\clubsuit}$} & { \scalebox{1.3}{$\blacktriangle$}} \\
						\\ \\
					\end{tabular} 
					\\ \hline
					\begin{tabular}[l]{@{}l@{}}How could we quantify the integration and coordination gains?\end{tabular}       
					& \begin{tabular}[c]{@{}l@{}}
						$\bullet$ Communication-assisted sensing \cite{cui2021integrating}  \\
						$\bullet$ Sensing-assisted communications \cite{yuan2021integrated,liu2020radar,huang2021mimo}\\
						$\bullet$ Performance gain and S$\&$C subspaces \cite{lu2022performance}
					\end{tabular}                          
					
					& \begin{tabular}[c]{@{}l@{}}
						$\bullet$ Proper measures for integration and \\~~coordination gains\\ $\bullet$ Geometric measure from the perfor-\\~~mance region\end{tabular} 
					& \begin{tabular}{c|c}
						\\   {$\boldsymbol{\odot}$} &  { $\bigstar$} \\ 
						 {$\boldsymbol{\clubsuit}$} \\ \\
					\end{tabular}                                                      
					\\ \hline
					\begin{tabular}[l]{@{}l@{}} How could we deal with the clock synchronization  and phase \\  offset both in the bistatic and distributed deployments? \end{tabular}                                                             
					& \begin{tabular}[c]{@{}l@{}}
						$\bullet$ GPS-based time $\&$ phase synchronization \cite{wang2009gps}\\
						$\bullet$ Reference clock, compensation and etc \cite{zhang2022integration} \\
						$\bullet$ Ultra-wideband technique  \cite{xue2021wicsync,perez2022time} \end{tabular}                                     
					
					& \begin{tabular}[c]{@{}l@{}}
						$\bullet$ Further reduction of clock synchro-\\~~nization errors\\
						$\bullet$ Security of time information in signals \end{tabular} 
					& \begin{tabular}{c|c}
						\\  {$\boldsymbol{\clubsuit}$} &
						 {\scalebox{1.3}{$\blacktriangle$}} \\
						&  {$\blacksquare$} \\
					\end{tabular}                                                                                        \\ \hline
					\begin{tabular}[c]{@{}c@{}}How far are the existing designs from the Pareto-optimal boundary?  \end{tabular}                   & \begin{tabular}[c]{@{}l@{}}
						$\bullet$ Pulse interval modulation method \cite{mealey1963method}\\ 
						$\bullet$ OFDM used for target detection \cite{sturm2011waveform}\\ 
						$\bullet$ Waveform Design \cite{liuxiang2020joint, liurang2021dual,8386661,2021chenli_Pareto}\end{tabular}                                          
					
					& \begin{tabular}[c]{@{}l@{}}
						$\bullet$ The essential tradeoff in ISAC\\ 
						$\bullet$ Pareto boundary-based the unified\\~~waveform\end{tabular}
					& \begin{tabular}{c|c}
						\\  {$\boldsymbol{\odot}$} & {  
							\scalebox{1.3}{$\blacktriangle$}} \\ \\ 
					\end{tabular}                                                       \\ \hline
					\begin{tabular}[l]{@{}l@{}}How could we improve the sensing performance by the emerging \\ super-resolution methods? \end{tabular}              
					& \begin{tabular}[c]{@{}l@{}}
						$\bullet$ CA technique for angle estimaton \cite{CA2014,CA2021} \\ 
						$\bullet$ Sparse array methods and algorithms \cite{Nested2010,Coprime2015,ni2023waveform} \\
						$\bullet$ Sparse signal reconstruction \cite{Ongrid2005,Offgrid2011,Gridless2015} \\
						$\bullet$ Non uniform sparse arrays \cite{sarangi2022super,sarangi2023super}
					\end{tabular}                                                         
					& \begin{tabular}[c]{@{}l@{}}
						$\bullet$ Carrier aggregation assisted sensing\\ 
						$\bullet$ Bistatic radar sensing\\ 
						$\bullet$ Other super-resolution algorithms\end{tabular} 
					& \begin{tabular}{c|c}
						\\  {$\boldsymbol{\odot}$} &  {\scalebox{1.3}{$\blacktriangle$}} \\
						&  {$\blacksquare$} \\ \\
					\end{tabular}                                                                        \\ \hline
					\begin{tabular}[l]{@{}c@{}}What is the potential cellular architecture of future network of sensing? \end{tabular}             
					& \begin{tabular}[c]{@{}l@{}}
						$\bullet$ Soft fractional frequency reuse and etc \cite{7096298}\\
						$\bullet$ User-centric C-RAN  \cite{pan2018user}\\ 
						$\bullet$ Cell-free for wireless network \cite{ngo2017cell,bjornson2013optimal}\end{tabular} 
					& \begin{tabular}[c]{@{}l@{}}
						$\bullet$ Cell-free architecture for ISAC network\\ 
						$\bullet$ Multi-station multi-cell cooperation \\~~scenario \end{tabular}                                                          & \begin{tabular}{c|c}
						\\  {$\boldsymbol{\clubsuit}$} &
						 {\scalebox{1.3}{$\blacktriangle$}} \\
						\\
					\end{tabular}                                                  \\ \hline 
					\begin{tabular}[l]{@{}l@{}}How could we conceive resource management schemes and protocols \\ specifically tailored for the ISAC network?\end{tabular} 
					
					& \begin{tabular}[c]{@{}l@{}}
						$\bullet$ PHY layer resource allocation in ISAC \cite{c1}\\ $\bullet$ Cellular cross-layer optimization \cite{c2, c3, c4, c6} \end{tabular}                      
					& \begin{tabular}[c]{@{}l@{}}
						$\bullet$ Channel access algorithm in MAC\\~~ layer \\
						$\bullet$ Queuing theory in Network layer \\
						$\bullet$ Adaptive source data encoding in \\~~Application layer\end{tabular}                                                         & \begin{tabular}{c|c}
						\\ \\  {\scalebox{0.95}{$\boldsymbol{\clubsuit}$}} & 
						 {\scalebox{1.2}{$\blacksquare$}} \\ \\ \\
					\end{tabular} 
					\\  \hline
					\begin{tabular}[l]{@{}l@{}} {How could we provide the security and privacy guarantees within  ISAC } \\  {applications?} \end{tabular}    
					& \begin{tabular}[c]{@{}l@{}}
						$\bullet$ Secure ISAC \cite{Onur2022,Onur2023}  \\
						$\bullet$ Radar privacy protection in shared spectrum\\ scenarios \cite{dimas2017spectrum,al2019mimo} \\
						$\bullet$ WiFi sensing security \cite{jiao2021openwifi} 
						\\ $\bullet$ Optimal pilots for anti-eavesdropping \cite{Channel_Estimation_zhu2020optimal} \\
						$\bullet$ Actual face protection by Doppler radar \cite{wang2022deeplearning_gender}
					\end{tabular}                                                                             & \begin{tabular}[c]{@{}l@{}}
						$\bullet$ Information theory perspective\\ 
						$\bullet$ Leakage reduce in PHY layer\\ 
						$\bullet$ Access control in MAC layer\end{tabular}                                                                       & \begin{tabular}{c|c}
						\\ \\  {$\boldsymbol{\odot}$} 
						&  {$\bigstar$} \\ 
						 {$\boldsymbol{\clubsuit}$}  &
						 {\scalebox{1.3}{$\blacktriangle$}} \\ \\ \\
					\end{tabular}
					\\  \hline
					\begin{tabular}[l]{@{}c@{}}How could we achieve MOMT recognition by relying on wireless signals?\end{tabular}       
					& \begin{tabular}[c]{@{}l@{}}
						$\bullet$ Human identify and recognition \cite{li2019survey,zhao2021human}\\ 
						$\bullet$ Radar- and WiFi-based sensing \\ \cite{qian2017widar,zhang2022integration,wu2021wifi,chen2018wifi} \end{tabular}    & \begin{tabular}[c]{@{}l@{}}
						$\bullet$ Improvement on the spatial resolution \\~~of wireless sensing\\ 
						$\bullet$ Multi-object-multi-task sensing\end{tabular}                                                             & \begin{tabular}{c|c}
						\\  {$\boldsymbol{\odot}$} &  {\scalebox{1.3}{$\blacktriangle$}} \\
						 {$\boldsymbol{\clubsuit}$} &
						 {$\blacksquare$} \\ 
					\end{tabular} 
					\\  \hline 
				\end{tabular} 
				\begin{tablenotes}    
					\item[1] In order to clearly indicate the above questions may arise in which kinds of ISAC scenarios, we artificially categorize them into two generic types, e.g., $\boldsymbol{\odot}:$ monostatic deployment and $\boldsymbol{\clubsuit} :$  bistatic /distributed deployments.          
					\item[2] In order to qualitatively express the above questions may fall into which focus of future works, we artificially categorize them into three types, e.g., $\bigstar :$ fundamental and theoretical investigation, \scalebox{1.3}{$\blacktriangle$}: signal processing algorithm and $\blacksquare:$ practical system design.  
				\end{tablenotes}            
		\end{threeparttable} }
	\end{table*}
	
	\begin{table*}[ht]
		\renewcommand\arraystretch{1.4}
		\centering
		\caption{Existing overview papers on ISAC}
		\resizebox{\textwidth}{!}{
			\begin{tabular}{|l|c|c|c|c|c|c|c|c|c|}
				\hline \hline Existing Works & {\cite{sturm2011waveform}} & {\cite{hassanien2016signaling}} &  {\cite{tan2018exploiting}} & {\cite{zhengle2019RCC}}& \cite{cui2021integrating} & {\cite{Andewzhang2021overview}} & {\cite{liu_an2022survey}} & {\cite{JSAC_liu2022integrated}} & This paper \\
				\hline Year & 2011 & 2016 & 2018 & 2019 & 2021 & 2021 & 2022 & 2022 & 2023 \\
				\hline Type & Tutorial & Tutorial & Overview & Survey & Survey & Overview & Survey & Survey & Survey \\
				\hline  Information-theoretic
				limits &  &  &  &  &  & $\checkmark$ & $\checkmark$ & $\checkmark$ & \CheckmarkBold \\
				\hline {Channel estimation from sensory data} &  &  & $\checkmark$ &  & & & & & \CheckmarkBold \\
				\hline {Integration $\&$ coordination
					gains} & &  & & & $\checkmark$ & & &  & \CheckmarkBold \\
				\hline {Clock synchronization $\&$ phase offset} & & & & &  & $\checkmark$ &  &  & \CheckmarkBold \\
				\hline Pareto-optimal boundary &  & & $\checkmark$ & $\checkmark$ & $\checkmark$ &  &$\checkmark$ & $\checkmark$ & \CheckmarkBold \\
				\hline Super-resolution algorithm & & & & & &$\checkmark$ &  & & \CheckmarkBold \\
				\hline {New cellular architecture} & & & & & & & & $\checkmark$ & \CheckmarkBold\\
				\hline  Resource management
				schemes $\&$ protocols  & & & & & & & $\checkmark$ &$\checkmark$ & \CheckmarkBold \\
				\hline {Security and privacy in wireless sensing} & & & & & & & & & \CheckmarkBold \\
				\hline MOMT by wireless signals & $\checkmark$ & $\checkmark$ & $\checkmark$ & $\checkmark$ & & & & $\checkmark$ & \CheckmarkBold \\
				\hline
		\end{tabular}}
		\label{Comparison}
	\end{table*}
	
	First of all, the fundamental theories of radar sensing and wireless communication are challenging to be unified under the joint concept of ISAC systems, since radar and wireless communication systems glean different information from the received signals \cite{YuanShen2010, IMMSE_guo2005mutual, kobayashi2018joint, liuyao2022generalized, xiong2022flowing}. Specifically, radar systems focus on how to minimize the uncertainty concerning the target environments based on the received echo signals. By contrast, communication systems concentrate on how to minimize the uncertainty related to the transmitted random signals in order to recover useful information  \cite{hua2022mimo,CRBrate_ren2022fundamental,ouyang2022performance,ouyang2022integrated}. In this spirit, it is of pivotal significance to resolve what the prime objective of ISAC systems is from an information-theoretic perspective. This will provide reliable theoretical guidance for the design of ISAC systems. Therefore we commence with the salient question \textbf{Challenge 1) What are the information-theoretic limits of ISAC systems?} Then we will discuss the intricate relationships among those limitations. The ultimate long-term objective would be to formulate pertinent multi-component objective functions unifying both communications and radar metrics and finding the optimal non-dominated Pareto front \cite{wang2020thirty}.
	Moreover, the sensory data might also be used for estimating the channel information to reduce the pilot overhead so that the ISAC systems can harvest sensing-assisted performance gains for communications \cite{du2022integrated,yuan2021integrated,huang2022joint}. As a step further, we attempt to ask and answer the question \textbf{Challenge 2) How much channel information can be inferred from the sensory data?} Finally, to quantify the mutual performance gains of ISAC systems \cite{lu2022performance}, we conclude the theoretical foundation of ISAC systems part by discussing the question \textbf{Challenge 3) How could we quantify the integration and coordination gains?}
	
	Secondly, towards system design, typical methods originally conceived for cellular networks might be adapted to the needs of ISAC systems, as exemplified by service requests and resource scheduling in the physical (PHY) layer. Nevertheless, when the sensing functionality is integrated into wireless systems as an extra basic service \cite{wukai2022integrating,c1}, the need for providing tangible sensing performance guarantees brings about several emerging challenges. For example, the synchronization requirement of sensing is significantly tighter than for communications, because the resultant timing offset may inflict interference upon the sensing functionality. To avoid phase offset and hence provide high-quality sensing services (e.g., super-resolution sensing), ISAC systems require precise clock synchronization between the transmitter and receiver \cite{zhang2022integration,xue2021wicsync,perez2022time}. Furthermore, with the potential integration of S$\&$C, both sensing-centric as well as communication-centric and joint designs have been proposed for ISAC signal processing \cite{JSAC_liu2022integrated}, which can be harnessed for striking a flexible performance tradeoff between S$\&$C. Even though numerous tradeoff designs have been proposed for maximizing the communication (or sensing) performance subject to sensing (or communication) requirements \cite{lu2022degrees,liu2020radar,liurang2021dual, huang2021mimo,liuxiang2020joint}, it remains unclear, what the Pareto-optimal signaling strategy is. Moreover, for considering radar sensing as an inherent service in future networks, it is quite important to improve the sensing resolution to meet the requirements of emerging applications such as V2X and IoT \cite{liu_an2022survey,JSAC_liu2022integrated}. As a result, the corresponding ISAC system design faces the following open questions: \textbf{Challenge 4) How could we deal with the clock synchronization and phase offset both in the bistatic and distributed deployments? Challenge 5) How far are the existing designs from the Pareto-optimal boundary? Challenge 6) How could we improve the sensing performance by exploiting the emerging super-resolution methods?}
	
	Thirdly, we note that the existing cellular architecture and the classic resource management schemes have to be fine-tuned in support of ISAC networks \cite{liu_an2022survey,cui2021integrating}. More specifically, the classical communication-only cellular architecture treats inter-cell interference as a harmful factor that has to be reduced. However, the seamless sensing service of the future might require a rich set of echo signals to fully reconstruct the surrounding environments. In this context, the inter-cell signals are considered as sensing-friendly signal contributions rather than as hostile interference. This motivates the conception of novel schemes reminiscent of the popular cooperative multi-cell processing (CoMP) concept for ISAC networks. Furthermore, resource management constitutes a major challenge for efficiently responding to sudden sensing requests. Accordingly, one has to address the following pair of open challenges in ISAC networks: \textbf{Challenge 7) What is the potential cellular architecture of the future network of sensing? Challenge 8) How could we conceive resource management schemes and protocols specifically tailored for the ISAC network?}
	
	Finally, it is worth pointing out that ISAC systems and networks may not be implemented until the security and privacy issues of sensing have been addressed \cite{li2021semisupervised, di2018wifi,lixinyu2022humanactivity}. For example, in human action sensing scenarios \cite{zeng2020multisense}, the radar/WiFi-based sensing \cite{jiao2021openwifi} technique tends to incur a potential illegitimate interception of a target, due to the open nature of the wireless sensing medium \cite{furqan2021wireless,overview_OJCOMS_osorio2022towards}. Moreover, the CSI might contain certain information of private nature concerning both the transmitters and targets \cite{zhang2022practical}, which has to be settled to avoid being eavesdropped upon simultaneously. On the other hand, as a typical use case in ISAC applications, multi-object multi-task (MOMT) recognition aims to identify multiple targets and recognize their behaviors simultaneously. Substantial efforts have been dedicated to the MOMT recognition in wireless human sensing relying on WiFi signals \cite{gao2021towards,yu2021wifi}, albeit with more emphasis on single-person sensing. Clearly, the more challenging high-accuracy multi-person sensing problem relying on wireless signals still remains open, especially when the reflected echoes are buried in clutter and interference. Based on the above brief discussions, we pose the final pair of cardinal questions: \textbf{ {Challenge 9) How could we provide security and privacy guarantees within ISAC applications?} Challenge 10) How could we achieve MOMT recognition by relying on wireless signals?}
	
	\subsection{Existing Efforts and the Scope of This Paper} 
	 {
		In this paper, we critically appraise the recent advances and formulate ten open challenges in ISAC systems, some of which have already had some initial progress, while others are still in the exploratory phase. Following the narrative of ``Theory-System-Network-Application'', we summarize the structure of this paper as well as the aforementioned open questions and the relevant ISAC scenarios in Fig. \ref{PaperFramework}. As a benefit of concerted community effort, the ISAC philosophy has evolved from a compelling theoretical concept to a practical engineering challenge \cite{YuanShen2010, IMMSE_guo2005mutual, kobayashi2018joint, liuyao2022generalized, xiong2022flowing,c1,sturm2011waveform,8114253,8386661,zhang2022integration,jiao2021openwifi,li2019survey, zeng2020multisense,alkhateeb2021deepsense,wang2019first_demonstration, niu2022rethinking, Model_zhang2022joint}. To further pave the way for its successful evaluation, we critically appraise the recent advances and summarize the above ten questions in TABLE \ref{Framework}.}
	
	 {
		Indeed, there have been several pioneering tutorial/overview/survey papers on ISAC-related topics throughout the recent decade, e.g., \cite{sturm2011waveform} on the intelligent design of dual-functional waveforms, \cite{hassanien2016signaling} on radar signals embedded into communication signals, \cite{tan2018exploiting} on WiFi-based residential healthcare sensing, \cite{zhengle2019RCC} on radar and communication coexistence, \cite{cui2021integrating} on ISAC for IoT scenarios, \cite{Andewzhang2021overview} on signal processing techniques conceived for joint communication and radar sensing, \cite{liu_an2022survey} on the fundamental limits of ISAC as well as \cite{JSAC_liu2022integrated} on the ISAC concepts proposed for 6G and beyond. In contrast to previous works on the specifics of ISAC fundamental limits, IoT applications or dual-functional wireless networks, this paper adopts a broader perspective on recent advances as well as open questions in ISAC and serves as a complement of the existing efforts \cite{cui2021integrating,JSAC_liu2022integrated, sturm2011waveform, hassanien2016signaling, tan2018exploiting,Overview_fengzhiyong2020joint,liu_an2022survey,zhengle2019RCC,zhou2022integrated,mishra2019toward, Andewzhang2021overview, Overview_akan2020internet}. The authors of \cite{liu_an2022survey} have discussed the fundamental limits of ISAC commencing from both device-free and device-based sensing scenarios, followed by the joint-design-based ISAC and practical ISAC systems. Moreover, other existing works such as \cite{zhou2022integrated} and \cite{mishra2019toward} have discussed the waveform design by categorizing it into sensing-centric, communication-centric, and joint design methods. Moreover, the authors of [29-31] generally considered three categories, namely sensing-only, communication-only, and ISAC scenarios. In contrast to this, we aim to provide another holistic perspective spanning from fundamental theory to various applications of ISAC by categorizing the proposed open challenges into four parts, as illustrated in Fig. \ref{PaperFramework}. For clarity, we provide a detailed comparison between the existing contributions and this paper in terms of the aforementioned ten open questions in TABLE \ref{Comparison}, followed by a detailed discussion in the remainder of the paper.} 
	
	\begin{table}[htbp]
		\caption{List of Abbreviations} \label{Abbre} 
		\normalsize
		\begin{tabular}{p{1.8cm}|p{6.2cm}} 
			\toprule
			Abbreviation & Definition\\
			\hline
			ISAC & Integrated sensing and communications\\
			B5G & Beyond 5G\\
			S\&C & Sensing and communications\\
			MOMT & Multi-object multi-task\\
			V2X & Vehicle-to-everything\\
			UAV & Unmanned aerial vehicle\\
			IoT & Internet of things\\
			MI & Mutual information\\
			KLD & Kullback-Leibler Divergence\\
			CRB &  Cram\'{e}r-Rao bound\\
			PCRB/BCRB & posterior/Bayesian CRB \\
			CIR/TIR & Channel/target impulse response\\
			AWGN & Additive white Gaussian noise\\
			PDF & Probability density functions \\
			SPEB & Squared position error bound\\
			MMSE & Minimum mean squared error\\
			SNR & Signal-to-noise ratio\\
			FIM & Fisher information matrix\\
			BS & Base station\\
			CSI & Channel state information\\
			V2I & Vehicle-to-infrastructure  \\
			MIMO & Multiple-input-multiple output\\
			EKF & Extended Kalman Filter\\
			OTFS & Orthogonal time frequency space\\
			OFDM & Orthogonal frequency division multiplexing\\
			TMO & Timing offset\\
			CFO & Carrier frequency offset\\
			TOA & Time-of-arrival\\
			TDOA & Time-difference-of-arrival\\
			GPS & Global positioning system\\
			UWB & Ultra-wideband\\
			PIM & Pulse interval modulation\\
			ASK/PSK & Amplitude/phase shift keying\\
			LFM & Linear frequency modulation\\
			SINR & Signal-to-interference plus noise ratio\\
			MSE & Mean square error\\
			DOA & Direction of arrival\\
			OF & Objective functions \\
			BER & Bit error rate\\
			CA & Carrier aggregation\\
			ULA & Uniform linear array\\
			RAN & Radio access networks\\
			C-RAN & Cloud RAN\\
			RRHs & Remote radio heads\\
			APs & Access points\\
			OSI & Open system interconnections\\
			MAC & Media access control\\
			RTS & Request to send\\
			CTS & Clear to send\\
			QoS & Quality of service\\
			DNN & Deep neural network\\
			FMCW & Frequency modulation wave\\
			PRI & Pulse repetition interval \\
			FFT & Fast Fourier transform\\
			3D & 3-dimensional\\
			\hline
		\end{tabular}
	\end{table}
	
	\vspace{-1em}
	\subsection{Organization of This Paper}
	The rest of this paper is organized into five sections as seen in Fig. \ref{PaperFramework}. In Section \ref{IF}, we introduce the basic performance metrics for both S$\&$C, discuss the associated channel estimation issues and provide a metric for quantifying the integration gain and coordination gain. Section \ref{System_Design} discusses clock synchronization, Pareto-optimal signaling strategies, and super-resolution sensing in the context of physical-layer system design. In Section \ref{ISAC_Network}, we briefly discuss the potential cell architectures and cross-layer protocols of ISAC networks. In Section \ref{Security_Wifi}, we shed light on sensing security and privacy issues as well as MOMT, while relying on wireless signals. Finally, Section \ref{Conclusion} concludes the paper. For crisp and convenient clarity, we organize all sections into three subsections, namely Background, Existing Literature as well as Future Directions and Potential Solutions. The related abbreviations of this paper are given in TABLE \ref{Abbre}.

	\section{Theoretical Foundations of ISAC Systems}\label{IF}
	In this section, we explore \textbf{\emph{Challenge 1-3}} of Fig. \ref{PaperFramework}. we first shed light on a range of measures proposed for ISAC systems from an information-theoretic perspective. Then, we highlight the associated channel overlap between S$\&$C and discuss how much channel information can be inferred from the sensory data.  {Finally, we propose a {\it ``simple but intuitive''} metric to grasp for assessing the integration and coordination gains based on the three categories of S\&C channels, namely, uncorrelated, moderately correlated, and strongly correlated scenarios. This metric is informally defined as the ratio of the area in the mutual-boost region for moderately correlated S\&C scenarios to the area in the non-boost region for uncorrelated S\&C scenarios.}

	\vspace{1em}
	\noindent
	\textbf{\emph{Challenge 1: What Are the Information-Theoretic Limits of ISAC Systems?}}
	
	{\it \textbf{1) Background:}} Information theory is critical and fundamental for evaluating the performance limits of ISAC systems \cite{JSAC_liu2022integrated,liu_an2022survey}. Although the integration of S$\&$C is indeed promising as a benefit of utilizing a common hardware platform and a common transmitted signal, their basic information theory has both connections and distinctions. For example, both S$\&$C systems focus on the mutual information (MI) maximization. Specifically, one might always try to increase the achievable rate or ergodic rate by maximizing the MI between the transmitted and received signals in wireless communication systems. The sensing tasks, on the other hand, mainly rely on estimation and detection. As for parameter estimation, maximizing the MI between the target impulse response and received echo signals usually leads to minimizing the minimum mean-square error (MMSE) \cite{MMSE_MIyang2007mimo,IMMSE_guo2005mutual}. For target detection, maximizing the Kullback-Leibler divergence (KLD) between the pair of scenarios when the target is present and absent asymptotically leads to a problem reminiscent of detection probability maximization \cite{kailath1967divergence,tangbo2015relative,thomas2006elements}.
	
	In what follows, we elaborate on the family of MI-oriented ISAC systems, and then we introduce other metrics for quantifying the sensing performance, such as the KLD and the  {Cram\'{e}r-Rao bound (CRB)}. Furthermore, we briefly discuss several intrinsic connections among these ISAC metrics.
	
	$\bullet$ {\it MI-Oriented ISAC Systems:} The classical information measure in modern wireless systems is the MI \cite{goldsmith2005wireless}. Let us consider a generic linear Gaussian model for both S\&C, which is formulated as
	\begin{subequations}\label{linear_model}
		\begin{align}
			\label{linear_comm_model}
			\bm{Y}_c =\bm{H}_c \bm{X}+\bm{Z}_c, \\
			\label{linear_sens_model}
			\bm{Y}_s =\bm{H}_s \bm{X}+\bm{Z}_s,
		\end{align}
	\end{subequations}
	where $\bm{Y}_c$ and $\bm{Y}_s$, $\bm{H}_c$ and $\bm{H}_s$, $\bm{Z}_c$ and $\bm{Z}_s$ represent the received signal, channel/target impulse response (CIR/TIR), and the additive white Gaussian noise (AWGN) at the communication and sensing receivers, respectively. Finally, $\bm{X}$ denotes the transmit signal. On one hand, with the goal of improving the performance of communication systems, one has to design the optimal $\bm{X}$ in order to maximize the conditional MI $I(\bm{X}; \bm{Y}_c | \bm{H}_c)$ \cite{thomas2006elements}, which quantifies the maximum achievable point-to-point rate, in the face of practical constraints such as the total power budget. On the other hand, radar systems aim for the maximum conditional MI $I(\bm{H}_s ; \bm{Y}_s \vert \bm{X})$ to quantify the sensing performance \cite{MMSE_MIyang2007mimo,MI_Sensing_yuan2020spatio, MI_Sensing_tang2010mimo}.  {Explicitly, we have the following pair of optimization problems toward communication-only or sensing-only scenarios: 
		\begin{subequations}\label{MIP}
			\begin{align}
				\max &~~ I(\bm{X} ; \bm{Y}_c | \bm{H}_c)~\mathrm{or}~I(\bm{H}_s ; \bm{Y}_s \vert \bm{X})\\
				~~~\mathrm{s.t.} &~~ \mathrm{Other~Specific~Practical~Constraints}.
			\end{align}
		\end{subequations}
		By appropriately optimizing the signaling strategies, such as symbol- and block-level transmit precoding \cite{liurang2021dual,li_ang2022practical}, we notice that the aforementioned optimization problem \eqref{MIP} leads to an MI-based tradeoff between S$\&$C, } {leading to the following weighted MI-oriented maximization problem in ISAC scenarios:
		\begin{subequations}\label{MI_Pareto}
			\begin{align}
				\max &~~ \rho I(\bm{X} ; \bm{Y}_c | \bm{H}_c)+ (1-\rho)I(\bm{H}_s ; \bm{Y}_s \vert \bm{X})\\
				~~~\mathrm{s.t.} &~~ \mathrm{Other~Specific~Practical~Constraints},
			\end{align}
		\end{subequations}
		where the weighting factor $\rho \in \left[ 0,  1\right] $ controls the priorities assigned to S\&C functionalities.
	}
	This means that one can find the Pareto-optimal MI boundary in ISAC systems to illustrate the MI tradeoff between S$\&$C \cite{ouyang2022integrated,ouyang2022performance,MI_Sensing_yuan2020spatio,lijinMI2022framework}.
	
	Indeed, the MI measure is suitable for radar sensing, but it might not always be precise. This is because the MI does not have an explicit operational interpretation in radar sensing systems, which is in stark contrast to wireless communication systems. More specifically, we can readily quantify how much useful information can be inferred by using \textit{bits} in communication systems, but it remains unclear how many bits one can obtain by observing the trajectory of the target, including its velocity and angle estimation.  {Moreover, it is worth noting that another drawback of using the MI for sensing is the difficulty in characterizing it by a closed-form formulation in general.}
	
	\begin{figure}[t!]
		\centering
		\epsfxsize=1\linewidth
		\includegraphics[width=\columnwidth]{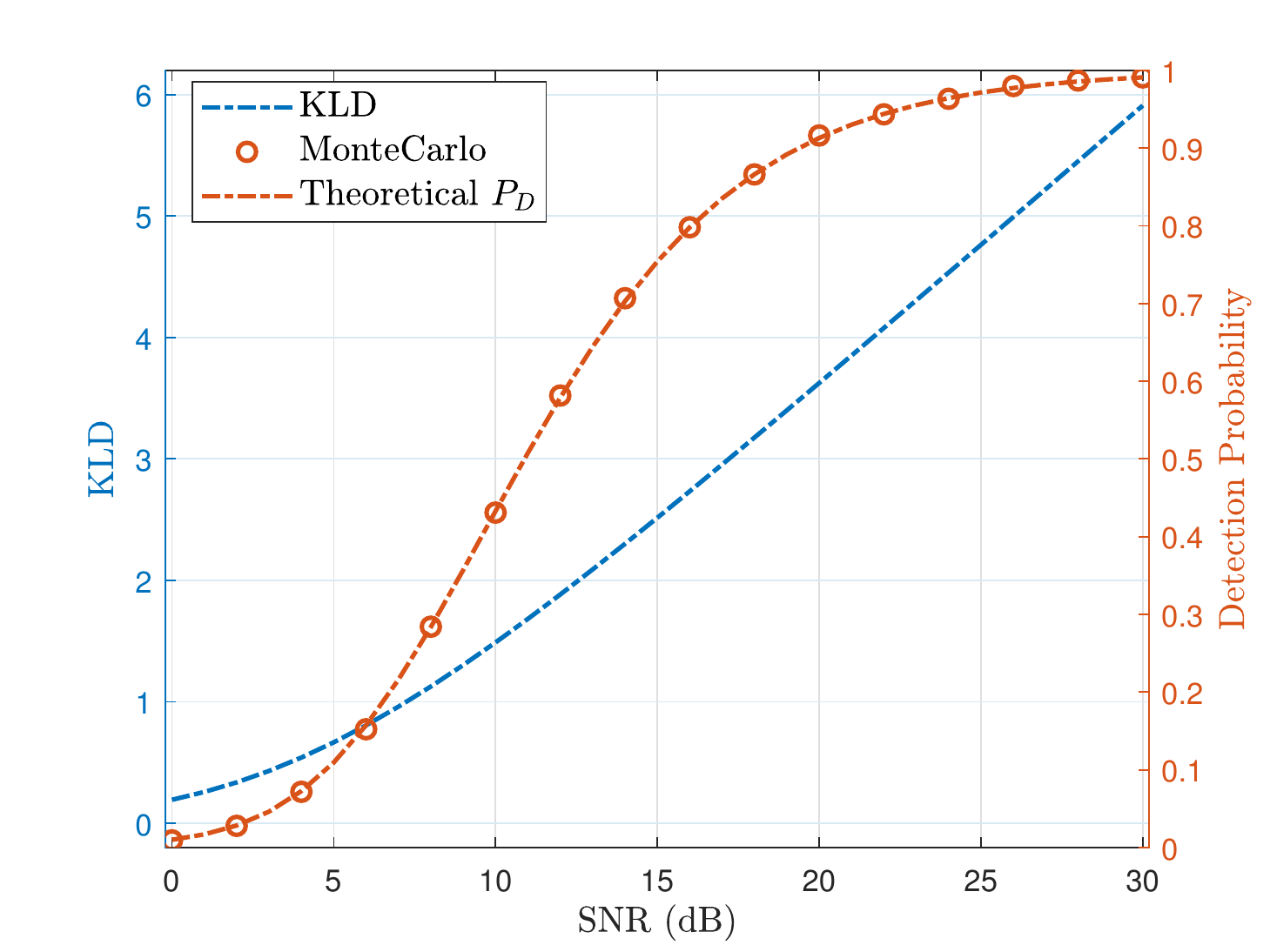}
		\caption{KLD and detection probability versus SNR.}
		\label{PD_KLD}
	\end{figure} 
	
	$\bullet$ {\it Target Detection Metric: KLD.} The KLD metric, a.k.a. the relative entropy, can be used for quantifying the difference between two distributions having the probability density functions (PDF) of $p_0$ and $p_1$, which is defined as 
	\begin{align}
		D\left(p_{0} \| p_{1}\right) & =\int p_{0}(x) \log \frac{p_{0}(x)}{p_{1}(x)} \mathrm{d} x \nonumber \\
		& = \mathbb{E}_{p_0} \left[\log  \frac{p_{0}(X)}{p_{1}(X)}\right],
	\end{align}
	where $\mathbb{E}_{p_0}(\cdot )$ represents the expectation under distribution $p_0$. In a statistical sense, the KLD quantifies the distance between a pair of probability distributions \cite{kailath1967divergence}. When the two distributions are identical, the corresponding KLD is zero. In particular, for the  detection problem (ignoring clutter) in monostatic scenarios, we seek to choose between two hypotheses, i.e. $\mathcal{H}_1$ representing that the target is present, or $\mathcal{H}_0$, the target is absent, which can be formulated as
	\begin{align}\label{BinaryDet}
		\bm{Y}_s =\left\{\begin{array}{l}
			\mathcal{H}_{1}: \bm{H}_{s} \bm{X}+\bm{Z}_{s}, \\
			\mathcal{H}_{0}: \bm{Z}_{s}.
		\end{array}\right.
	\end{align}
	Let us denote the PDF of $\bm{Y}_s$ under $\mathcal{H}_1$ and $\mathcal{H}_0$ as $p_1(\bm{Y}_s)$ and $p_0(\bm{Y}_s)$, respectively. Then, upon following the Chernoff-Stein lemma \cite{thomas2006elements}, the KLD can be used for quantifying the target detection probability, expressed as \cite{kailath1967divergence,tangbo2015relative,thomas2006elements}
	\begin{align}\label{KLD_PD}
		D\left[p_0(\bm{Y}_s) \| p_1(\bm{Y}_s)\right]=\lim _{N \rightarrow \infty}\left(-\frac{1}{N} \log \left(1-P_{D}\right)\right),
	\end{align}
	where $N$ is the number of observations and $P_D$ is the target detection probability. More explicitly, Equation \eqref{KLD_PD} suggests that upon increasing $N$, maximizing the KLD is asymptotically equivalent to maximizing the detection probability. 
	
	To elaborate a little further for lifting the KLD into target detection problem, let us consider the basic example of \eqref{BinaryDet} in the single-antenna scenario, having the following representation
	\begin{align}\label{Det_model}
		{y}[\ell] =\left\{\begin{array}{l}
			\mathcal{H}_{1}: h x[\ell]+z[\ell], \ell = 0,1,\dots,L-1, \\
			\mathcal{H}_{0}: z[\ell],\ell = 0,1,\dots,L-1, 
		\end{array}\right.
	\end{align}
	where $h\sim \mathcal{CN}(0,\sigma_h^2)$ is the reflecting coefficient, $x[\ell]$ is the transmit signal and $z[\ell]\sim \mathcal{CN}(0,\sigma_z^2)$ is the AWGN, respectively. By stacking all the received signals $y[\ell]$ into $\bm{y} = [y[0],\dots,y[L-1]]^T$, the PDFs of $\bm{y}$ under $\mathcal{H}_{0}$ and $\mathcal{H}_{1}$ can be expressed as 
	\begin{subequations}\label{y_pdf}
		\begin{align}
			p_0(\bm{y}) &= \frac{1}{\pi^L \sigma_z^{2L}} \mathrm{exp}\left[-\frac{1}{\sigma_z^{2}} \bm{y}^H\bm{y}\right],\\
			p_1(\bm{y}) &= \frac{1}{\pi^L \det(\bm{C}_s + \sigma_z^2 \bm{I})} \mathrm{exp}\left[-\bm{y}^H (\bm{C}_s + \sigma_z^2 \bm{I})^{-1} \bm{y} \right],
		\end{align}
	\end{subequations}
	where $\bm{x} = [x[0],\dots,x[L-1]]^T$ represents the transmitted signals and $\bm{C}_s = \sigma_h^2 \bm{x}\bm{x}^H$ denotes the covariance matrix of $h\bm{x}$, while $\bm{I}$ is the $L$-dimension identity matrix, respectively. Then the log-likelihood ratio can be derived as 
	\begin{align}
		\log \frac{p_1(\bm{y})}{p_0(\bm{y})} &= -\bm{y}^H \left[(\bm{C}_s + \sigma_z^2 \bm{I})^{-1} - \frac{1}{\sigma_z^2} \bm{I} \right]\bm{y} \nonumber \\
		&- \log \det(\bm{C}_s + \sigma_z^2 \bm{I}) + \log \det \sigma_z^{2L},
	\end{align}
	Following the Neyman-Person lemma, the detection probability is \cite{kay1993fundamentals}
	\begin{align}\label{P_D}
		P_D = P_{FA}^{\frac{1}{1+\frac{P \sigma_h^2}{\sigma_z^2} }},
	\end{align} 
	where $P=\bm{x}^{H} \bm{x}$ denotes the total transmit power and $P_{FA}$ represents the given false alarm probability.  {We remark that it is difficult to design the transmit signal $\bm{x}$ by \eqref{P_D} in ISAC scenarios where $\bm{x}$ must be tailored to achieve specific communication performance optimization goals, such as reducing the symbol error rate (SER) \cite{liurang2021dual}. This is primarily due to the term $\bm{x}^H\bm{x}$, which is independent of the power budget $P$. As an alternative manner to tackle this issue, one may choose to maximize the KLD which is a function of the transmitted signals, as detailed below.}
	
	Let us denote the KLD between $\mathcal{H}_{0}$ and $\mathcal{H}_{1}$ by $D(p_0\| p_1)$, which can be calculated by 
	\begin{align}\label{KLD_Toy}
		& D(p_0\| p_1) =  \int p_{0}(\bm{y}) \log \frac{p_{0}(\bm{y})}{p_{1}(\bm{y})} \mathrm{d} \bm{y} \nonumber\\
		&= - \mathbb{E}_{p_0(\bm{y})} \left[ \log \frac{p_1(\bm{y})}{p_0(\bm{y})} \right] \nonumber \\
		&= \mathbb{E}_{p_0(\bm{y})} \left[ \bm{y}^H \left((\bm{C}_s +  \sigma_z^2\bm{I})^{-1} - \frac{1}{\sigma_z^2}\bm{I} \right)\bm{y} \right] + \phi  \nonumber \\
		& = \mathrm{tr}\left[ \sigma_z^2 (\bm{C}_s +  \sigma_z^2\bm{I})^{-1} - \bm{I} \right] + \phi, 
	\end{align}
	where we have $\phi = \log \det(\bm{C}_s + \sigma_z^2 \bm{I}) - \log \det \sigma_z^{2L}$. Then, the Chernoff-Stein lemma indicates that the missed detection probability of $1-P_D$ becomes exponentially small, with an exponential rate $D(p_0\| p_1)$ given above. Finally, we identify the relationship between the KLD and the detection probability versus transmit $\mathrm{SNR} = P/\sigma_z^2$ in Fig. \ref{PD_KLD}. It can be observed that upon increasing the signal-to-noise ratio (SNR), the KLD is also increased, which results in improved detection performance.
	
	\begin{figure}[t!]
		\centering
		\epsfxsize=1\linewidth
		\includegraphics[width=\columnwidth]{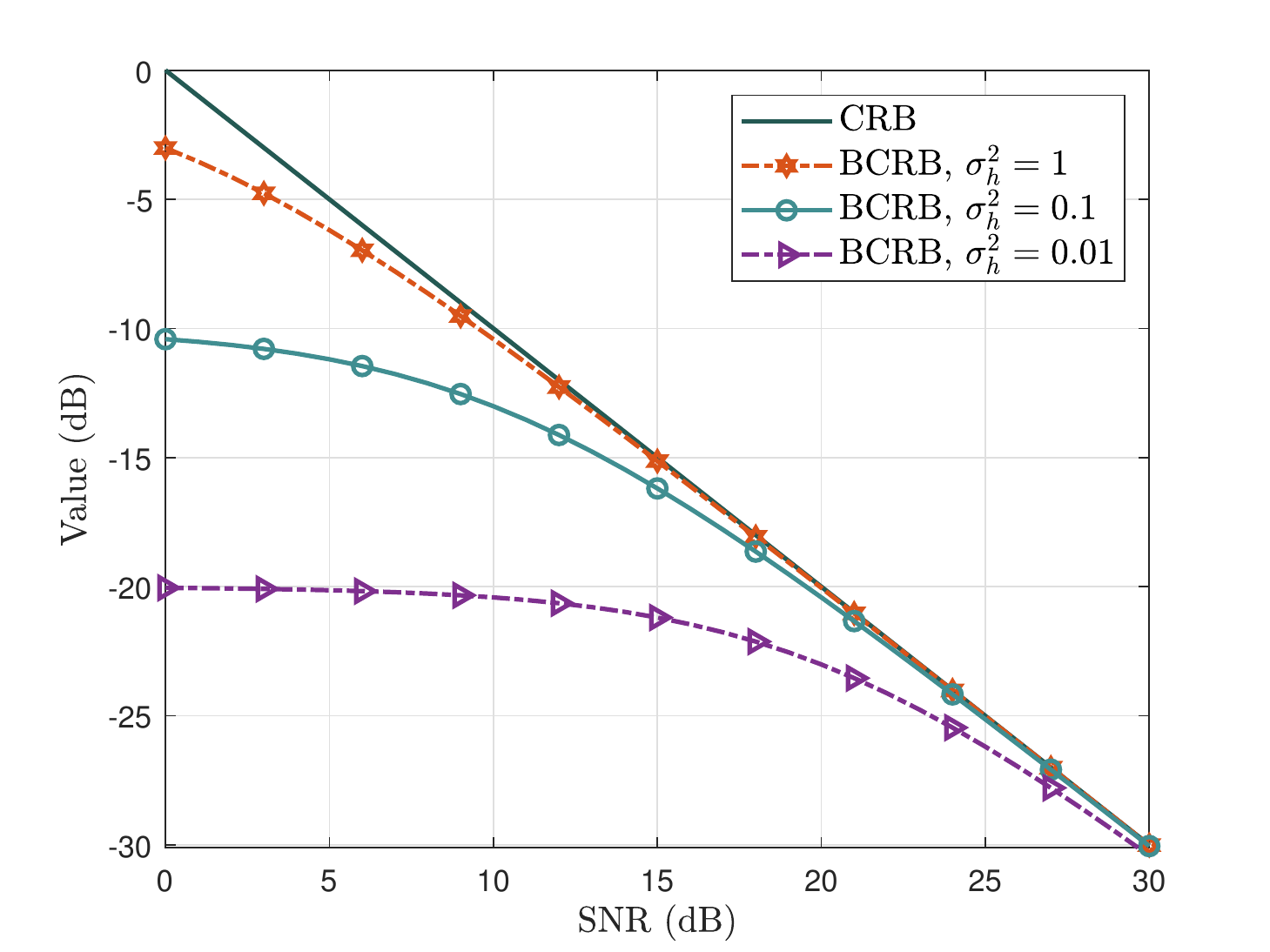}
		\caption{CRB and BCRB versus SNR.}
		\label{P_BCRB}
	\end{figure} 
	
	 {It is worth pointing out that the specific choice of sensing metrics may be determined by the application scenarios and design principles. We remark that the detection probability $P_D$ in the binary test \eqref{Det_model}, as expressed in \eqref{P_D}, typically lacks a straightforward link with the transmit signals, making the design of the transmit signal $\bm{x}$ a complex task. Another viable approach is to use the KLD expression, which offers more flexibility in designing the transmitter's covariance matrix, such as $\bm{C}_s$ in \eqref{KLD_Toy}. This increased flexibility is particularly valuable in ISAC scenarios, where the transmit waveform design must satisfy stringent communication requirements.}

	$\bullet$ {\it Parameter Estimation Metric: CRB.} We assume that the vector $\boldsymbol{\theta}$ denotes the parameters to be estimated, and its estimate is ${\boldsymbol{\hat{\theta}}}$. Typically, we are concerned with unbiased estimation scenarios, which indicates $\mathbb{E}(\boldsymbol{\hat{\theta}}) = \boldsymbol{\theta}$. Then the estimation performance can be quantified by the well-known CRB, which serves as the lower bound on the mean square error (MSE) of any ${\boldsymbol{\hat{\theta}}}$ over $\boldsymbol{\theta}$. Let $\bm{J}(\boldsymbol{\theta})$ denote the Fisher information matrix (FIM). Then the CRB matrix can be written as \cite{kay1993fundamentals} 
	\begin{align}
		\bm{M}_{\mathrm{CRB}}( \boldsymbol{\theta} )= \bm{J}^{-1}(\boldsymbol{\theta}), 
	\end{align}
	where $[\bm{J}(\boldsymbol{\theta})]_{i j}=-\mathbb{E}\left[\frac{\partial^2 \log p(\bm{y} | \boldsymbol{\theta})}{\partial \theta_{i} \partial \theta_{j}}\right]$ represents the $i,j$-th element of the FIM $\bm{J}(\boldsymbol{\theta})$ and $\log p(\bm{y}| \boldsymbol{\theta})$ is the log-likelihood function of the observed data $\bm{y}$, given $\boldsymbol{\theta}$. The above expression represents the perspective of the Frequentists who assert that the $\boldsymbol{\theta}$ to be estimated is an unknown but deterministic parameter \cite{bayarri2004interplay}. Accordingly, the CRB of the deterministic parameter $\bm{\theta}$ is
	\begin{align}\label{CRB}
		\mathrm{CRB}(\bm{\theta}) = \mathrm{tr}\left[\bm{J}^{-1}(\boldsymbol{\theta})\right]. 
	\end{align}

	On the other hand, the Bayesians state that $\boldsymbol{\theta}$ represents unknown random variables having {\it a-priori} distribution of $\pi (\boldsymbol{\theta})$. Therefore, the joint PDF of $\boldsymbol{\theta}$ and $\bm{y}$ is formulated as
	\begin{align}\label{joint_PDF}
		p (\boldsymbol{\theta} , \bm{y}) = p(\bm{y} | \boldsymbol{\theta}) \pi (\boldsymbol{\theta}).
	\end{align}
	According to \eqref{joint_PDF}, the {\it a-posteriori} FIM of $\boldsymbol{\theta}$ can be written as \cite{YuanShen2010}
	\begin{align}
		\bm{J}_{\mathrm{p}}(\boldsymbol{\theta}) & =-\mathbb{E}\left(\frac{\partial^2 \log p (\boldsymbol{\theta} , \bm{y})}{\partial \boldsymbol{\theta}^2 }\right) \nonumber\\
		& =\underbrace{-\mathbb{E}\left(\frac{\partial^{2} \log p\left(\bm{y}| \boldsymbol{\theta} \right)}{\partial \boldsymbol{\theta}^{2}}\right)}_{\text {Observed Fisher Information }} \underbrace{-\mathbb{E}\left(\frac{\partial^{2} \log \pi \left(\boldsymbol{\theta}\right)}{\partial \boldsymbol{\theta}^{2}}\right)}_{\text {Prior Fisher Information }} \nonumber \\
		& \triangleq \bm{J} + \bm{J}_{\mathrm{prior}},
	\end{align}
	which leads to the Bayesian/{\it a-posteriori} CRB (BCRB/PCRB) of
	\begin{align}
		\mathrm{BCRB}( \boldsymbol{\theta} )= \mathrm{tr}\left[\bm{J}_{\mathrm{p}}^{-1}(\boldsymbol{\theta})\right].
	\end{align}
	
	To provide further intuition about the estimation performance, let us recall the basic linear model of the single-antenna scenario in \eqref{BinaryDet}, having the following representation
	\begin{align}
		{y}[\ell] = h x[\ell]+z[\ell], \ell = 0,1,\dots,L-1, 
	\end{align}
	where $h$ is the reflecting coefficient having the {\it a-priori} distribution $\mathcal{CN}(0,\sigma_h^2)$ to be estimated.  {For ease of illustration, we assume that the noise is AWGN, where $z[\ell]\sim \mathcal{CN}(0,\sigma_z^2)$ and $\sigma_z^2$ denotes the variance. We would like to emphasize that incorporating colored noise might ameliorate this basic example, but in a first approximation, one can simply replace the white noise variance matrix $\sigma_z^2 \bm{I}$ with the colored noise variance matrix.} Denote the real part and imaginary part of $h$ by $h_R$ and $h_I$. Then the joint PDF of $\boldsymbol{\theta} = [h_R, h_I]^T$ and $\bm{y}$ can be written as 
	\begin{align}
		p(\boldsymbol{\theta}, \bm{y}) = &\underbrace{\frac{1}{ \pi^{L} \sigma_z^{2L}} \mathrm{exp}\left[-\frac{1}{\sigma_z^2} (\bm{y}-h\bm{x})^H (\bm{y}-h\bm{x})\right]}_{p(\bm{y} | \boldsymbol{\theta})} \nonumber \\
		&\cdot \underbrace{\frac{1}{ \pi \sigma_h^2 } \mathrm{exp}\left[-\frac{1}{\sigma_h^2} |h|^2  \right]}_{ \pi (\boldsymbol{\theta})}. 
	\end{align}
	Therefore, the {\it a-posteriori} FIM of $\boldsymbol{\theta}$ can be formulated as
	\begin{equation}
		\bm{J}_{\mathrm{p}}(\boldsymbol{\theta}) = 2 \left[ \begin{array}{cc}
			\frac{\bm{x}^H\bm{x}}{\sigma_z^2} + \frac{1}{\sigma_h^2} & 0 \\
			0 & \frac{\bm{x}^H\bm{x}}{\sigma_z^2} + \frac{1}{\sigma_h^2}
		\end{array}
		\right].
	\end{equation}
	The corresponding BCRB of $\boldsymbol{\theta}$ is expressed as 
	\begin{equation}
		\mathrm{BCRB}(\boldsymbol{\theta}) = \frac{\sigma_h^2\sigma_z^2}{\sigma_h^2 \bm{x}^H \bm{x} + \sigma_z^2}.
	\end{equation}
	Additionally, if the {\it a-priori} distribution $\pi (\boldsymbol{\theta})$ is not available, the BCRB degenerates into the classical CRB in \eqref{CRB}, given as \cite{kay1993fundamentals}
	\begin{equation}
		\mathrm{CRB}(\boldsymbol{\theta}) = \frac{\sigma_z^2}{\bm{x}^H \bm{x}}.
	\end{equation}

	Notice that the BCRB usually provides additional Fisher information, which is also independent of the real value of the parameters to be estimated, resulting in better estimation performance than the classical CRB. We illustrate this point in Fig. \ref{P_BCRB}. It is observed that with the {\it a-priori} distribution at hand, the BCRB metric is attainable with better estimation performance than the classical CRB. We remark that the CRB and BCRB/PCRB represent different perspectives in statistics, which constitute fundamental philosophical and cognitive issues \cite{bayarri2004interplay}.

	\begin{figure}[!t]
		\centering
		\epsfxsize=1\linewidth
		\includegraphics[width=\columnwidth]{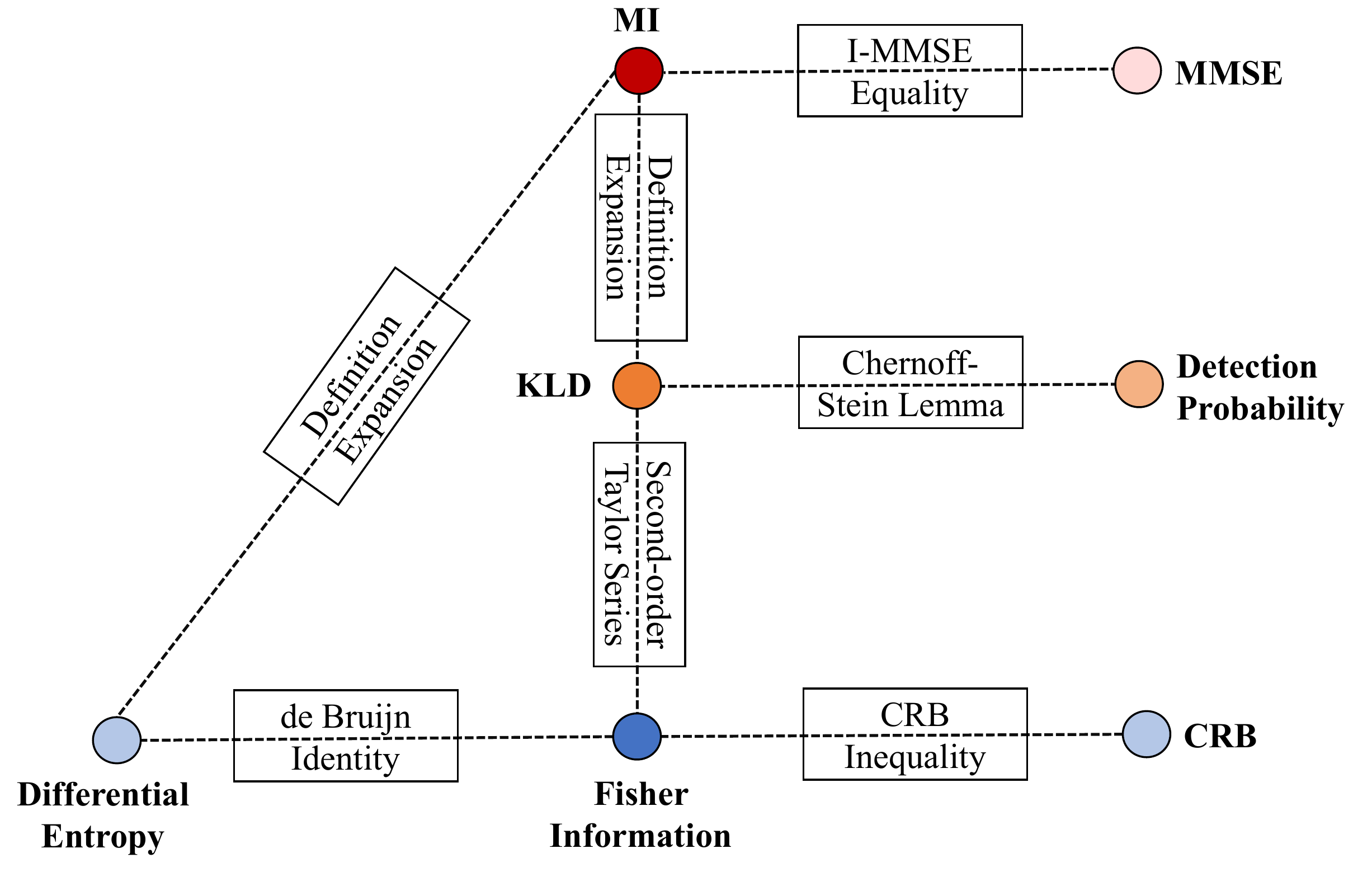}
		\caption{ {The connections among ISAC metrics.}}
		\label{Metrics}
	\end{figure} 
	
	 {As illustrated in Fig. \ref{Metrics}, there are several connections among the aforementioned metrics in general, which are detailed below. Furthermore, these metrics may also gradually find their way into the Pareto-optimization of ISAC systems. To elaborate briefly, it cannot be expected that an ISAC system having the lowest MMSE is also the best in terms of any of the other metrics. Hence if we want to improve any of the other metrics, we must accept an MMSE degradation for example. This Pareto-optimization effort may commence from a pair of metrics and then gradually extend to three or more of them. However, as the number of metrics is increased, the `search-space' may gradually become excessive hence requiring substantial future research.}
	
	$\bullet$ The I-MMSE equation \footnote{Briefly, this equation stated that the MMSE of communication signal $\bm{X}$ is equal to the derivative of the MI with respect to the SNR \cite{IMMSE_guo2005mutual}. However, in monostatic sensing systems, the ISAC base station already has the perfect knowledge of its own transmit signal $\bm{X}$. Then one only has to estimate the target parameters embedded in $\bm{H}_s$ instead of $\bm{X}$ in the I-MMSE equation.} relates the MMSE of communication signal $\bm{X}$ to the mutual information, as \cite{IMMSE_guo2005mutual}
	\begin{align}
		\frac{\mathrm{d}}{\mathrm{dSNR}} I(\bm{X} ; \bm{Y})=\frac{1}{2} \mathrm{MMSE}_{\bm{X}}(\mathrm{SNR}).
	\end{align}
	
	$\bullet$ The MI can be expressed also by KLD in the form of $I(\bm{X}; \bm{Y}_c) = D\left[p(\bm{X},\bm{Y}_c)  \| p(\bm{X})p(\bm{Y}_c) \right]$ \cite{thomas2006elements}.
	
	$\bullet$ The Fisher information characterizes the curvature of the KLD between two distributions \cite{valero2017generalization}. Let us assume that the log-likelihood function of $\boldsymbol{\theta} \in \boldsymbol{\Theta}$ is given by $\log p_0(\bm{y}| \boldsymbol{\theta})$ and  $\log p_1(\bm{y}| \boldsymbol{\theta}_1)$ for a given $\boldsymbol{\theta}_1 \in \boldsymbol{\Theta}$. The second-order Taylor series expansion with respect to $\boldsymbol{\theta}$ at $\boldsymbol{\theta} = \boldsymbol{\theta}_1$ may be written as 
	\begin{align}
		D(p_0 \| p_1 ) & = \frac{1}{2} \left(\boldsymbol{\theta}-\boldsymbol{\theta}_1\right)^{T} \bm{J}(\boldsymbol{\theta}_1)
		\left(\boldsymbol{\theta}-\boldsymbol{\theta}_1\right) \nonumber \\ 
		& + o \left( \| \boldsymbol{\theta}-\boldsymbol{\theta}_1 \| ^2 \right).
	\end{align}
	
	$\bullet$ The Fisher information can be linked to the differential entropy via the de Bruijn identity given by \cite{stam1959some} 
	\begin{equation}\label{Bruijn_Identity}
		\frac{\mathrm{d}}{\mathrm{d} t} h(\bm{H}_c \bm{X} + \sqrt{t} \bm{N}_0)=\frac{1}{2} \operatorname{tr}\{\bm{J}(\bm{H}_c \bm{X}+\sqrt{t} \bm{N}_0)\},
	\end{equation}
	where the parameter $t$ is assumed to be non-negative, $\bm{N}_0$ is a noise vector having independent standard Gaussian entries and it is assumed to be independent of $\bm{X}$. Here, $h(\cdot)$ represents the differential entropy and $\bm{J}({\cdot})$ is the Fisher information matrix. By following the linear Gaussian model of \eqref{linear_comm_model} and assuming $\bm{N}_c = t\bm{N}_0$, the de Bruijn identity states that the Fisher information of $\bm{Y}_c$ can be viewed as the curvature of differential entropy \cite{IMMSE_guo2005mutual}, which can be expressed as 
	\begin{equation}
		\frac{\mathrm{d}}{\mathrm{d} t} h(\bm{Y}_c)=\frac{1}{2} \operatorname{tr}\{\bm{J}(\bm{Y}_c)\}.
	\end{equation}

	{\it \textbf{2) Existing Literature:}} Given the evolving integration of S$\&$C, the information theories of S$\&$C may also be expected to be unified. As for the fundamental limits of sensing, a recent use case was reported in a wireless localization scenario \cite{YuanShen2010}. By introducing the squared position error bound (SPEB) to characterize the localization accuracy\cite{YuanShen2010}, one can analyze the fundamental limits of device-based localization. On the other hand, when considering the communication part, the early results on the information-theoretic connection between S$\&$C were disseminated in the well-known I-MMSE equation \cite{IMMSE_guo2005mutual}. 
	
\begin{figure}[!t]
		\centering
		\epsfxsize=0.8\linewidth
		\includegraphics[width=\columnwidth]{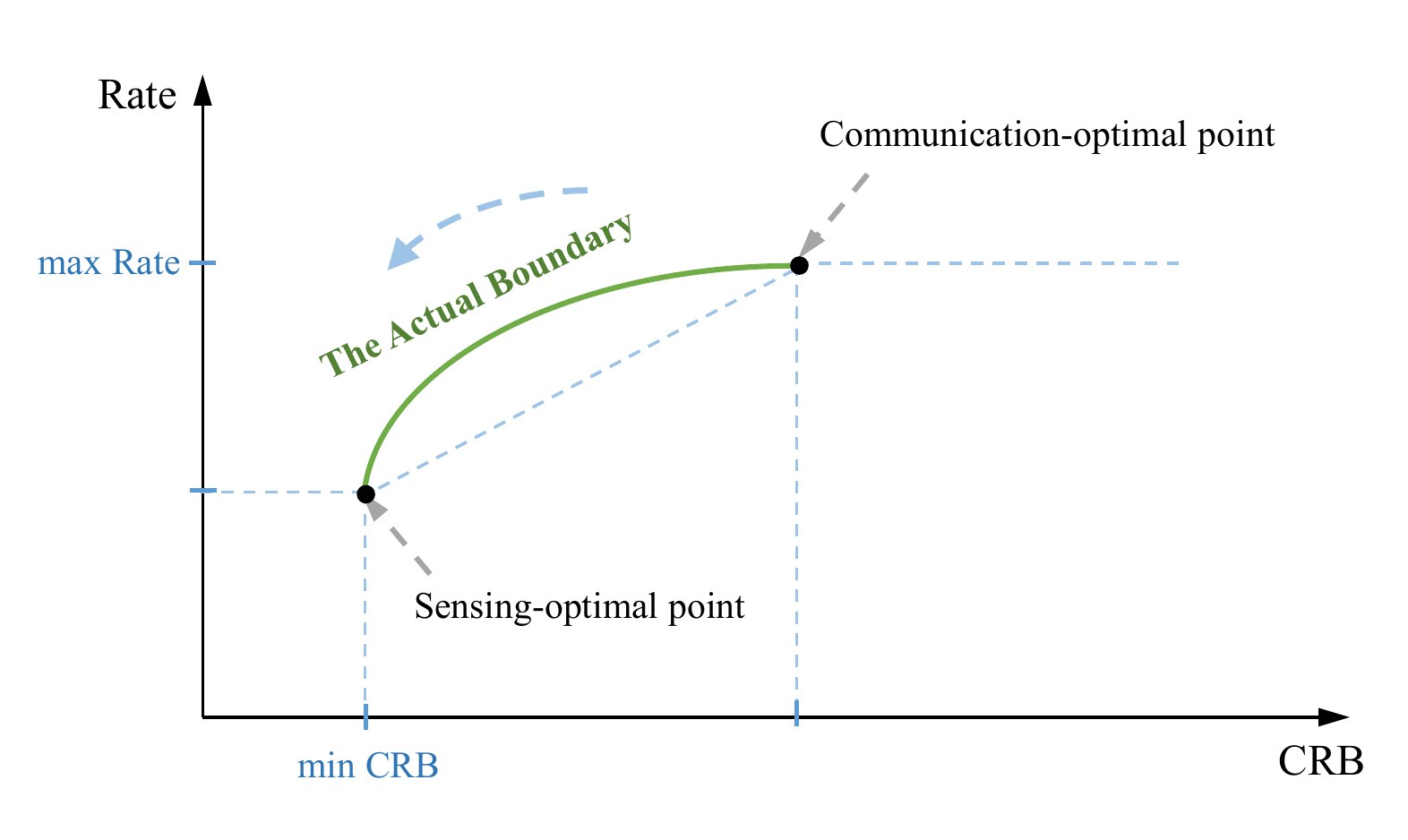}
		\caption{The pentagon inner bound of the CRB-rate region.}
		\label{Paretobound}
	\end{figure}

	Another information-theoretic perspective is represented by the capacity-distortion tradeoff model, where the target response (e.g., radar echo signals) is modeled as a delayed feedback channel \cite{kobayashi2018joint}. In this setup, the state-dependent channel outputs the signal to the receiver and returns the state feedback to the transmitter for estimating the channel it state. Following the spirit of \cite{kobayashi2018joint}, the authors of \cite{liuyao2022generalized} extended to correlated sensing and channel states by considering a generalized channel model of multiple-access ISAC systems. More recently, the authors of \cite{xiong2022flowing} proposed a pentagon inner bound of the CRB-rate region to reveal the fundamental tradeoff between S$\&$C, as illustrated in Fig. \ref{Paretobound}. This model first revealed the information-theoretic connections among the classic communication capacity, the target channel states and the parameters to be estimated in ISAC systems. 
	
	{\it \textbf{3) Future Directions and Potential Solutions:}} Clearly, the information theoretic research of ISAC systems is still in its infancy. There are numerous open problems to be addressed in future work, some of which are listed as follows.
	\begin{itemize}
		\item  {The weighted optimization problem \eqref{MI_Pareto} leads to the Pareto-optimal MI boundary in ISAC systems \cite{ouyang2022integrated,ouyang2022performance,MI_Sensing_yuan2020spatio,lijinMI2022framework}, but different design goals and metrics should also considered in the context of the Pareto boundary, such as communication rate and the target detection probability.}
		\item Again, substantial community effort is required for quantifying the S$\&$C performance in terms of their MI. In simple terms, how many bits must be transmitted in different sensing scenarios, for example, for delay, angle and Doppler estimation.
		\item Identifying the most promising techniques (e.g., coding approach) capable of approaching the Pareto front remains widely unexplored. Although the authors of \cite{xiong2022flowing} invested substantial efforts into analyzing the two corner points, i.e., communication- and sensing-optimal performance in Fig. \ref{Paretobound}, how to attain the actual boundary of S$\&$C is still an open issue.
		\item There are still fundamental issues to be addressed both in the target detection and in the wireless communication components of ISAC systems from an information-theoretic perspective. Pioneering efforts have been made in joint communication and binary state detection in \cite{KLD_joudeh2022joint}. As a further step, it is necessary to clarify the connection between the KLD measure (in both single and multiple target(s) detection) as well as the mutual information (in wireless communication), which may help us to reveal the fundamental tradeoff between the target detection probability and communication rate. 
		\item  {To reveal the fundamental limits of ISAC systems, a potential solution is to establish the unified metrics for both S\&C. For example, the authors of \cite{Fuwang_SER_TVT} defined the sensing estimation rate (SER) to unify the information- and estimation-theoretic perspectives of ISAC systems. However, it is still challenging to carry out a unified performance evaluation in other ISAC scenarios.} 
		\item In ISAC systems, a promising technique is to perform sensing by utilizing communication signals. However, the signals dedicated to information transfer are inherently random. Therefore, the specific sensing metrics, such as the ergodic CRB or ergodic PCRB have to be tailored and well-defined for different use cases.
		\item The fundamental information theory for unveiling the attainable degrees of freedom, and of massive MIMO ISAC systems as well as of their networking issues is still open. 
	\end{itemize}  
	
	\vspace{1em}
	\noindent
	\textbf{\emph{Challenge 2: How Much Channel Information Can be Inferred From the Sensory Data?}}
	
	{\it \textbf{1) Background:}} In conventional wireless communication systems, the base station (BS) firstly transmits pilots to all the users via the downlink channel. The users then estimate the channel state information (CSI) and transmit their channel estimates back to the BS in the uplink channel. Naturally, the pilot symbols result in communication overheads and thus they limit the useful transmission rate \cite{liu2020radar}. In the context of ISAC, the sensory data that contains information about the surrounding environment can also be exploited to obtain channel information. It is envisioned that ISAC may offer a solution that either partially or fully eliminates the need for this feedback loop and thus boosts the communication performance \cite{du2022integrated}. 
	
	\begin{figure}[htbp]
		\centering
		\epsfxsize=1\linewidth
		\includegraphics[width=\columnwidth]{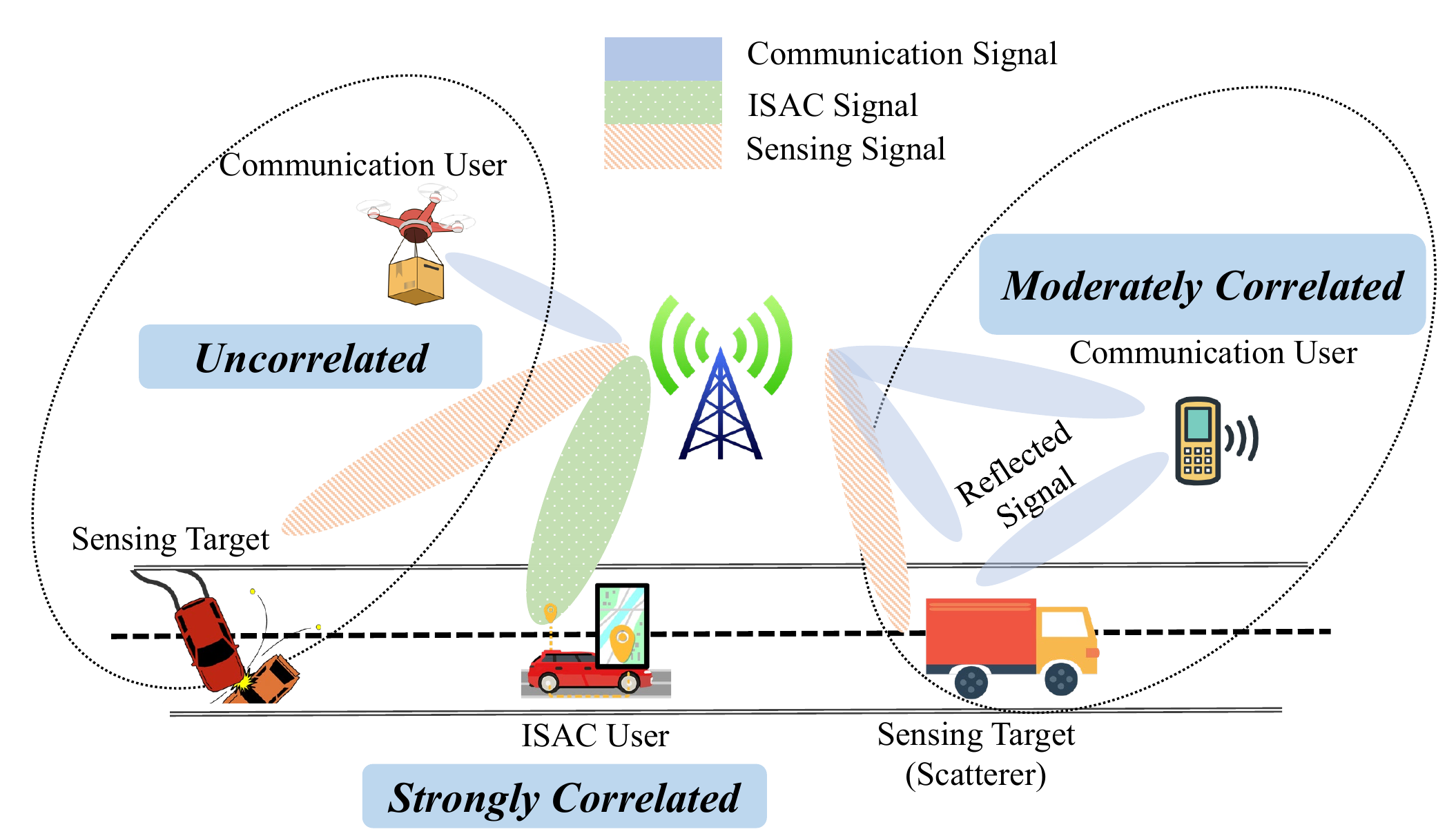}
		\caption{Three categories of S$\&$C environment with typical scenarios.}
		\label{Categories_ISAC}
	\end{figure}  
	
	The prerequisite of inferring CSI is that there is a sufficiently high correlation between the sensing and communication channels. Therefore, we firstly classify the correlation of S$\&$C channels into three categories, as shown in Fig. \ref{Categories_ISAC}.
	\begin{itemize}
		\item {\it Uncorrelated:} S$\&$C take place in different spatial environments and the channels of sensing and communication are completely uncorrelated or independent of each other. For example, the BS is delivering information to a UAV and simultaneously sensing a passing-by vehicle. Therefore, it is difficult to infer any useful channel information from the sensory data in an uncorrelated scenario.
		\item {\it Moderately correlated:} In this category, the channels of sensing and communication are partially correlated, when, for instance, the BS is communicating with a pedestrian and sensing a passing-by vehicle at the same time. The latter happens to be a scatterer of the communication signal, in which case the sensing path contributes partly to the communication channel. Accordingly, the S$\&$C channels are moderately coupled. Therefore, the ISAC BS may infer partial CSI from the strong echoes \cite{huang2022joint}.
		\item {\it Strongly correlated:} The strongly coupled scenario refers to the case when the sensing target is exactly the communication target and the channels of S$\&$C are strongly correlated. A typical example can be found in vehicle-to-infrastructure (V2I) communication systems \cite{liu2020radar}, where the roadside unit (RSU) can exploit the echo signals to predict and track the vehicular target, so as to assist the downlink communication and to reduce the pilot overhead.
	\end{itemize}  
	
	{\it \textbf{2) Existing Literature:}} In massive multiple-input-multiple-output (mMIMO) systems, the transmitted beampattern is chosen by tentatively harnessing all the beams from a pre-defined codebook, which leads to high beam training overhead in practical scenarios \cite{zhang2020beam,zhang2019codebook}. The sensory data containing specific features such as velocity, location and angle could provide useful information both for beam tracking and beam prediction, hence reducing the pilot and feedback overheads to a certain extent. Some Bayesian filtering methods, such as extended Kalman Filters (EKF) \cite{liu2020radar} can be employed for improving the tracking accuracy and reducing the pilot overhead. Furthermore, deep learning algorithms can be leveraged to map the features from the sensory data to the optimal beam, thus further reducing the communication overheads \cite{demirhan2022radar,OJCOMS_wu2022blockage,alkhateeb2021deepsense}. 
	
	{\it \textbf{3) Future Directions and Potential Solutions:}} The correlation of S$\&$C channels can also be considered as a parameterization model, where the three categories of Fig. \ref{Categories_ISAC} can be modeled by the correlation between channel parameters. To better motivate the technical discussions in this part, we first recall the general linear Gaussian model of \eqref{linear_model} and add the parameters $\boldsymbol{\eta}_{c}$ and $\boldsymbol{\eta}_{c}$, yielding
	\begin{subequations}\label{P1}
		\begin{align}
			\bm{Y}_{c} & =\bm{H}_{c}(\boldsymbol{\eta}_{c} ) \bm{X}+\bm{Z}_{c},\\
			~~~\bm{Y}_{s} & =\bm{H}_{s}(\boldsymbol{\eta}_{s} ) \bm{X}+\bm{Z}_{s},
		\end{align}
	\end{subequations}
	where $\bm{H}_{c}(\boldsymbol{\eta}_{c} )$ and $\bm{H}_{s}(\boldsymbol{\eta}_{s} )$ denote the communication and sensing channels that fluctuate with the variation of the channel parameters $\boldsymbol{\eta}_{c}$ and $\boldsymbol{\eta}_{s}$, respectively. 
	\begin{figure}[t!]
		\centering
		\epsfxsize=1\linewidth
		\includegraphics[width=\columnwidth]{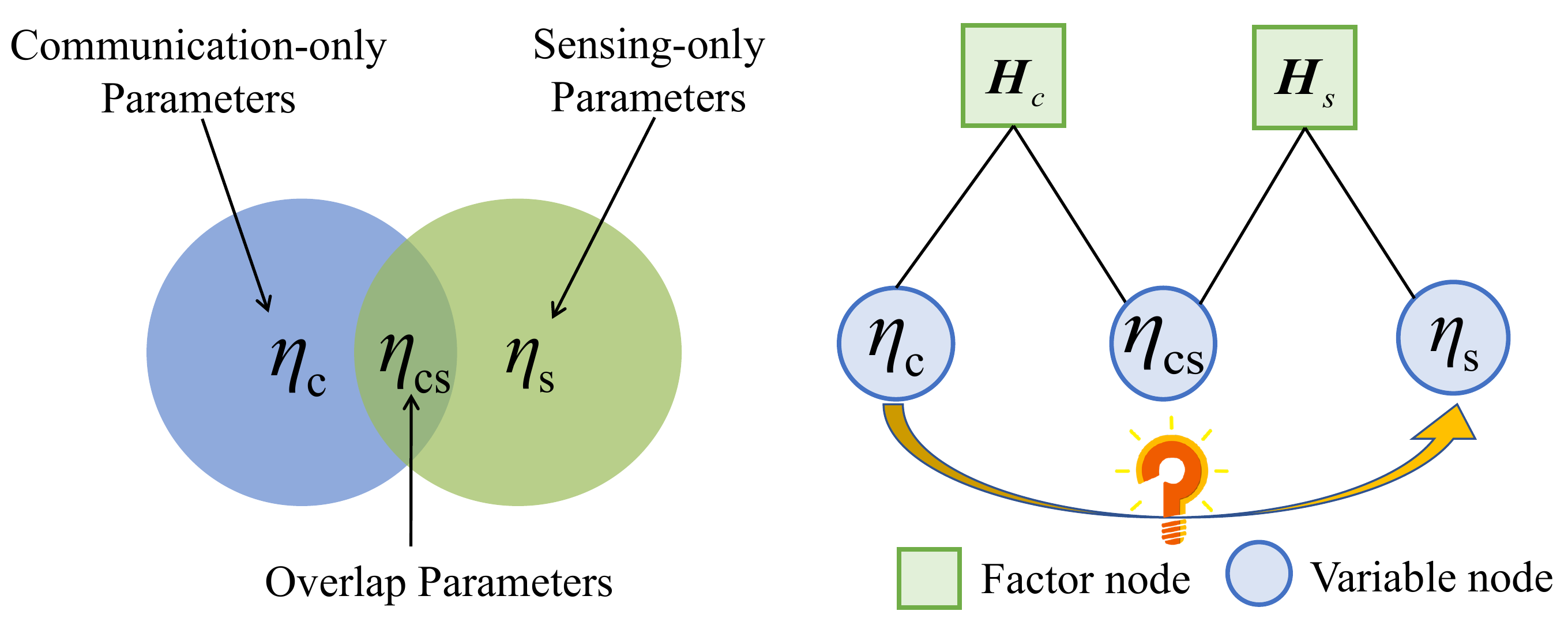}
		\caption{ {Channel parameters between S$\&$C with the corresponding factor graph.}}
		\label{Para_Cate_ISAC}
	\end{figure}
	
	The relationship between the parameters of interest in sensing and communication can be illustrated in the Venn diagram of Fig. \ref{Para_Cate_ISAC}, where $\boldsymbol{\eta}_{cs}$ represents the correlation parameters in both the S$\&$C channels. Following the three categories of ISAC channels in Fig. \ref{Categories_ISAC}, $\boldsymbol{\eta}_{cs}$ may be a null set in the uncorrelated scenario. Otherwise, $\boldsymbol{\eta}_{cs}$ will reflect the correlation of both S$\&$C channels. The factor graph of Fig. \ref{Para_Cate_ISAC} encapsulates the relationship between the parameters, intimating how we can infer channel information from sensing data. More particularly, the mutual information ${I}\left(\bm{H}_{c}; \bm{H}_{s} | \bm{Y}_s, \bm{X} \right)$ between $\bm{H}_{c}$ and $\bm{H}_{s}$ can be applied to estimate how much information can be inferred from the sensory data $\bm{Y}_s$. Generally, ${I}\left(\bm{H}_{c}; \bm{H}_{s} | \bm{Y}_s, \bm{X} \right)$ will be higher in strongly coupled scenarios than that in the uncorrelated scenario. This information-theoretic perspective provides some theoretical insights concerning how to glean some channel information from the sensory data.
	
	 {Finally, to harness the sensory data to infer CSI, one has to concentrate on the modeling and algorithmic design methods in the future. Specifically, one has to precisely model the relationships between the communication channels $\bm{H}_{c}$ and the sensing channels $\bm{H}_{s}$ through factor graph relationships, especially in complex wireless environments, while considering spectrum interference, the presence of clutter sources and so on. Meanwhile, meeting the real-time requirements of sensing presents a significant challenge in designing efficient practical algorithms suitable for diverse scenarios, such as dynamic V2I networks. To this end, it would be promising to establish corresponding data sets of complex environments and explore machine learning methods for utilizing huge sensory data to infer the CSI.}
	
	\vspace{1em}
	\noindent
	\textbf{\emph{Challenge 3: How Could We Quantify the Integration and Coordination Gains?}}
	
\begin{figure}[t!]
		\centering
		\epsfxsize=1\linewidth
		\includegraphics[width=0.9\columnwidth]{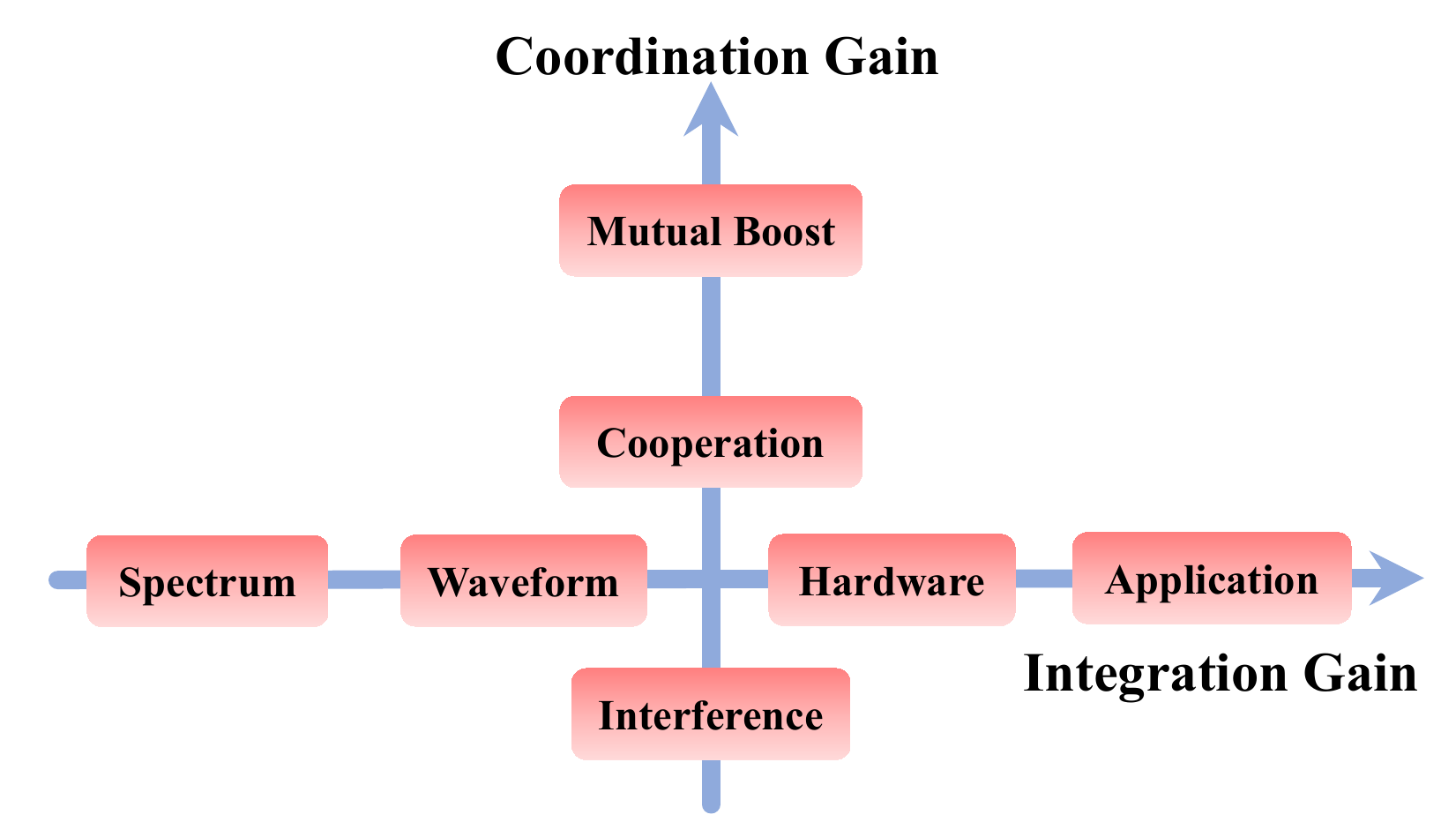}
		\caption{ {The stylized interplay of integration and coordination gains in ISAC.}}
		\label{Gains_Relation}
	\end{figure}

	{\it \textbf{1) Background:}} Sharing the same spectrum is only the first step in moving from complete separation toward the integration of sensing and wireless communication systems. As a benefit of the commonalities between the S$\&$C systems in terms of their hardware architecture and signal processing algorithms, ISAC systems aim for a higher degree of integration, which we refer to as the {\it integration gain} \cite{cui2021integrating}. On the other hand, with the aid of mutual assistance, we can attain beneficial {\it coordination gain}, hence boosting the ISAC performance \cite{cui2021integrating}. The stylized relationship of the {\it integration gain} and {\it coordination gain} is portrayed in Fig. \ref{Gains_Relation}, which manifest themselves in terms of spectrum sharing \cite{zhengle2019RCC}, waveform design \cite{xiaozhiqiangSI2022waveform}, interference management \cite{xu2021rate,qian2023enhancement}.
	
	\begin{figure}[t!]
		\centering
		\epsfxsize=1\linewidth
		\includegraphics[width=0.8\columnwidth]{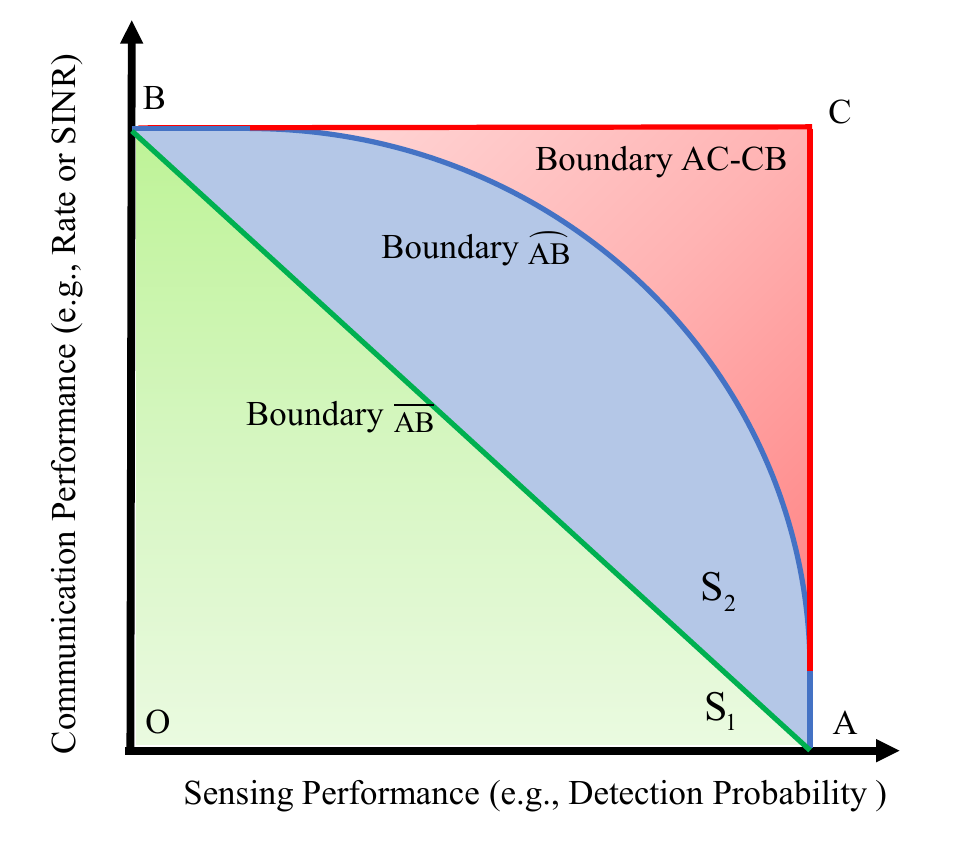}
		\caption{ {Graphical illustration of various integration and coordination scenarios.}}
		\label{Gains_quantify}
	\end{figure}
	
	 {To illustrate the main ideas, let us consider a sensing-assisted V2I communication scenario reported in \cite{liu2020radar}. More specifically, the radar echos received at the RSU are leveraged to predict a vehicle's trajectory for beam alignment in the next round of communication. Therefore, downlink communications can benefit in terms of reduced pilot overhead \cite{liu2020radar}.} In this setup, by utilizing both a common platform as well as the same waveform, the ISAC system can perform dual-functional S$\&$C for improving the hardware-, spectrum-, and energy-efficiencies. This naturally attains both integration and coordination gains in ISAC systems at the same time.

	{\it \textbf{2) Future Directions and Potential Solutions:}} 
	
	On one hand, if orthogonal resources are scheduled for S$\&$C functionalities, in the context of either time- or frequency-division schemes, this implies that no integration gain is achieved, since no resources are shared between S$\&$C. On the other hand, a beneficial integration gain can be acquired when the sensing and communication channels are highly overlapped \cite{lu2022performance}. For example, if the communication user is also the target to be sensed, the frequency- or time-domain resources are shared and the channels are correlated, as illustrated in Fig. \ref{Categories_ISAC}. In such a case, both the S$\&$C functionalities glean benefits. Therefore, we conclude that the integration gain depends on the correlation of the S$\&$C channels (i.e., uncorrelated and strongly correlated as the boundary AC-CB shows in Fig. \ref{Gains_quantify}). These scenarios have to be discussed on a case-by-case basis. 
	
	Before exploiting the benefits of integrated S$\&$C, we have to define metrics for quantifying the performance gains of ISAC systems \cite{lu2022performance}. We attempt to highlight the integration and coordination gains in Fig. \ref{Gains_quantify}, where points A, B, and C stand for the sensing-optimal, communication-optimal, and S$\&$C-optimal performance, respectively. In what follows, we first elaborate further on the different degrees of the channel overlap and the corresponding performance bounds. Then we discuss how to quantify the performance gains in ISAC.
	 {
		\begin{itemize}
			\item {\it Uncorrelated:} The inner bound, i.e., boundary $\overline{\mathrm{AB}}$ of Fig. \ref{Gains_quantify}, refers to the scenario where the S$\&$C fail to boost each other. This is the case upon allocating orthogonal resources between S$\&$C, hence resulting in no integration gain. A basic example can be found in Fig. \ref{Categories_ISAC}, where the BS is sensing a passing-by vehicle while simultaneously communicating with a UAV. This case indicates that S\&C take place in different spatial environments and the wireless resources are allocated in an orthogonal spatial region.
			\item {\it Moderately correlated:} In most practical scenarios, the S$\&$C components may benefit from each other. Then the achievable performance can be illustrated by the arc-shaped boundary $\wideparen{\mathrm{AB}}$ of Fig. \ref{Gains_quantify}. Again, as illustrated in Fig. \ref{Categories_ISAC}, the passing-by vehicle to be sensed is also a scatterer of the communication signal. This case means that some of wireless resources, e.g., the transmit power, can be jointly reused by S\&C systems.
			\item {\it Strongly correlated:} The upper bound, i.e., the boundary AC-CB of Fig. \ref{Gains_quantify} represents the scenario, where the wireless channels of S$\&$C are fully aligned. In other words, the system can always operate at both the sensing- and communication-optimal points without any performance erosion on either side \footnote{ {It is worth pointing out that there is a two-fold tradeoff in ISAC systems, namely, the subspace tradeoff (ST) and deterministic-random tradeoff (DRT) \cite{xiong2022flowing}. In this paper, we mainly focus on the ST from the perspective of S\&C channels, while we refer readers to \cite{xiong2022flowing} for more details on DRT. To avoid out-of-scope digression, we discuss the S\&C performance tradeoffs in the specific context where the data frame length of ISAC signals is sufficiently large and the impact of DRT on the S\&C system is negligible \cite{xiong2022flowing,lu2023random,xiong2023torch}.}}. As illustrated in Fig. \ref{Categories_ISAC}, when the sensing target is also playing the role of a communication user, the wireless resources are jointly exploited by S\&C, leading to high integration and coordination gains.
		\end{itemize}    
	}
	
	It is plausible to find that the blue area $\mathrm{S}_{2}$ represents the performance gain, when both S$\&$C are mutually reinforced by each other, while the green area $\mathrm{S}_{1}$ denotes the non-cooperative case. Following the graphical illustration of various possible integration and coordination gains, we can visualize both the integration and coordination gains in the ISAC systems by the ratio between the areas of the above performance regions, which can be informally expressed as 
	\begin{align}
		\mathrm{Integration~ \&~Coordination~Gains} \varpropto \mathrm{Area}\left(\frac{\mathrm{S}_{2}}{\mathrm{S}_{1}} \right).
	\end{align}
	
	 {
		We remark that this is merely an intuitive representation for the sake of illustration, rather than a rigorous mathematical description. Usually, it is challenging to quantify $\mathrm{S}_{1}$ and $\mathrm{S}_{2}$, which has to be analyzed based on the specific conditions.  As a preliminary work, the authors of \cite{lu2022performance} attempted to quantify the integration gain of ISAC systems in a single-target single-antenna communication user scenario. This was attempted by defining the subspace ``correlation coefficient'' to attain S2 to investigate the coupling effect between S\&C channels. The insight is that a larger ``correlation coefficient'' of S\&C may lead to a larger performance region S2, resulting in higher performance gain. 
	}

	\section{Physical-Layer System Design}\label{System_Design}
	In this section, we tackle \textbf{\emph{Challenges 4-6}} of Fig. \ref{PaperFramework} and present three key aspects of physical-layer system design, with an emphasis on clock synchronization, Pareto-optimal-oriented signaling strategies, and super-resolution methods conceived for the Sub-6G bands. In particular, we commence by briefly discussing the synchronization and phase offset issues of both bistatic and distributed scenarios. Then, inspired by the joint design of the ISAC signal, how to attain Pareto-optimal signaling strategies is discussed as well. Finally, we briefly introduce the family of super-resolution methods designed for wireless sensing networks and propose potential research directions for achieving super-resolution sensing.
	
	\vspace{1em}
	\noindent
	\textbf{\emph{Challenge 4: How Could We Deal With the Clock Synchronization and Phase Offset in the Bistatic and Distributed Deployments?}}
	
	{\it \textbf{1) Background:}} 
	Asynchronous operation imposes widely recognized challenges on collaborative communication networks. Hence sophisticated techniques have been proposed for the conventional communication systems to tackle this problem, e.g., broadcast-based infrastructure synchronization \cite{cena2015implementation}. However, these methods may not be applicable to ISAC systems, since their synchronization requirement related to the sensing function is generally more tight than that of the communication component. For example, the synchronization requirement between BSs relying on different techniques typically ranges from $0.2 ~\mu s$ to $12.8 ~\mu s$ \cite{levesque2016survey}. By contrast, for meter-level positioning accuracy, the synchronization requirement is at the nanosecond/sub-nanosecond level, since the time-of-arrival difference at $0.3 \times 10^9~m/s$ and $0.3~m$ distance is $1 ~\mu s$. Thus, the clock synchronization problem should be reconsidered in ISAC systems, which will be discussed in this subsection.

\begin{figure}[!t]
		\centering
		\epsfxsize=1\linewidth
		\includegraphics[width=\columnwidth]{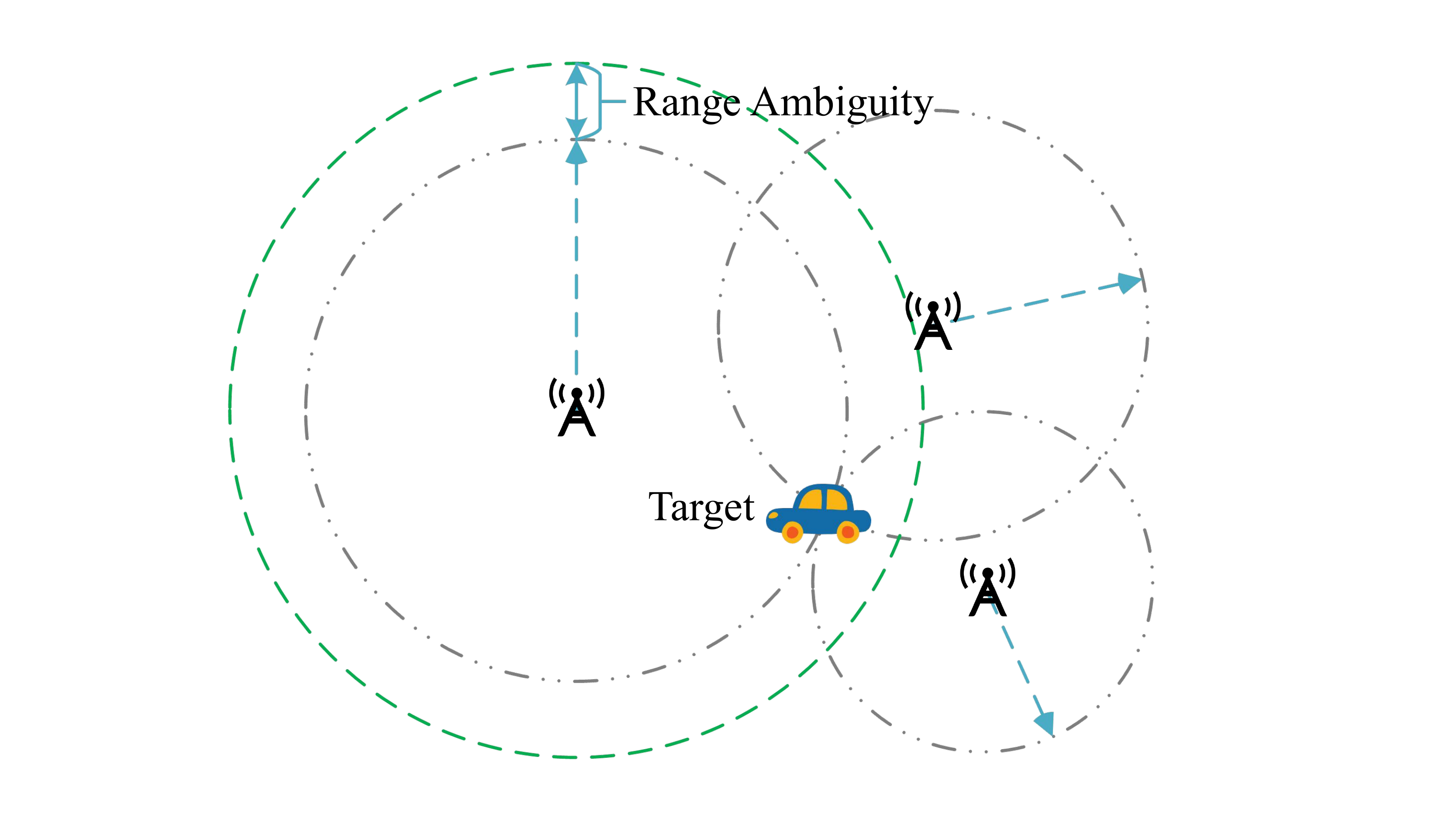}
		\caption{Range bias caused by TMO.}
		\label{TMO}
	\end{figure}

	{\it \textbf{2) Influence of Asynchronous Clock:}} 
	There are several issues related to asynchronous clocks, such as timing offset (TMO), carrier frequency offset (CFO), and random phase shift. Specifically, TMO mainly influences the range estimation accuracy by incurring range bias in time-related estimation algorithms, such as the family of time-of-arrival (TOA) and time-difference-of-arrival (TDOA) based schemes. As shown in Fig. \ref{TMO}, due to the existence of TMO, an extra error is imposed on estimating the propagation delay, and accordingly a range bias is introduced. In the case of TDOA, tight clock synchronism between the receivers is also required. Given the fact that the signal travels at the speed of light, even a tiny estimation error in signal delay, e.g., at the level of nanoseconds, could lead to a range error of meters, which may prevent accurate navigation. The CFO brings about ambiguity when estimating the Doppler frequency, thereby exacerbating the error imposed on the estimation of the target speed. Apart from TMO and CFO, the presence of random phase shifts prevents us from coherently aggregating measurements at different timeslots/packets \cite{zhang2022integration}. Since clock asynchronism may gravely degrade a system's sensing performance, one of the major tasks in designing the ISAC system is that of ensuring tight clock synchronization between the active nodes or, alternatively compensating for the synchronization errors.
	
\begin{figure}[!t]
		\centering
		\epsfxsize=1\linewidth
		\includegraphics[width=\columnwidth]{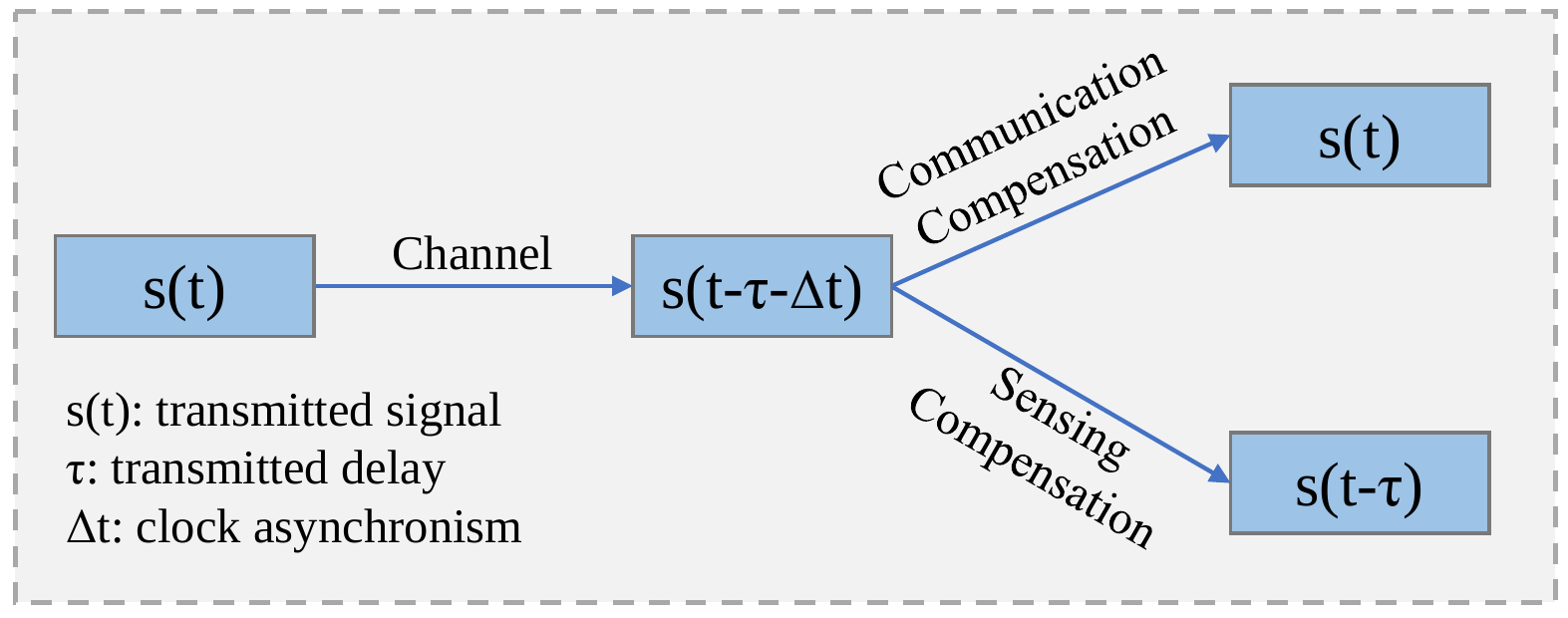}
		\caption{Different compensation techniques in ISAC systems.}
		\label{comp}
	\end{figure}

	To elaborate briefly, communication receivers harness channel estimation which is also capable of simultaneously canceling out both the propagation delay and clock offset. By contrast, an ISAC receiver should perform compensation for different functions in different ways, as shown in Fig. \ref{comp}, where the sensor should extract the target information from the received signal. Therefore, only the clock offset can be removed. This leads to a difference between the synchronization of ISAC and communication-only systems.
	
	{\it \textbf{3) Existing Literature:}} 
	A whole suite of sophisticated methods capable of achieving high-accuracy synchronization have been proposed, for example, by applying a common reference clock \cite{wang2009gps}. The common reference clock may be broadcast by a dominant node of the ISAC system or by the global positioning system (GPS). Then the active nodes of the system may simply lock on to this reference clock to synchronize with each other. As a result, the hardware cost will be significantly reduced, since only a few active nodes should have a high-precision clock. Specifically, by extracting the time information from the GPS signal, a high clock synchronization accuracy can be obtained at the nanosecond level\cite{wang2009gps}. In \cite{xue2021wicsync}, a clock synchronization scheme based on the ultra-wideband (UWB) technique has been proposed. The theoretical results show that the method of \cite{xue2021wicsync} achieves a high clock accuracy (less than 3ns) for 100 nodes. The synchronization is realized by exchanging packets having timestamps among the active nodes. With the help of the UWB technique, high-resolution timestamps can be obtained, and accordingly, high-precision synchronization between the active nodes is achieved. 
	
	For an ISAC receiver equipped with multiple antennas, the clock asynchronism can be canceled out with the aid of signal processing, since the asynchronous clock could be regarded as adding some phase-shifted terms to the expression of channel states. Since the phase-shifted terms imposed by the asynchronous clock across different antennas are approximately the same, mathematical manipulations can be performed to remove these terms. Hence synchronization between active nodes can be readily achieved. We refer readers to \cite{zhang2022integration} for further details concerning the above solutions.
	
	{\it \textbf{4) Future Directions and Potential Solutions:}} 
	Again, although numerous methods have been proposed to address the clock synchronization issues in various application scenarios, they may not be directly applicable to ISAC systems. The GPS signal-based methods require a long synchronization time, which limits their application in high-mobility scenarios. As for signal processing aided techniques, they generally require a multiple antenna receiver, and the synchronization accuracy is lower than that of the GPS-based method. Therefore, how to adapt these schemes to the ISAC system is still an open question.
	
	Applying UWB techniques in ISAC systems is indeed a possible solution. Firstly, UWB schemes can help the ISAC signal obtain high-resolution time stamps to realize a high-precision measurement. Moreover, the UWB technique fits naturally into ISAC systems, since a high carrier frequency and a wide operating bandwidth will be used in future ISAC systems to increase the data rate. However, although UWB ISAC obtains excellent clock synchronization performance, it prevents the application of algorithms based on the narrow-band assumption. Furthermore, the synchronization scheme based on UWB requires two rounds of signal trips between transceivers\cite{xue2021wicsync}, which is also unsuitable for high-mobility scenarios. These issues still have to be circumvented.
	
	The security of time information is also an open research topic. As mentioned above, even a tiny synchronization error will cause a considerable range estimation perturbation. In the case of tracking automotive vehicles, the ghost targets caused by asynchronous timing may lead to accidents. Therefore, further research is required to secure the timing information in the transmitted signals.  {We remark that in the near-field model, the TOA estimation may be eliminated by exploiting the distance information embedded in the array response \cite{yuanwei_OJCOMS,chen20236g,an2023toward}. In ISAC systems, if the targets are also communication users in location division multiple access (LDMA) \cite{wuzilong_LDMA}, their location can also be extracted from the communication unit. It is envisaged that near-field ISAC has promising prospects \cite{wangzhaolin_NFISAC,yang2023enhancing}.}

	\vspace{1em}
	\noindent
	\textbf{\emph{Challenge 5: How Far Are the Existing Designs From the Pareto-Optimal Boundary? }}
	
	{\it \textbf{1) Background:}}
	 {In general, the Pareto-optimal boundary can be interpreted as the optimal case in the ``moderately correlated'' scenarios of S\&C channels of Fig. \ref{Gains_quantify}. Finding suitable waveform designs and signaling strategies for approaching the Pareto-optimal boundary is of considerable importance in ISAC systems.} At the time of writing, research has primarily been focused on optimizing either radar performance metrics under communication constraints or vice versa~\cite{liu2021cramer}. Nevertheless, due to the absence of a Pareto optimization framework for ISAC systems, attaining an optimal performance trade-off between the two functionalities remains challenging. The underlying problem is to find the Pareto-optimal boundary~\cite{2021chenli_Pareto} and to measure the distance between the existing designs and the Pareto-optimal boundary.
	
	{\it \textbf{2) Existing Literature:}} Generally, the waveform design in the existing ISAC research can be split into three categories: radar-centric design \cite{radar_C2019mmwave}, communication-centric design \cite{comm_c2006combining} and joint waveform design \cite{joint_Ckumari2017performance}. The radar-centric approaches are implemented based on the radar probing signals, such as that of pulse interval modulation (PIM) and index modulation to map the communication symbols onto the radar pulses \cite{mealey1963method,Overview_huang2020majorcom}. Similar methodologies, such as the combination of amplitude/phase shift keying (ASK/PSK) and linear frequency modulation (LFM) signals were also proved to be effective for carrying information based on radar probing signals \cite{LFMnowak2016co,LFMzhipeng2015communication}. As for the communication-centric approaches, they are designed relying on existing communication signals and protocols. One can extract the target information of Doppler and delay parameters from the transmitted-and-received signals by classical signal processing techniques, such as the fast Fourier transform (FFT) and the fractional Fourier transform (FrFT). In addition, the IEEE 802.11ad protocol has also been adopted for radar sensing in vehicular networks \cite{8114253}.
	
	Compared to the approaches based on existing radar or communication waveforms, the joint waveform design takes both the S\&C performance into consideration, which is deemed to be a promising direction in ISAC systems as a benefit of its capability of striking a favorable tradeoff between S\&C.  {We remark that sensing MI does not have an exact operational interpretation and MI-oriented S\&C tradeoffs based on \eqref{MIP} may not be suitable for investigating Pareto-optimal design. In contrast to the MI-oriented S\&C tradeoffs in \eqref{MIP}, we highlight another representative formulation as \cite{8386661}
	\begin{subequations}
		\begin{align}
			\min_{ \bm{X} }&~~ \rho \left\| \bm{H}_c\bm{X}-\bm{S}\right\|_{F}^2 + (1-\rho) \left\| \bm{X}-\bm{X}_0\right\|_{F}^2, \label{eq_pareto}\\
			~~~\mathrm{s.t.} &~~ \mathrm{Specific~Practical~Constraints},
		\end{align}
	\end{subequations}}
	where $\bm{H}_c$, $\bm{X}$, $\bm{S}$, and $\bm{X}_0$ denote the communication channel matrix, the transmitted signal matrix, the desired constellation symbol matrix and the ideal radar waveform, respectively. Thus, the first term in \eqref{eq_pareto} represents the multi-user interference encountered in downlink communication, and the second term forces $\bm{X}$ to approach a well-designed pure sensing waveform. The weighting factor $\rho \in \left[ 0,  1\right] $ controls the priorities assigned to S\&C functionalities.
	
\begin{figure}[!t]
		\centering
		\epsfxsize=1\linewidth
		\includegraphics[width=0.8\columnwidth]{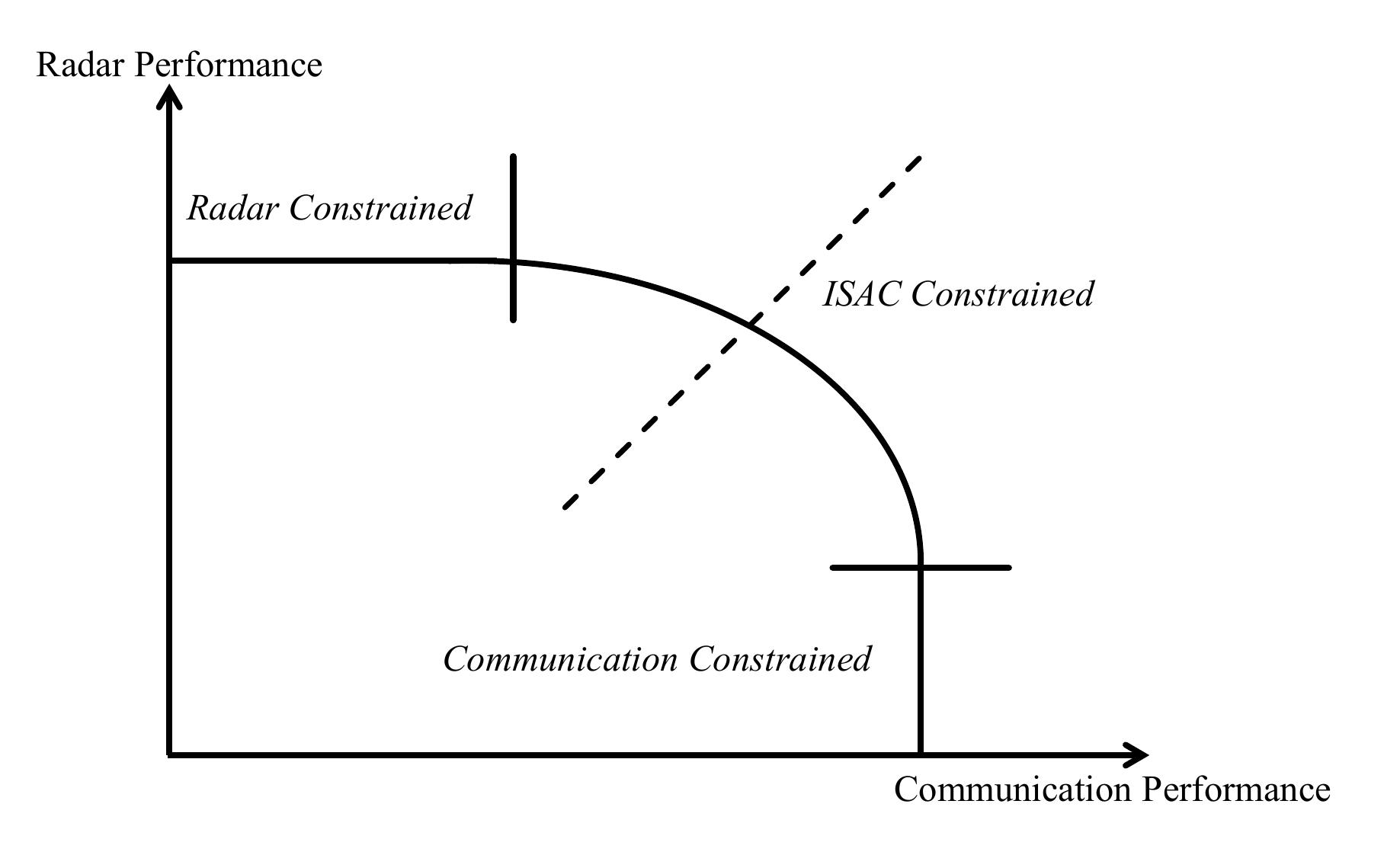}
		\caption{Illustration of the Pareto boundary of achievable performance region.}
		\label{pareto}
	\end{figure}

	{\it \textbf{3) Future Directions and Potential Solutions:}}
	While the ISAC waveform design has been extensively studied in the recent literature \cite{radar_C2019mmwave,comm_c2006combining,joint_Ckumari2017performance}, the Pareto-optimal signaling strategies still remain open challenges. First of all, it is unclear how to define the joint performance metrics for the ISAC system, despite the abundance of existing radar-only and communication-only metrics \cite{wang2020thirty,xu2019sixty}. For instance, the signal-to-interference plus noise ratio (SINR), the achievable sum rate, etc, are widely used in communication-only waveform design. There are also popular metrics such as the CRB, MMSE, and MI for radar-only design. Secondly, the optimal waveform design is rather challenging due to the design conflicts between radar and communication waveform optimization, especially in light of the increased complexity introduced by both the communication and radar constraints. The Pareto boundary of the achievable performance region is shown in the stylized Fig.~\ref{pareto} \cite{2021chenli_Pareto}. In a nutshell, there are still several open questions to be explored, some of which are closely related to the fundamental limits of ISAC, such as,
	
	$\bullet$ Where is the Pareto boundary and how far are the existing designs from the boundary of optimal design?
	
	$\bullet$ Is there any opportunity to attain the Pareto-optimal performance between sensing and communication based on a unified waveform?
	
	These questions may only be answered on a step-by-step basis, commencing from simple twin-component objective functions (OF) based on the bit error rate (BER) of communication and the target detection probability of radar for example. As a next step further, metrics may be added to the OF one-by-one, which of course, expands the search-space. Hence sophisticated reduced-scope search techniques must be conceived.
	
	\vspace{1em}
	\noindent
	\textbf{\emph{Challenge 6: How Could We Improve the Sensing Performance by the Emerging  Super-Resolution Methods?}}
	
	{\it \textbf{1) Background:}} Resolution is one of the most important factors determining the sensing performance, especially for the tasks of localization, imaging, and recognition. Most current cellular networks (e.g., 4G and 5G) operating in sub-6G bands can only provide meter-level accuracy sensing due to the low range and angle resolutions, which are unable to meet the requirements of demanding applications, such as V2X, IoT, etc \cite{JSAC_liu2022integrated}.  
	
	Although the recently proposed millimeter wave and THz communication systems can potentially provide high-resolution sensing service within their limited coverage area, achieving super-resolution sensing in the sub-6G bands is still of significance to reuse the current cellular network for sensing. It is widely exploited in radar signal processing that the distance resolution is determined by the bandwidth $B$ of the transmitted waveforms \cite{richards2014fundamentals}, i.e. $\Delta R = c/2B$, where $c$ is the speed of light, and the angular resolution depends on the size of the array aperture $W_\text{NN}$ \cite{van2002optimum}, i.e., $\Delta \theta = 2\arcsin (0.446\lambda/W_\text{NN})$, where $\lambda$ represents the wavelength of the transmitted signal. Therefore, it is an appealing idea for super-resolution methods to extend both the signal bandwidth and the array aperture in the sub-6G bands.      
	
	\begin{figure}[!t]
		\centering
		\epsfxsize=1\linewidth
		\includegraphics[width=\columnwidth]{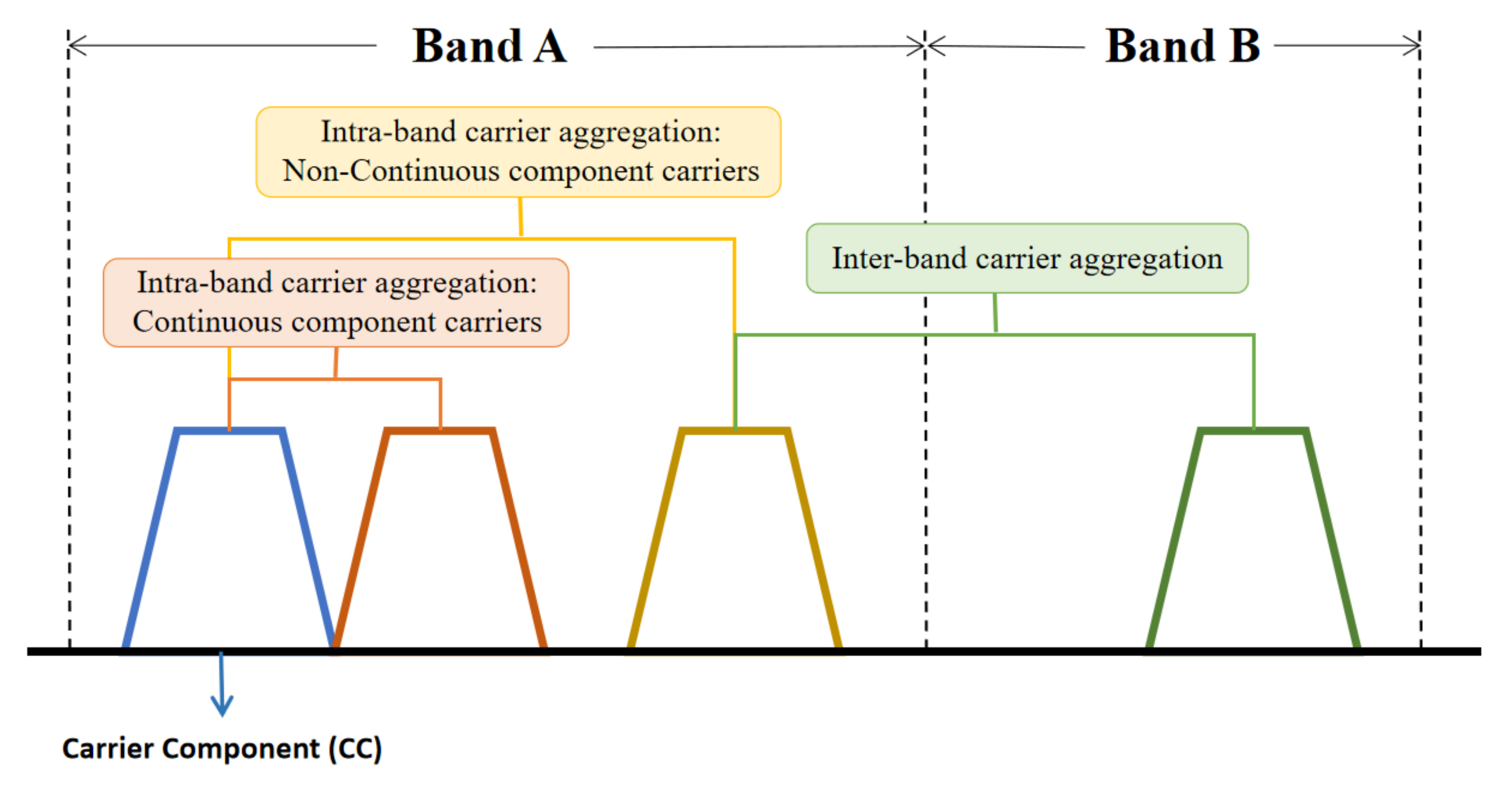}
		\caption{Harnessing carrier aggregation.}
		\label{CA_Scheme}
	\end{figure}

	{\it \textbf{2) Existing Literature:}}
	
	$\bullet$ {\it Carrier Aggregation Assisted Sensing:} Carrier aggregation (CA) is a powerful technique of harnessing multiple component carriers across the available spectrum bands. As illustrated in Fig. \ref{CA_Scheme}, CA can be mainly classified into three types, i.e., intra-band contiguous CA, intra-band non-contiguous CA, and inter-band non-contiguous CA \cite{CA2014}. On the one hand, the communication throughput and peak data rate can be significantly improved through CA techniques \cite{CA2021}. As a further benefit, the bandwidth aggregating several discrete carriers is expected to achieve an improved distance resolution. However, there are also several challenges to be addressed. For instance, the propagation path loss and Doppler shift will be quite different for the non-contiguous carrier components, hence a well-designed carrier selection scheme and a matching resource allocation method are required for circumventing the degradation in S$\&$C performance. Nevertheless, although the CA technique has already been applied in operational communication systems, there are numerous new challenges when considering sensing functionality. For instance, the non-contiguous CA may cause high sidelobes in the ambiguity function of sensing, since the initial phases of discrete carriers are not necessarily continuous, which leads to significant sensing performance degradation.     
	
\begin{figure}
		\centering
		\epsfxsize=1\linewidth
		\includegraphics[width=\linewidth]{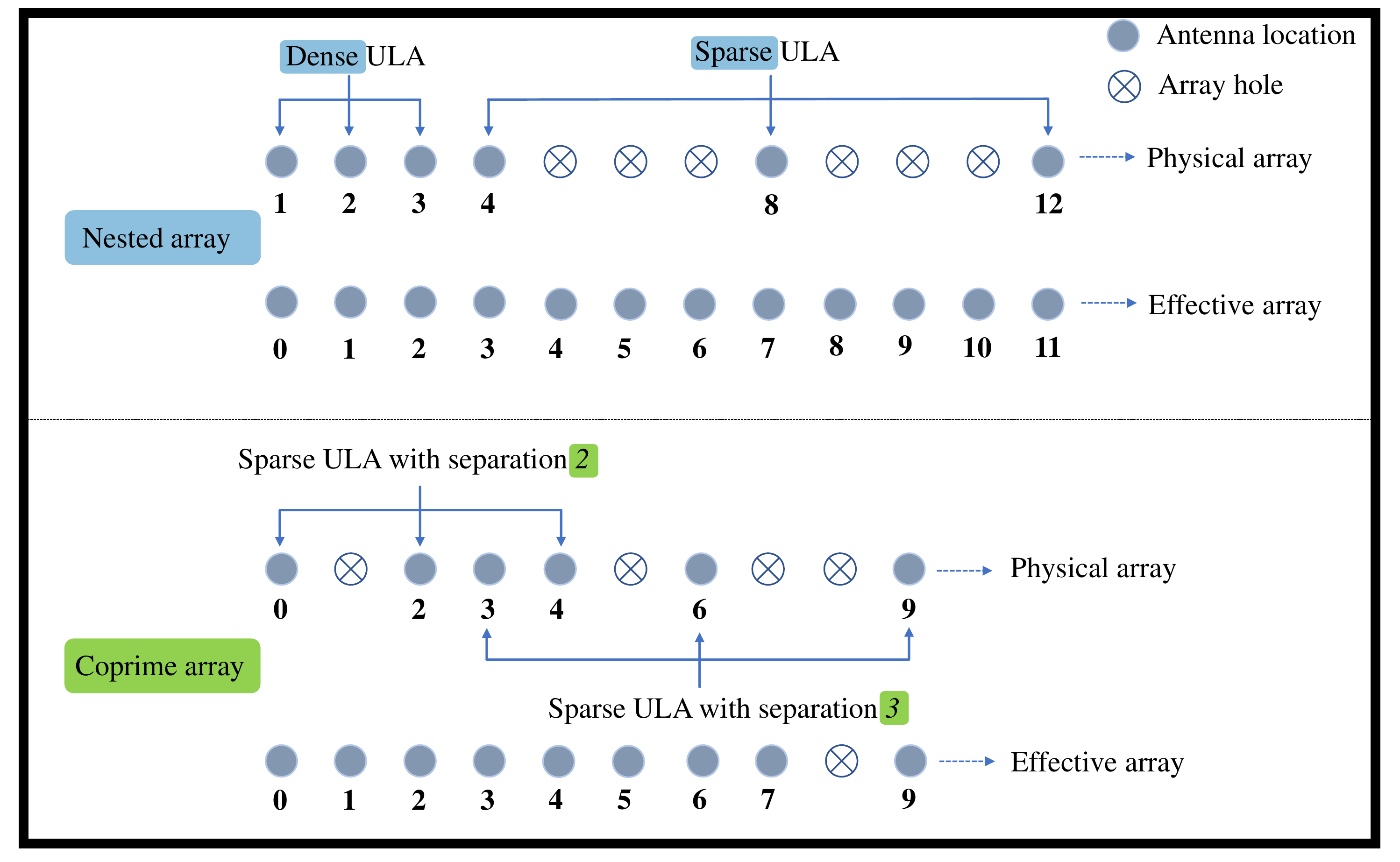}
		\caption{Array geometry of the 6-antenna nested array and coprime array.}
		\label{Sparse_Array}
	\end{figure}

	$\bullet$ {\it Sparse Array Based ISAC Platform:} Compared to the conventional uniform linear array (ULA), the core idea of sparse arrays is converting the sample covariance matrix of the sensor outputs into the so-called difference coarray domain by combing two or more ULAs, which have increased inter-sensor spacing, hence enlarging the virtual array aperture\cite{liu2017cramer}. Specifically, let $\mathbb{S}$ be an integer set of the sensor locations. Then the difference set can be defined as $\mathbb{D}=\{n_1-n_2|n_1, n_2 \in \mathbb{S} \}$. One of the typical sparse array geometries is the nested array \cite{Nested2010}, which consists of a dense ULA with separation 1 (in unit of $\lambda /2$), and a sparse ULA with sensor separation of $(N_1+1)$. The associated sensor locations can be expressed by $\mathbb{S}=\{n|n=1,\cdots,N_1\} \cup \{m(N_1+1)|m=1,\cdots,N_2\}$. As illustrated in Fig. \ref{Sparse_Array}, there are 6 physical sensors, with $N_1=N_2=3$ forming a difference coarray associated with integers spanning from 0 to 1, which significantly enlarges the array aperture compared to the conventional ULA. Another popular array geometry is coprime array \cite{5609222}, which consists of two sparse ULAs with sensor separation determined by a coprime pair of integers $N$ and $M$. The sensor location set is defined as $\mathbb{S}=\{nM|n=0,\cdots,N_1\} \cup \{mN|m=1,\cdots,2M-1\}$. Fig. \ref{Sparse_Array} shows a coprime array with $M=2$ and $N=3$ forming a difference coarray containing consecutive lags from 0 to 7. Although lag 8 is missing (usually called a `hole'), coprime arrays are capable of reducing the mutual coupling effect compared to nested arrays. By using the above sparse arrays, the angular resolution is expected to be substantially improved thanks to the increased virtual array aperture size. Moreover, from the perspective of communication systems, using a sparse array will not bring about intractable challenges for the transmission and reception of communication symbols. Therefore, using a sparse array constitutes a promising technique of acquiring super-resolution at sub-6G bands for ISAC systems.  
	
	$\bullet$ {\it Super-resolution Algorithms for ISAC:} There are two main categories of super-resolution algorithms for array signal processing, namely the subspace-based angle estimation algorithms, such as the multiple signal classification (MUSIC) \cite{bienvenu1983optimality} and the estimation of signal parameters via rotational invariance techniques (ESPRIT)\cite{roy1989esprit}. These kinds of methods are based on the data signal's covariance and hence they are very sensitive to noise, data snapshots, and source correlations. Another family is constituted by the sparse signal reconstruction frameworks, including on-grid \cite{Ongrid2005}, off-grid \cite{Offgrid2011}, and gridless \cite{Gridless2015} algorithms. These methods exploit the spatially sparse nature of the received signals rather than their second-order statistics, where the latter is applicable for limited snapshots, for an unknown number of sources, etc. Furthermore, combining the sparse array geometry with the super-resolution algorithms is also a promising scheme for improving the resolution of ISAC systems. 
	
	{\it \textbf{ 3) Future Directions and Potential Solutions:}} The aforementioned solutions rely on state-of-the-art techniques for improving both the distance and angular resolutions of ISAC systems. However, the unique challenges and opportunities introduced by the ISAC techniques are still widely unexplored. For example, the well-designed dual-functional radar and communication signal is expected to improve joint bandwidth utilization. Furthermore, high-resolution sensing can be achieved by establishing a communication-assisted sensing framework for cooperative networks.

	\section{ISAC Networks \& Cross-Layer Design}\label{ISAC_Network}
	In this section, we mainly explore \textbf{\emph{Challenges 7-8}} of Fig. \ref{PaperFramework} and elaborate on a pair of key concepts in networked sensing, including the cellular architecture as well as cross-layer resource management and protocols. We remark that there is a paucity of research focusing on these two concepts, hence this section is more of a philosophical discussion of potential questions and an outlook on future directions.
	
	\vspace{1em}
	\noindent
	\textbf{\emph{Challenge 7:What is the Potential Cellular Architecture of Future Network of Sensing?}}
	
	{\it \textbf{ 1) Background:}}
	For future radio access networks (RAN), sensing is envisioned as a key functionality to be integrated into the dense cellular architecture, which facilitates the construction of network sensing based on sharing the dual-functional transmitted signals and hardware implementation. In this way, both the user equipment and cellular network can sense the surroundings and enable a variety of applications such as environmental monitoring and target detection.
	
	{\it \textbf{ 2) Existing Literature:}}
	In the 5G cellular network, multi-station cooperation based on the cloud RAN (C-RAN) concept is widely used \cite{pan2018user}. For communication-only systems, the inter-cell interference of remote radio heads (RRHs) can be mitigated by specifically designed interference management strategies, such as soft fractional frequency reuse or coordinated multi-point techniques~\cite{7096298}. Furthermore, cooperative transmission in the scenario of user-centric C-RAN has also been investigated, where usually global channel state information is required \cite{pan2018user}. In general, inter-cell interference is decked to be a harmful factor that has to be reduced.

	\begin{figure}[htbp]
		\centering
		\epsfxsize=1\linewidth
		\includegraphics[width=\columnwidth]{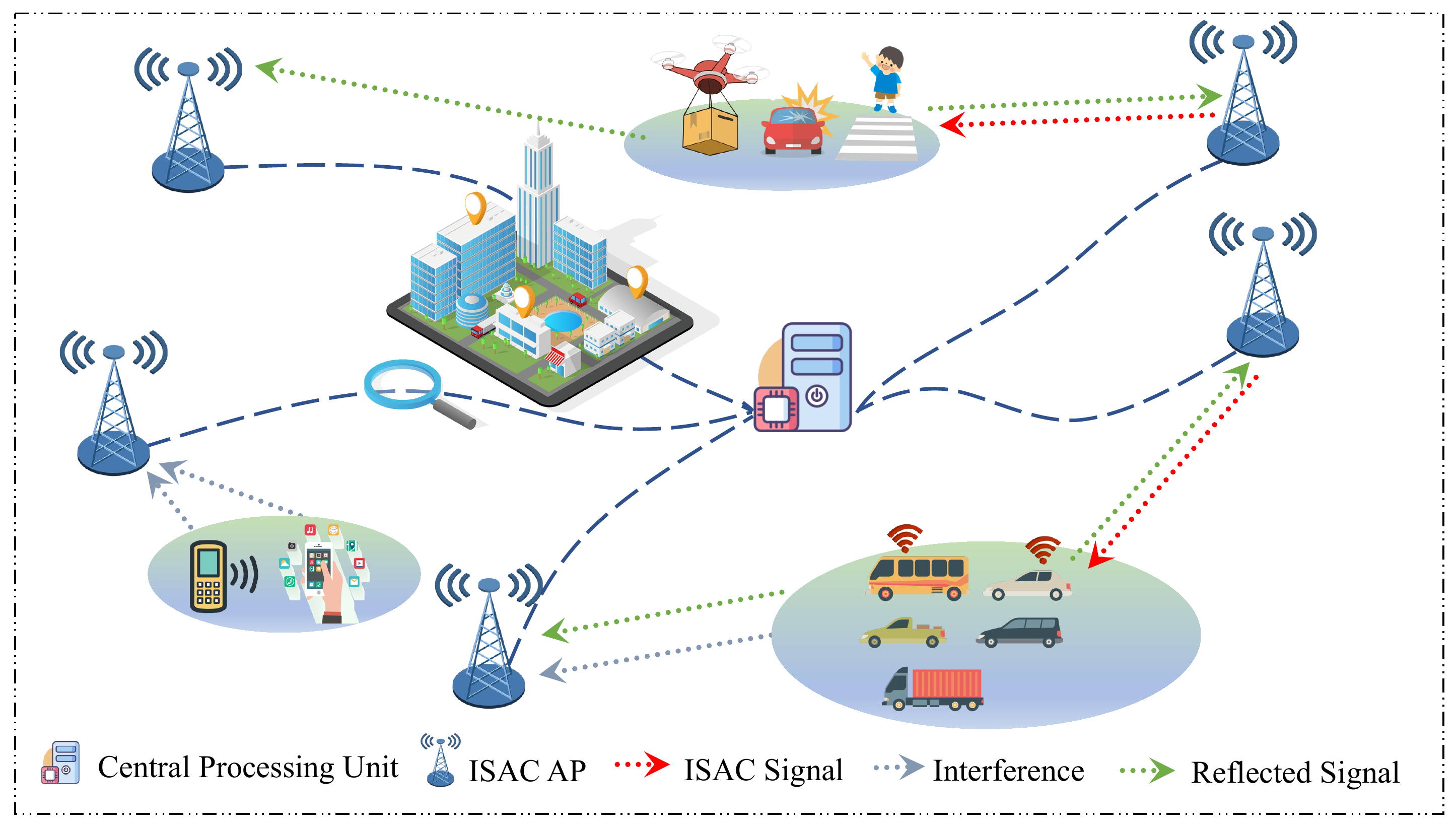}
		\caption{Potential cellular architecture in ISAC systems.}
		\label{cell}
	\end{figure}
	
	Multi-station cooperation is potentially capable of expanding the detection range and/or improving the SNR as well as the detection probability. A typical multi-station-multi-cell cooperation scenario is illustrated in Fig.~\ref{cell}. Specifically, the single target (e.g., UAV \cite{meng2022uav}) can be cooperatively detected by multiple stations from different viewing angles, which can provide accurate overall sensing of the detected target. In this scenario, the cooperation of cells and stations is of vital importance. However, the inter-cell interference in ISAC networks may be rather different from that of communication-only networks. Inter-station interference may contain useful information with respect to the targets of interest, which has to be exploited for enhancing the sensing performance, rather than being canceled. In addition to receiving the echo signal originating from the monostatic sensing operation, each station may also receive ISAC interference transmitted by other BSs or UEs.
	
	{\it \textbf{ 3) Future Directions and Potential Solutions:}}
	Both the cell architecture and the station location are important factors. The cellular architecture of ISAC maybe expected to rely on the existing station location resources, which may be a convenient way of realizing the integration of communication and sensing networks. However, the communication-only design principle may not be appropriate for frequency reuse in terms of functionalities and locations.
	
	The cell-free massive MIMO architecture~\cite{ngo2017cell,bjornson2013optimal,OJCOM_cellfree_d2020analysis}, on the other hand, may be a promising one in support of ISAC systems. In cell-free massive MIMO, the access points (APs) can be distributed over a larger area and provide flexible load-balancing, where a group of users may be supported by a group of APs. A larger number of communication users and targets can be cooperatively served or detected simultaneously at a high S$\&$C performance.
	
	 {
	It is important to note that networked sensing holds significant promise in terms of achieving seamless sensing in the future, particularly with the proliferation of ISAC base stations. In this context, these distributed base stations have the capability to accumulate vast amounts of sensory data, which is then relayed via backhaul links to the computational processing unit for extensive data analysis and for the subsequent extraction of sensing information. This integration often demands the seamless amalgamation of multistatic sensing, cooperative communications, and computational elements \cite{zhuguangxu2022ISCC}. However, networked sensing also has its own challenges that require immediate attention. For example, the collaboration in multistatic systems necessitates precise clock synchronization, and the presence of clutter interference in complex environments gravely impacts the overall efficacy of networked sensing.}
	
	\vspace{1em}
	\noindent
	\textbf{\emph{Challenge 8: How Could We Conceive Resource Management Schemes and Protocols Specifically Tailored for the ISAC Network?}}
	
	{\it \textbf{1) Background:}} In ISAC network of the future, it can be predicted that the request for sensing services arises usually randomly and unexpectedly. Based on Fig. \ref{cell}, we have to conceive effective resource management schemes in response to these sensing requests. In this spirit, the frame structure and the resource scheduling algorithms have to be tailor-made in support of flawless ISAC services.
	
	{\it \textbf{2) Existing Literature:}} 
	The authors of \cite{c1} designed a sophisticated PHY layer as well as a resource allocation strategy and then evaluated the performance tradeoffs between S$\&$C. However, if the sensing services are heavily congested, the ISAC-BS cannot respond to all the requests in time, since the PHY layer resources, such as the bandwidth or time slots, are limited. To satisfy the emerging burst of sensing requests in ISAC networks, efficient cross-layer resource management techniques have to be conceived. The cross-layer optimization of traditional wireless communication has been widely investigated \cite{c2, c3, c4, c6}, but the cross-layer optimization of ISAC networks is in its infancy, which may involve several open system interconnections (OSI) layers \cite{c3}. We commence our discourse with the PHY, followed by the media access control (MAC), network, and application layer. 
	
	We briefly highlight the open problems in the cross-layer optimization of ISAC networks in TABLE \ref{cross_layer_optimization}, and expound on them in the following. 
	
	\begin{table}[t!]
		\caption{Resource management schemes and protocols for ISAC network.}
		\centering
		\begin{tabular}{c}
			\includegraphics[width=\columnwidth]{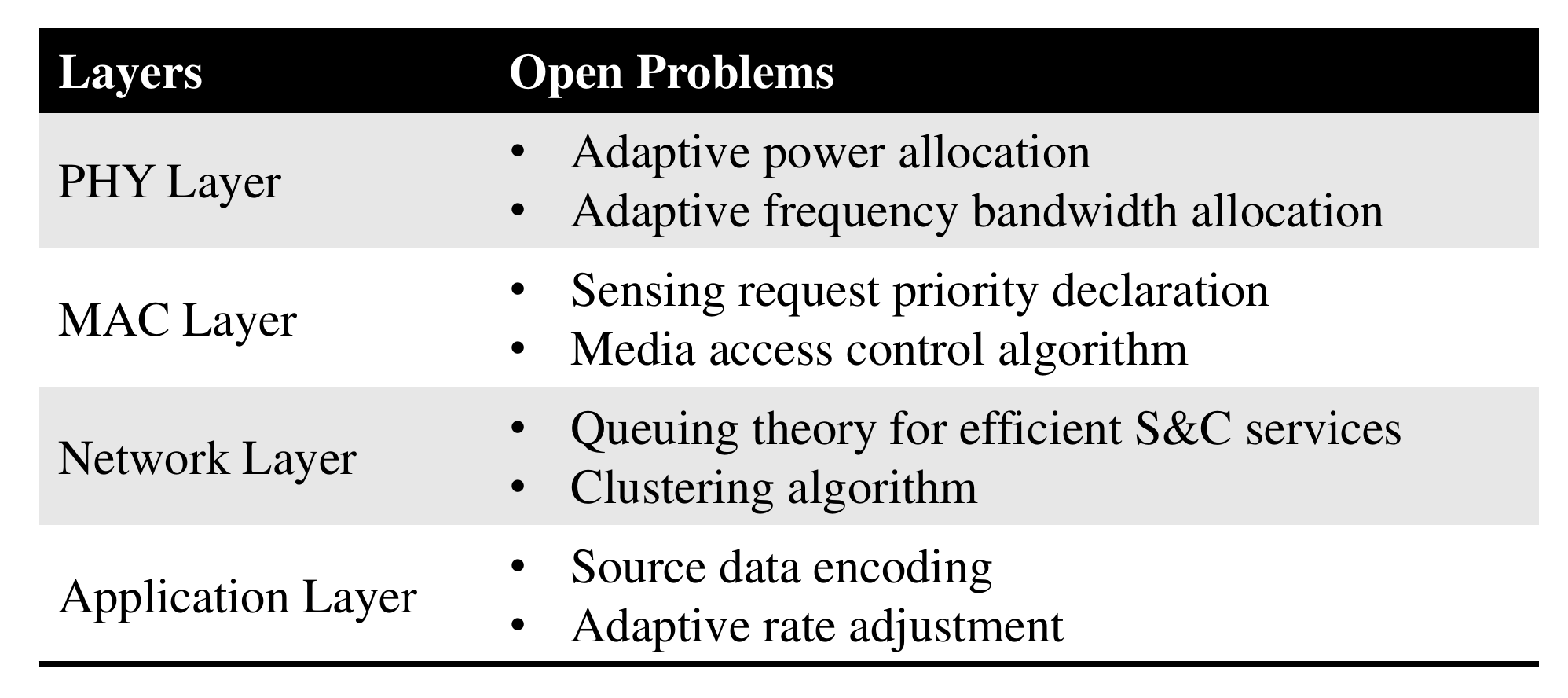}
		\end{tabular}
		\label{cross_layer_optimization}
	\end{table}
	
	$\bullet$ {\it PHY layer:} Although noise and interference reduction have been widely studied in the context of dedicated communication or sensing functionality, handling the interference caused by echo waves has become a challenging new problem in the ISAC scenarios. Due to the shared use of limited resources, such as spectrum and hardware, ISAC is expected to enhance both the S$\&$C performance. However, the interference caused by echo waves plays different roles in communication and sensing services. For instance, in wireless communication scenarios, all propagation paths can be leveraged to improve the communications performance, as valuable signal energy may be gleaned from every path. Conversely, for sensing services, only the specific paths reflected by the targets of interest are desired. Therefore, to support ISAC, it is essential to distinguish sensing echoes from the received multipath signal, which remains an open issue that needs to be resolved.
	
	$\bullet$ {\it MAC layer:} To handle the burst of sensing requests, a feasible strategy is to prioritize S$\&$C requests based on their specific requirements. For instance, the sensing requests of high-velocity vehicles should receive higher sensing priority and better Quality of Service (QoS) than those of stationary objects. For sensing requests with a higher priority, the BS should respond promptly and allocate more resources to deliver high-specification services. To implement this feature, the control frame structure of the Media Access Control (MAC) layer has to be designed for low latency. As the ISAC network simultaneously provides both S$\&$C services, the channel access procedure has evolved from that of traditional wireless networks. For example, the Request To Send / Clear To Send (RTS/CTS) channel access mechanism is commonly used in 802.11 wireless networks to avoid frame collisions. However, in the ISAC network, the control frames can also serve sensing tasks. In a scenario where there is a bistatic sensing link in the ISAC network, the sensing initiator sends an RTS frame that has both S$\&$C functions. Subsequently, the sensing receiver has to send the sensing result back, while the communication receiver sends the CTS back to the sensing initiator. To avoid a collision between the feedback of sensing results and wireless communication control frames, the sequence of control frame interactions in the channel access procedure has to be redesigned.
	
	$\bullet$ {\it Network layer:} Compared to traditional wireless communication networks, the PHY layer of the ISAC network has additional sensing tasks. To evaluate the QoS of both sensing and communication, additional metrics such as sensing delay, bandwidth utilization, and sensing accuracy must be considered. In traditional wireless networks, typically queuing theory is used for efficiently scheduling communication. However, the challenge in ISAC is how to incorporate sensing metrics into queuing theory. In addition to queuing theory, clustering algorithms can also be used to support S$\&$C services. In contrast to traditional sensing networks, users in ISAC networks may have both S$\&$C service requests, and may only be in communication with a sensing target for a short period. Therefore, the sensing clusters in ISAC evolve rapidly, making traditional clustering algorithms unsuitable. To address this problem, agile adaptive clustering algorithms must be developed. For example, the BS and users that have sensed the same target can automatically be grouped into a cluster, expediting the clustering process.
	
	$\bullet$ {\it Application layer:} In the ISAC network, transmitting results sensed by the BS to the user is a critical operation. To improve transmission efficiency, effective source encoding mechanisms can be harnessed for reducing the physical resource requirements. As a benefit, more sensory data can be transmitted within a given bandwidth. When a user sends a sensing request to the BS, it can also provide the target class, indicating whether it is a vehicle or a pedestrian. This is advantageous because different types of sensing targets require varying amounts of sensory data. Vehicles or pedestrians have a higher priority in sensing services, since they have to signal their velocity, movement direction, and location to describe their states. In contrast, stationary objects such as trees or buildings require less data, perhaps only their location. Based on this difference, we can employ a more complex source data encoding scheme for high-priority objects to improve the sensory data transmission efficiency, while simultaneously giving cognizance to the compression delay and complexity. By contrast, low-priority objects can be encoded by low-complexity source encoding schemes, thereby conserving computational resources.
	
	{\it \textbf{3) Future Directions and Potential Solutions:}}
	The above discussions highlighted the potential problems in the resource management of ISAC network. The optimization strategies were introduced layer-by-layer. However, since every optimization action has the potential to increase the cross-layer interaction, it is important to consider their long-term effects right across the entire framework\cite{c5}. Moreover, in future research on ISAC techniques, we have to formulate bespoke optimization problems. We conclude by highlighting some of the open problems at a glance in TABLE \ref{cross_layer_optimization} again.
	
	\section{ISAC Applications}\label{Security_Wifi}
	 {In this section, we critically appraise \textbf{\emph{Challenges 9-10}} of Fig. \ref{PaperFramework} and first highlight that the security and privacy issues of wireless sensing systems have to be addressed before integrating sensing into existing cellular networks and supporting ISAC applications.} Then, we consider a representative use case related to the sensing of human activities by utilizing wireless signals. Besides, we elaborate on a suite of open questions and their potential solutions.
	
	\vspace{1em}
	\noindent
	\textbf{\emph{ {Challenge 9: How Could We Provide Security and Privacy Guarantees Within ISAC Applications?}}}
	
	{\it \textbf{1) Background:}} In the next-generation era, sensing may be viewed as a compelling service in support of emerging applications, such as the smart home and V2I networks \cite{c1,JSAC_liu2022integrated}. However, due to the open nature of the wireless sensing medium, the sensing services incur potential security risks and privacy issues \cite{furqan2021wireless}. Specifically, malicious entities may overhear the CSI of the targets that itself may be confidential target-related information \cite{wang2022deeplearning_gender}. Malicious agents may also misuse localization information \cite{ dimas2017spectrum,al2019mimo} as well as the information they inferred concerning target movements \cite{wang2022deeplearning_gender}. As a result, those malicious entities may become capable of breaching the users' privacy, or of contaminating the legitimate reception. We refer to contamination of the legitimate signal as a security problem and to privacy breaches as privacy problems (mainly relevant to the legitimate target), respectively.
	\begin{figure}[!t]
		\centering
		\epsfxsize=1\linewidth
		\includegraphics[width=\columnwidth]{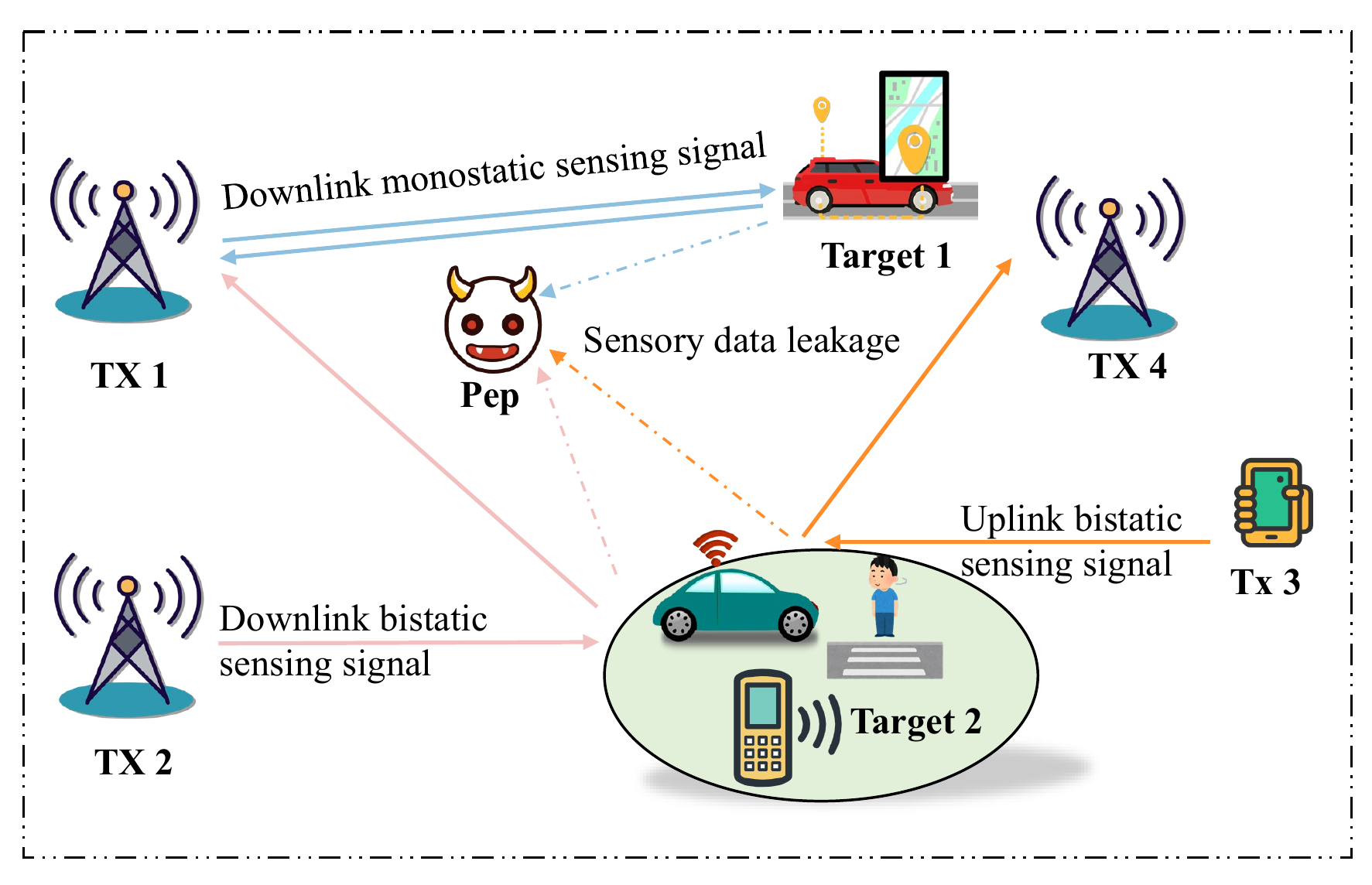}
		\caption{Sensing security and privacy scenarios.}
		\label{Sensing_Privacy}
	\end{figure}
	
	{\it \textbf{2) Existing Literature:}} In radar-communication spectrum sharing scenarios, one has to guarantee sensing security for the transmitter, since the transmit precoding matrix assigned to the communication system contains implicit information about the radar \cite{dimas2017spectrum,al2019mimo}. Hence, the authors of \cite{dimas2017spectrum} investigated different precoder matrices to simulate an adversary inference attack and characterized the associated risks. As a further advance, the authors of \cite{al2019mimo} proposed another precoder design by considering the tradeoff between the interference power and radar security. The authors of \cite{jiao2021openwifi} proposed to apply a so-called CSI ``fuzzer'' to enhance the privacy of WiFi CSI applications. On the other hand, although wireless sensing is capable of potentially avoiding the direct leakage of the targets' images relying on visual acquisition \cite{wang2022deeplearning_gender}, the channel impulse response might also contain private information concerning the targets, such as peoples' movement in human action sensing \cite{zeng2020multisense}.
	Recent advances have revealed for example optimal pilot designs conceived for eavesdropping-resistant channel estimation \cite{Channel_Estimation_zhu2020optimal}, as a potential solution for protecting the security of sensing services. 
	
	To provide further intuitions about the issues of sensing security and privacy, let us elaborate further with the aid of Fig. \ref{Sensing_Privacy}. For example, in a monostatic downlink sensing scenario, the transmitter (Tx 1) wishes to sense a legitimate target (Target 1). It was reported that the precoding matrix might contain the location of the radar transmitter \cite{dimas2017spectrum}. Due to the leakage of the standardized transmit waveform, a potential unauthorized entity, a.k.a. a peeper (Pep) wishes to intercept the designed waveform and contaminate it by malicious spatial interference. Meanwhile, the sensing information that contains the presence, the location, and the behavior of the targets may be reflected by Target 2 and intercepted by Pep, in which case the target's privacy is leaked (see the dashed line in Fig. \ref{Sensing_Privacy}). It may be anticipated that more and more sensing security and privacy problems will occur in wireless sensing, as illustrated for downlink/uplink bistatic sensing in Fig. \ref{Sensing_Privacy}.
	
	{\it \textbf{3) Future Directions and Potential Solutions:}} In contrast to classical pure communication security, the sensing systems seek to convert the TIR $\bm{H}_s$ instead of the transmit signals $\bm{X}$. While some of the existing communication security approaches may be borrowed to secure the sensing network \cite{zou2016survey}, there are still quite a lot of open issues to be addressed. In the sequel, we briefly discuss potential solutions from different perspectives.
	\begin{itemize}
		\item {\it Information theoretic perspective:} The leakage of target-related information, which is usually carried by the transmitted waveform, can be mitigated by optimizing the transmit signals in order to maximize the MI $I(\bm{Y}; \bm{H}_s | \bm{X})$ of the transmitter while minimizing the MI at Pep. 
		\item {\it Physical layer perspective:} To reduce sidelobe leakage of the sensing beam, one can minimize the energy radiation in the Pep's direction in the spatial/angular domain by utilizing transmit beamforming techniques, so that the Pep cannot intercept the waveform leaked. If the Pep is located within the spatial/angular beam of the target, another potential solution to avoid interception is to add artificial noise. More explicitly, one may specifically condition the artificial noise that is only known by the transmitter, while contaminating the Pep's reception.
		\item {\it MAC layer perspective:} In typical wireless standards, both the payload waveform and the pilot organization are openly accessible, which may however cause privacy leakage in sensing systems. Hence innovative techniques are required for protecting the S$\&$C privacy and security by conceiving sophisticated authentication and access control solutions. By contrast, the legitimate transmitters/receivers can be identified through location-, angle-, Doppler- and even channel-aware secret key generation, while protecting them from unauthorized parties.
	\end{itemize}
	
	\vspace{1em}
	\noindent
	\textbf{\emph{Challenge 10: How Could We Achieve MOMT Recognition by Wireless Signals?}}
	
	{\it \textbf{1) Background:}} Identifying human targets and recognizing their actions is essential for numerous context-aware applications, such as assisted living, health monitoring, and intelligent transportation. Compared to the commonly used camera-based human sensing solutions, the wireless signal-based approach is more robust to environmental variations and its coverage tends to be more ubiquitous. Camera-based and beamforming-aided techniques may also be beneficially combined \cite{zheng2023vision}.
	
	\begin{figure}
		\centering
		\includegraphics[width=1\linewidth]{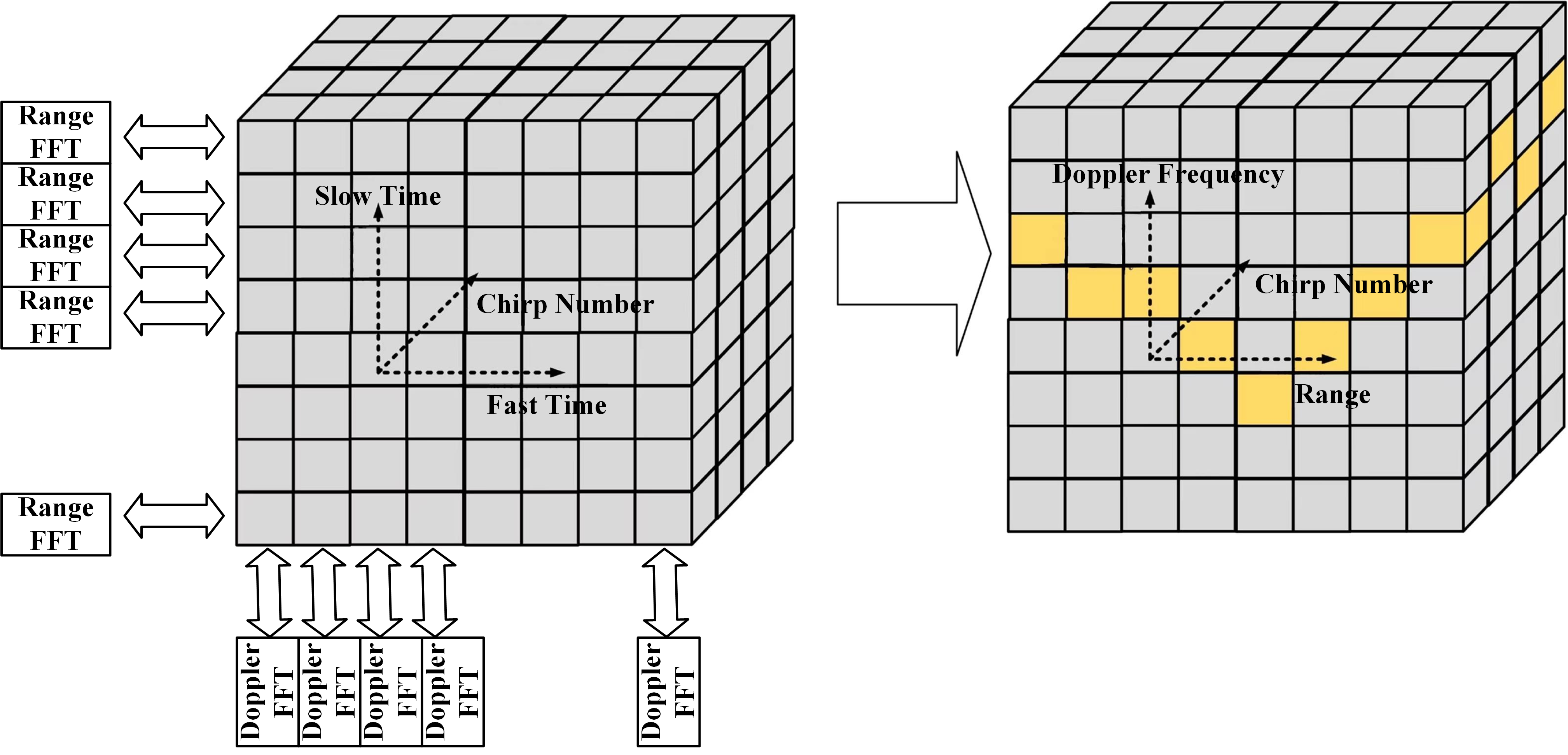}
		\caption{Radio signal processing pipeline for human sensing.}
		\label{Fig1}
		\label{RDS_Processing}
	\end{figure} 
	
\begin{figure}[t!]
		\centering
		\includegraphics[width=1\linewidth]{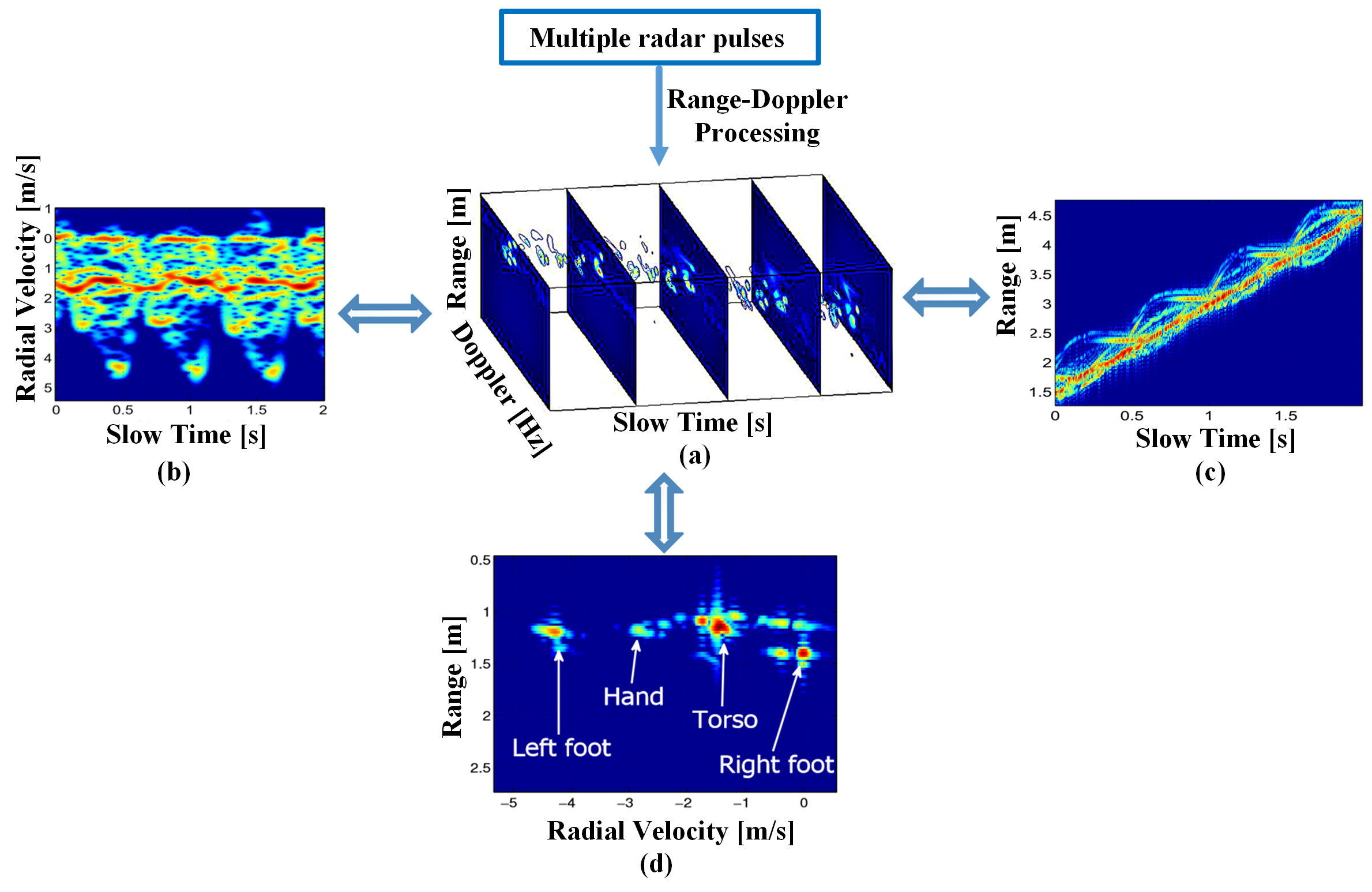}
		\caption{Sensing measurements that contain different information \cite{li2019survey}. (a) 3D time-Range-Doppler map, (b) 2-dimensional (2D) time-Doppler map, (c) 2D time-range map, and (d) 2D range-Doppler map.}
		\label{Fig2}
	\end{figure}

	There are two main categories of wireless human sensing methods, i.e. radar-based and WiFi-based human sensing. As a benefit of its ability to estimate speed and range (time delay), radar has natural advantages in terms of near-field human sensing and has been increasingly employed in miniaturized portable form. Given a common high-accuracy clock signal, radar systems are capable of achieving precise transceiver synchronization even under bistatic and networked deployments, facilitating accurate human-related feature extraction. Furthermore, compared to WiFi systems, radar has higher signal bandwidth and hence has finer range resolution in separating multiple targets or components. Radar-based human sensing is achieved by measuring the movement parameters of human targets with the aid of the received echo signals, such as range, Doppler speed, and angle. The general radar signal processing pipeline of human sensing is illustrated in Fig. \ref{RDS_Processing}. As shown in this figure, when considering a continuous frequency modulation wave (FMCW) based radar as an example, the reflected echo signals can be transformed into a time-varying sequence of ``fast time-slow time'' snapshots. By performing the fast Fourier transform (FFT) along the fast-time and the slow-time dimensions, respectively, the 2D radar snapshots can be converted into a ``range-Doppler frequency'' map. In this way, a series of ``fast time-slow time'' data matrices can be transformed into a 3D "time-range-Doppler frequency" data cube, as shown in Fig. \ref{Fig2}(a). The time-varying range and Doppler speed parameters of human targets can be estimated With the aid of the 3D data cube. In addition, by compressing one of the three dimensions, three types of radar maps can be obtained. For instance, by summing the 3D data cube along the range domain, a 2D ``time-Doppler frequency'' spectrogram (see \ref{Fig2}(b)) is acquired, which depicts the variations of human moving speed versus time. By compressing the 3D cube along the time domain, a ``range-Doppler frequency'' map (see \ref{Fig2}(d)) could be obtained, which can be employed for distinguishing different targets according to their range and speed information. Furthermore, with the aid of an antenna array at the receiver, angular human target information can be estimated using algorithms like multiple signal classification (MUSIC) \cite{wang2022single}. 
	
	\begin{figure}[t!]
		\centering
		\includegraphics[width=1\linewidth]{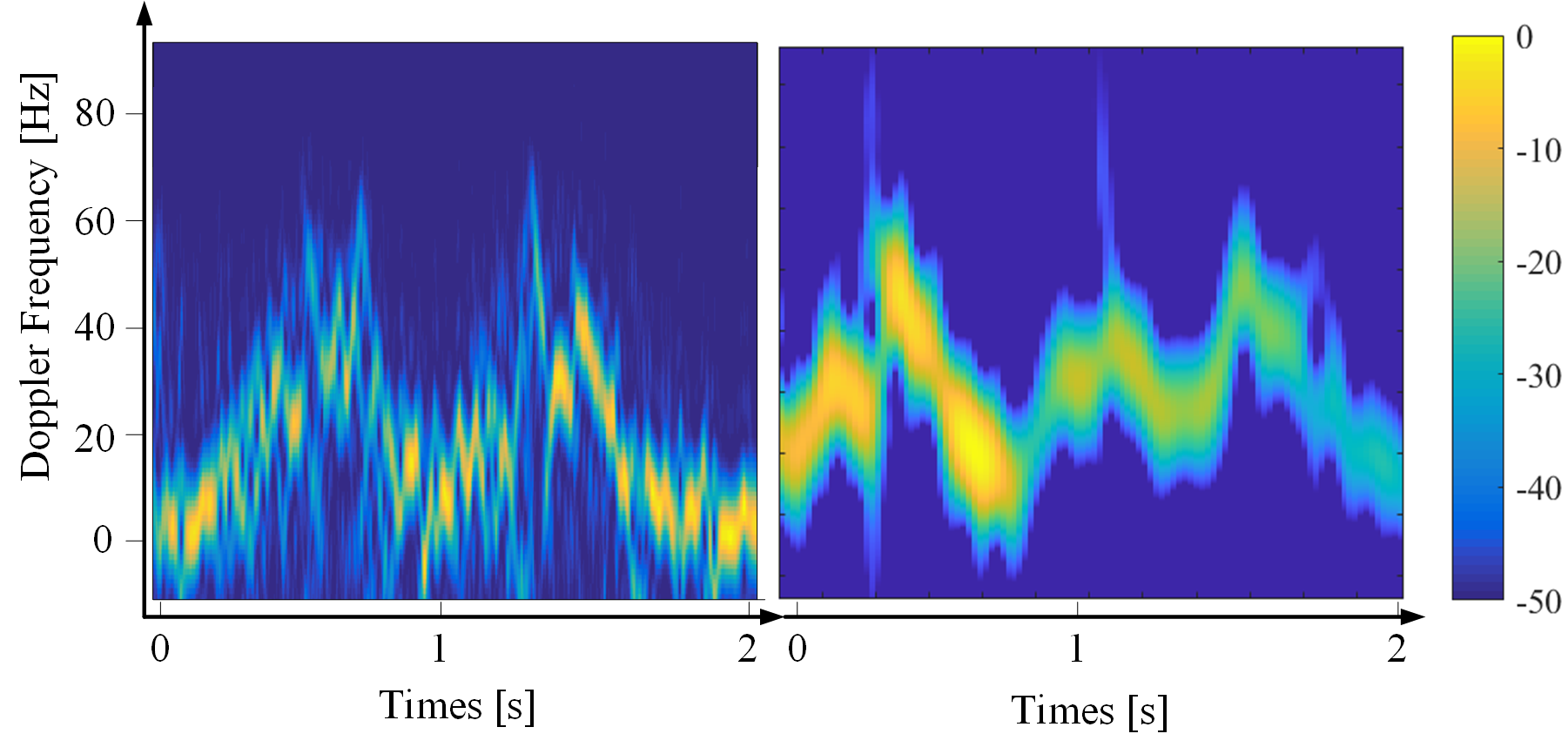}
		\caption{Time-Doppler frequency maps from a WiFi system and a 5G NR system, respectively.}
		\label{Fig3}
	\end{figure}

	{\it \textbf{2) Existing Literature:}} Again, in ISAC systems, both the pilot signals and data frames can be employed for sensing. For sensing relying on their data frames, the transmitted communication information has to be first demodulated from the received signals, and then the demodulated data frame can be exploited for sensing. Another option is to exploit the CSI estimated with the aid of pilot symbols, which is by far the most appealing solution. The CSI captures how human targets and other objects in the environment affect the propagation of ambient wireless signals. In this context, the classic MIMO-aided OFDM technology and its diverse multicarrier relatives can provide a 3-dimensional (3D) matrix of complex values, showing the variations of multipath channels in the presence of moving human targets in the time, frequency, and spatial domains. Due to the low distance-range resolution caused by the narrow bandwidth, ISAC systems cannot directly estimate the target range. Hence the existing solutions tend to generally estimate the relative range of the target \cite{qian2017widar}. Furthermore, since no common clock signal is available in the current communication systems, and the WiFi systems are mostly bistatic, there is undesirable timing offset, carrier frequency offset, and sampling offset in the received CSI measurements, substantially degrading the performance of human sensing tasks \cite{zhang2022integration}. As a result, phase offset removal is a vital step for human sensing. In addition to constructing a reference signal, the commonly utilized single-node solutions can be classified into cross-antenna 
	cross-correlation (CACC) \cite{10.1145/3210240.3210314} and cross-antenna signal ratio (CASR) aided approaches \cite{10.1145/3351279}. Based on these phase removal solutions, both modeling-assisted \cite{wu2021wifi} and learning-enhanced techniques \cite{chen2018wifi} have been proposed for estimating human-related parameters and used for different human sensing tasks.

	 {In the realm of human activity recognition (HAR) using radio signals, the process can be broadly categorized into feature extraction and activity identification, both of which benefit from machine learning (ML) techniques used for enhancing the sensing performance. In what follows, we briefly discuss the ML methods of HAR: }
	
	 {
		$\bullet$ {\it Feature Extraction}: Radio features are obtained through manual feature engineering or automatic machine learning (ML) algorithms \cite{lixinyu2022humanactivity}.Manual feature engineering typically relies on the amplitude and phase of received signals, as they are influenced by human activity  \cite{lixinyu2022humanactivity,li2019survey}. Time-varying Doppler/micro-Doppler frequency shifts are effective for single-person activity recognition, representing radial velocities of various body parts \cite{1pegoraro2020multiperson}. Additionally, the domain-independent body-coordinate velocity profile (BVP) combines Doppler frequency, orientation, and location to characterize human gestures \cite{2zheng2019zero}. In multi-person scenarios, the spatial parameters like range and angle are essential for distinguishing signals from different targets. ML, especially deep learning (DL), automates HAR-related feature extraction \cite{3zheng2021enhancing}. While early research combined convolutional neural networks (CNNs) with radio-based HAR feature extraction, this approach can cause temporal information loss. To address this issue, memory-enabled recurrent neural networks (RNNs) are often adopted, despite their high computational complexity, resulting in significantly improved performance \cite{li2019survey}.}
	
	 {
		$\bullet$ {\it Activity Recognition}: HAR algorithms fall into two categories: model-based and learning-based \cite{4wu2017device}. Model-based approaches, like the Fresnel Zone and Angle of Arrival models, mathematically define the relationship between human activity and signal variations. These approaches quantify human movement-related parameters through signal dynamics and hold the potential for fine-grained activity recognition and exploring sensing limits. Learning-based algorithms focus on mapping sensing measurements to human activity labels using pre-extracted features \cite{zeng2020multisense,6jokanovic2017fall}. Both manually and automatically extracted features can be used, but learning-based methods often combine ML-based feature extraction and classification for end-to-end activity classification. Common ML classifiers include Hidden Markov Models (HMM), K-nearest neighbors (kNN), and support vector machines (SVM), while DL classifiers encompass fully connected neural networks, CNNs, RNNs, and more.}
	
	{\it \textbf{3) Future Directions and Potential Solutions:}} In practice, both WiFi and cellular systems can be utilized for sensing. The former is more suitable for indoor human sensing, since the hardware is simple and the system's power is lower, while the latter is more promising for outdoor sensing tasks. We used a pair of wireless communication systems, including i) a WiFi system associated with 40
	MHz bandwidth, 5.0 ms pulse repetition interval (PRI), and 5.8 GHz central frequency and ii) a 5G new radio (NR) BS system with 100 MHz bandwidth, 10.0 ms PRI, and 3.6 GHz central frequency, to collect sensing measurements. Both the WiFi and NR systems collect signal reflections from separately deployed receiver antennas, so they are working in the bi-static sensing mode. Then, the time-Doppler maps obtained from a WiFi system and a 5G NR system are shown in Fig. \ref{Fig3}~\cite{lixinyu2022humanactivity}, respectively. The maps describe the variations of Doppler frequencies, in which a human jumps twice consecutively. It can be seen that the intensity of the NR spectrogram is strong, indicating that the NR system may be deemed more robust to interference and has a wider coverage range for human sensing. Furthermore, the lower PRI of WiFi results in a finer time-resolution at the left of Fig. \ref{Fig3}~\cite{lixinyu2022humanactivity}.

	\begin{figure*}[t!]
		\centering
		\epsfxsize=1\linewidth
		\includegraphics[width=1\textwidth]{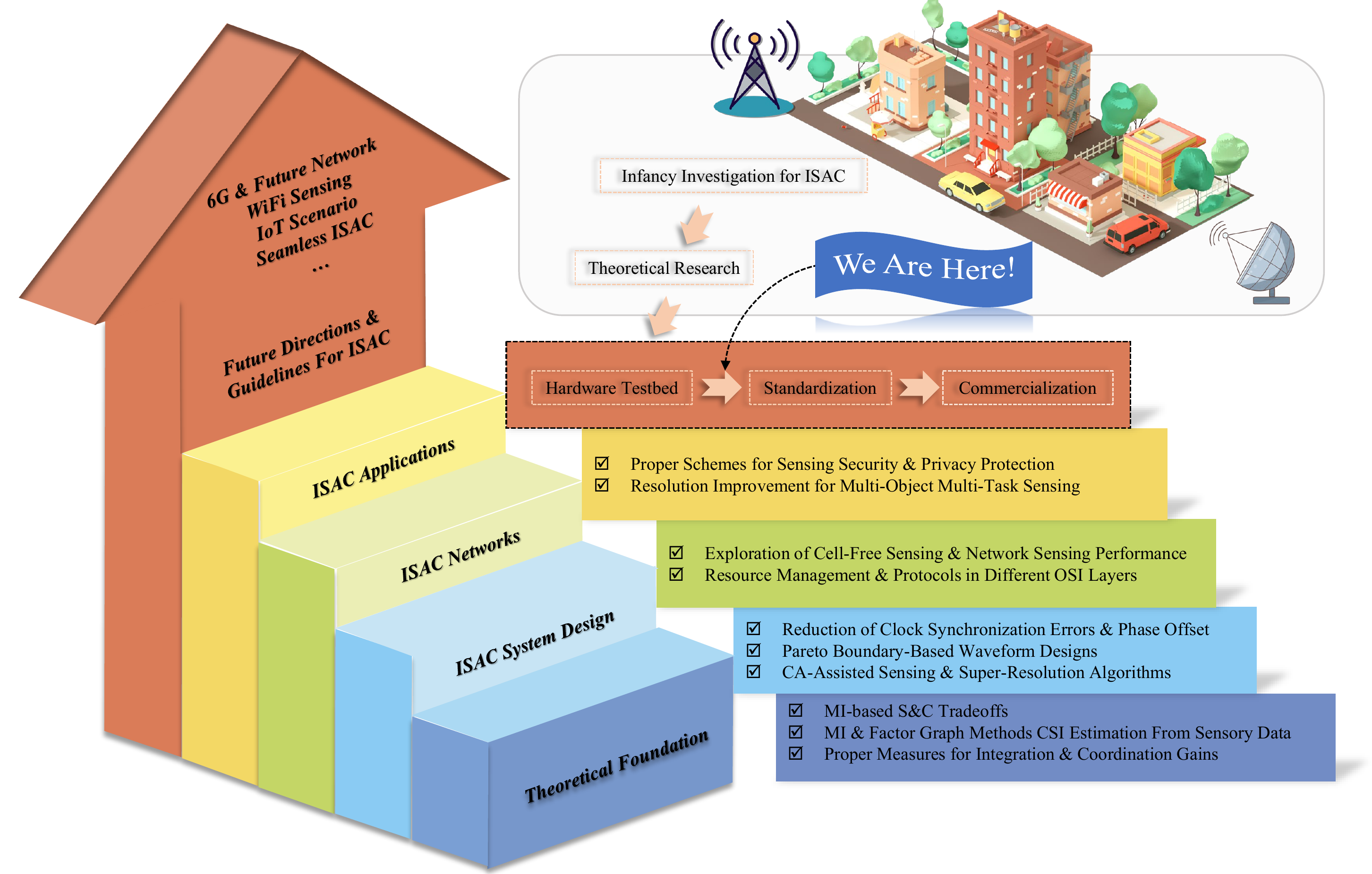}
		\caption{Design guidelines and future research directions for ISAC systems.}
		\label{Guidelines}
	\end{figure*}
	
	Although a wide variety of solutions have been proposed for wireless human sensing \cite{lixinyu2022humanactivity,10.1145/3210240.3210314,10.1145/3351279,chen2018wifi,wu2021wifi}, there are still numerous open problems to be solved. For instance, most existing work merely focuses on single-person sensing \cite{wang2022single}. Initial attempts concerning multi-person sensing tasks were carried out under the assumption that the signals reflected from different persons are independent and separable \cite{zeng2020multisense}. In cellular networks, beamforming is capable of suppressing clutter \cite{lu2022degrees}, while carrier aggregation technology is promising in terms of increasing the range resolution, hence it is promising for improving the performance of multi-person sensing. Additionally, device-free and device-based solutions \footnote{Device-free sensing involves detecting changes in the environment caused by the presence of people or objects without the use of dedicated sensing devices. On the other hand, device-based sensing involves using dedicated sensing devices, such as cameras or sensors, to directly measure the properties of the environment.} can be combined for sensing multiple persons. Specifically, by associating the device-free reflected signals and the device-based tag information (e.g., the ID of mobile phones), multiple persons can be separated more easily. Furthermore, the low spatial resolution of the current ISAC devices (e.g., WiFi and portable radars) limits the multi-target separation capability of these wireless sensing systems. In future research, we need to improve the spatial resolution of wireless sensing systems and devise solutions for MOMT sensing.  {For example, we have to investigate the potential communication-assisted sensing by utilizing CSI to boost the sensing performance, where the receivers may analyze a variety of channel features to detect the presence and location of multiple targets.} In some applications, combining these solutions with cameras are expected to be beneficial \cite{zheng2023vision}.  {In B5G and 6G, one needs to investigate more advanced technologies to enable the emerging sensing scenarios where extracting human-related features and performing activity recognition are difficult, such as integrated human sensing and communications, ultra-reliable and low-latency human sensing \cite{li2019survey}, and non-line-of-sight human sensing \cite{lixinyu2022humanactivity}.}

	\section{Design Guidelines and a Brief Summary}\label{Conclusion}
	In this section, we first provide some general design guidelines and future directions for ISAC systems based on the ten challenges discussed in Fig. \ref{PaperFramework}, which are summarized in Fig. \ref{Guidelines}. Then, we conclude by highlighting a range of take-home messages.

	\begin{figure}
		\centering
		\epsfxsize=1\linewidth
		\includegraphics[width=\linewidth]{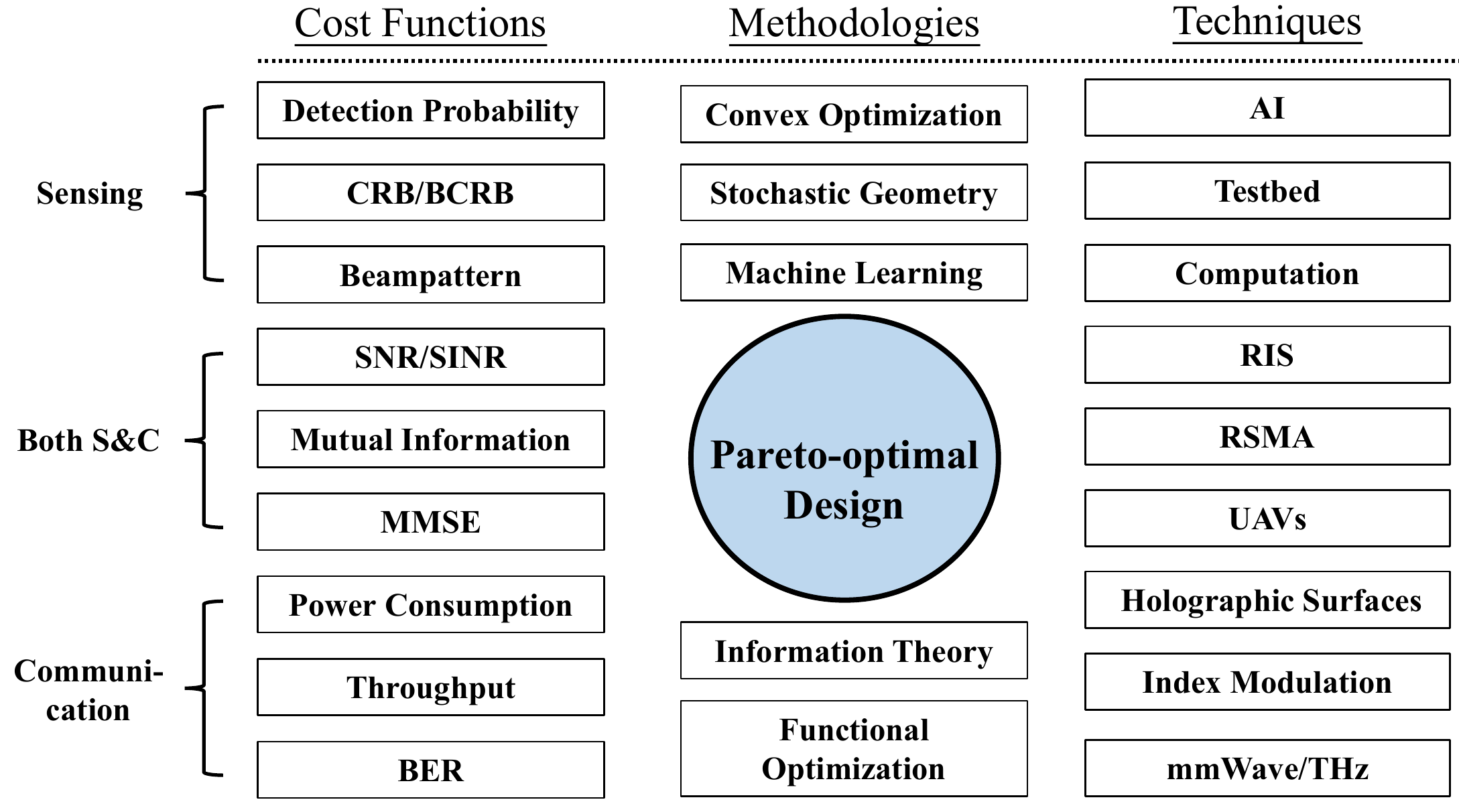}
		\caption{Stylized factors affecting the Pareto-optimal design of ISAC systems.}
		\label{Technique}
	\end{figure}  
	
	\subsection{Design Guidelines}
	As evidenced by more and more emerging 6G white papers and WiFi 7 
	papers, ISAC has drawn significant attention from major industrial enterprises, and its standardization is also under discussion in 3GPP \cite{Sta_tan2021integrated,Ericsson}. As shown in Fig. \ref{Guidelines}, we touched upon the most influential design factors based on the discussions in Section \ref{IF}-\ref{Security_Wifi}, ranging from the theoretical foundations of ISAC, ISAC system designs, ISAC networks, and ISAC applications. 
	
	$\bullet$ {\it Theoretical Foundations of ISAC:} The holistic design of ISAC systems has to take the theoretical foundations into consideration. The role of MI in radar sensing systems has to be clarified first, so that one may quantify the MI-based S$\&$C tradeoffs in the design of ISAC systems. As a further step, it is pivotal to infer the inner linkage between sensory data and communication CSI by relying on the MI-based approach. Furthermore, it is important to characterize both the integration and coordination gains by exploiting {\it ``simple but intuitive''} metrics.
	
	$\bullet$ {\it Physical-Layer System Design:} As presented in Section \ref{System_Design}, we have discussed three key aspects of ISAC system designs, including clock synchronization, Pareto-optimal signaling strategies and super-resolution methods. In order to reliably integrate the S$\&$C components, it is essential to reduce the clock synchronization errors and phase offset in distributed deployment scenarios in support of high-precision ranging \cite{zhang2022integration}. It is also important to explore the entire Pareto-front of optimal ISAC solutions, while gradually including an increasing number of S$\&$C metrics, as the technology evolves \cite{2021chenli_Pareto}. Finally, super-resolution algorithms, such as CA-assisted radar sensing, may be exploited for improving the sensing performance in a variety of emerging applications, such as V2X and IoT scenarios \cite{JSAC_liu2022integrated,IoTJwang2022integrated}.
	
	$\bullet$ {\it ISAC Networks \& Cross-Layer Design:} To incorporate an ISAC capability into the existing infrastructures, cell-free sensing has to be explored  \cite{cellfreebehdad2022power}, in which the sensing capability may be provided as a basic service. In contrast to the ubiquitous communication service, the sensing service tends to be performed in a more random bursty manner. Hence, the joint resource management schemes and protocols tailored for ISAC networks are highly desirable and are expected to lead to compelling services.
	
	$\bullet$ {\it ISAC Applications:} The ISAC network of the future has to be rolled out with security and privacy guarantees  \cite{weizhongxiang2022toward,su2020secure}. Some potential directions relying on the optimization of both the PHY and MAC layers have been discussed in Section \ref{Security_Wifi}. Furthermore, to realize practical ISAC services relying on commercial wireless devices, it is critical to improve the sensing resolution for supporting MOMT sensing, in both the smart home and in smart city applications \cite{IoTJwang2022integrated,li2019survey,zhao2021human}. 
	
	In addition to the proposed design guidelines and directions of ISAC systems, substantial efforts are required for promoting the standardization and commercialization of ISAC. As illustrated in Fig. \ref{Technique}, we highlight the general philosophy of the Pareto-optimal design for ISAC systems, with a glimpse of its pivotal factors. Specifically, we should choose multi-component cost functions bearing in mind the salient design perspectives as well as methodologies, including convex optimization \cite{liu2021cramer,su2020secure,liuxiang2020joint}, stochastic geometry \cite{chenwenrong2022isac}, machine learning and so on \cite{demirhan2022radar,OJCOMS_wu2022blockage,alkhateeb2021deepsense}. Furthermore, artificial intelligence (AI) and testbed based on practical data sets are required \cite{zhuguangxu2022ISCC,IoTJwang2022integrated,OJCOMS_wu2022blockage,alkhateeb2021deepsense}. Furthermore, in-depth over-the-air ISAC and computation investigations are necessitated \cite{lixiaoyang2022over}, while relying on the latest advances both in S\&C, such as reflecting intelligent surface (RIS)-assisted ISAC \cite{liurang2022integratedris}, rate-splitting multiple access (RSMA)-aided ISAC \cite{maoyijie2022rate}, UAVs equipped with ISAC in the sky \cite{meng2022uav}, and holographic ISAC \cite{zhang2022holographic}, just to name a few.
	
	\subsection{Summary}
	In this paper, we critically appraised the recent advances and formulated ten open challenges in ISAC systems, some of which have already made initial progress, while others are still in the exploratory phase. 
	Firstly, we introduced the theoretical foundations of ISAC systems, starting with an introduction to \textbf{\emph{Challenge 1}} on fundamental theory that concerns the limitations of S\&C performance. Then in \textbf{\emph{Challenge 2}}, we discussed how to infer the CSI from the sensory data and presented several potential solutions and future directions. In addition, we presented \textbf{\emph{Challenge 3}}, which focuses on integration and coordination gains, and proposed an informal metric to quantify them. 
	Furthermore, we have addressed the design issues of ISAC systems in the context of \textbf{\emph{Challenges 4-6}}. More specifically, we have elaborated on the open problems of clock synchronization, Pareto-optimal signaling strategies, and super-resolution methods, along with their potential solutions and future directions. 
	We then continued by shifting our focus to ISAC networks. By considering {\em sensing as a service} in the cellular network of the future, we investigated the potential cellular architectures as well as the cross-layer resource management and the protocol design of networked sensing in \textbf{\emph{Challenge 7}} and \textbf{\emph{Challenge 8}}, respectively. 
	Next, we concentrated our attention on attractive ISAC applications. We highlighted the associated sensing security and privacy issues, proceeding by presenting the corresponding future directions from the perspective of the PHY and MAC layer in \textbf{\emph{Challenge 9}}. Finally, we touched upon a human activity sensing scenario relying on wireless signals and discussed the open issues of MOMT sensing in \textbf{\emph{Challenge 10}}.
	
	\bibliographystyle{IEEEtran} 
	\bibliography{ref}
\end{document}